\documentclass[runningheads,envcountsame,envcountsect,orivec]{llncs}
\usepackage{cmap}
\usepackage[T1]{fontenc}
\usepackage{newtxtext}
\usepackage{dsfont}
\usepackage{environ}

\input{definitions.sty}

\renewcommand{\Pr}{\mathop{\bf Pr\/}}
\newcommand{\E}{\mathop{\mathbb{E}\/}}  %
\newcommand{\Var}{\mathop{\bf Var\/}}

\newcommand{\supp}{\textnormal{supp}}

\newcommand{\bone}{\boldsymbol{1}}

\def\<{\langle}
\def\>{\rangle}

\newif\ifshowcomments
\showcommentsfalse

\ifshowcomments
\newcommand{\jason}[1]{\textcolor{blue}{\textbf{[Jason: }#1]}}
\newcommand{\mike}[1]{\textcolor{red}{\textbf{[Mike: }#1]}}
\newcommand{\lioba}[1]{\textcolor[HTML]{4169E1}{\textbf{[Lioba: }#1]}}
\newcommand{\kushal}[1]{\textcolor{orange}{\textbf{[KB: }#1]}}
\else
\newcommand{\jason}[1]{\ignorespaces}
\newcommand{\mike}[1]{\ignorespaces}
\newcommand{\lioba}[1]{\ignorespaces}
\newcommand{\kushal}[1]{\ignorespaces}
\fi

\provideboolean{revisionversion}
\setboolean{revisionversion}{true}
\ifthenelse{\boolean{revisionversion}}{
  
  \newcommand{\delt}[1]{{\color{orange} \sout{#1}}}
}{
  
  \newcommand{\delt}[1]{\ignorespaces}
}

\newcommand{\defeq}{\ensuremath{\triangleq}}

\titlerunning{Bayesian MCP TFMs}
\authorrunning{}

\begin{document}

\title{Bayesian analysis of inclusion transaction fee mechanisms \\ with multiple concurrent proposers}

\author{Jason Milionis \and Kushal Babel}
\institute{Category Labs \\
    \email{\href{mailto:jm@cs.columbia.edu}{jm@cs.columbia.edu},
        \href{mailto:babel@cs.cornell.edu}{babel@cs.cornell.edu}}
}

\date{}
\maketitle

\thispagestyle{empty}

\begin{abstract}
The goal of this paper is to distill insight around the trade-offs present on Bayesian equilibria of transaction fee mechanism (TFM) design under $n$ multiple concurrent proposers (MCP).
We conceptualize a technical toolkit that unlocks the analysis of hard-to-characterize equilibria of finite Bayesian MCP TFM games by examining sequences of joint user--proposer mixed equilibria as the number of users increases and keeping the relative congestion $\rho$ that they cause fixed in the sequence limit.
This allows us to analyze both congested and uncongested regimes.
We first consider the popular mechanisms of ``pay-once'' (which charges transactions its bid only once regardless of the duplicate copies that end up being included by proposers) and ``pay-per-inclusion'' (PPI in short, which charges a transaction for each copy that appears).
We prove that PPI is efficient in the limit of equilibria sequences, and incentivizes users (who have free choice to choose the number of proposers they desire to route their transaction to) to route to one proposer.
We characterize $m$ users' asymptotic behavior in PPI even in finite markets varying with $O(\sqrt{m})$, where we show users in a $\Theta(n\sqrt{\rho(1-\rho)/m})$ window near an efficient cutoff exhibit fixed inclusion probability akin to a normal distribution's CDF.
We additionally show that on a value-dependent basis, pay-once exhibits a higher degree of duplication and lower probability of inclusion for higher-value transactions than PPI.
Pay-once's inefficiency persists even in the large market limit and even in regimes with ample blockspace where all transactions fit across all blocks.
We study interpolations between these two mechanisms proving that some higher coefficients (towards PPI) reduce duplication and increase inclusion probability.
We then use our tools to show that exclusive proposer payments yield inefficient equilibria (in fact, with maximal possible duplication), even if users still pay multiple times when included more than once and their payment is burnt when non-exclusive.
As an interesting special case of our full set of results, we show that a dynamic base fee that posts a price resulting in ample blockspace yields efficient and optimal transaction inclusion for the pay-per-inclusion MCP transaction fee mechanism with a positive duplicate base fee.
\end{abstract}

\section{Introduction}

Multiple concurrent proposer (MCP) blockchain systems are a core primitive at the forefront of recent efforts in blockchain infrastructure \citep{solana,monad,aptos,sui,sui_barnacle_no_tfm,sui_sedna_future_tfm}.
In these systems, as opposed to single-leader based protocols, there are $n>1$ proposers who \emph{simultaneously} propose transaction batches, the union of which proposals is then typically considered the ``block.''
This allows block building to escape the control of a single entity, instead being controlled by a decentralized set of participants.
One of the core components of any such system is its transaction fee mechanism (TFM). While they have garnered some recent attention \citep{matt_itcs,saraf}, the Bayesian analysis of such MCP TFMs has remained elusive, primarily due to its complexity in settings similar to those analyzed in the past by \citet{gafni_bayesian,chen_bayesian,wu_revenue,leaderless,matt}.
This paper examines and resolves the analysis and equilibria of the most widely considered MCP TFMs, as well as interpolations between them.
To do so, we introduce techniques of independent interest, in analyzing sequences of equilibria of finite games as a technical toolkit, and studying fixed relative congestion at the limit as these finite games tend towards a continuum of users.\footnote{We argue that analyses of the continuum sometimes exhibit different and non-overlapping equilibria (including structure of these), and as such, might not sufficiently indicate the real-world behavior of finite games; limits of these finite games that keep relative congestion fixed as the number of users' transactions increases may be more indicative of the actual behavior.}
These two key technical tools allow us to proceed to give the fundamental properties of the underlying MCP TFMs in a Bayesian setting.

\subsection{Our contributions}
We evaluate and formally derive equilibria under different regimes that cover both congested and uncongested blockspace. We use the standard Bayesian model of transactions' values, where all values are sampled independently and identically distributed according to a distribution.
Such settings have remained elusive with multiple concurrent proposers predominantly due to the technical difficulty of dealing with both finite games (the equilibria sequences of which we examine) and probabilistic mixed routing strategies (for example, past models \citep[such as][]{saraf} have only looked into the proposer side, while assuming exogenous aggregate bids not determined at equilibrium).
We center our work on corners of the design space that are under consideration in many blockchain deployments \citep{solana,monad}:
\begin{enumerate}
\item Pay-per-inclusion (PPI in short), where every transaction is being charged its full bid every time it is being included,
\item Pay-once, where, regardless of the duplicate times a transaction gets included, it only pays its bid once, and its payment is split among all including proposers, and
\item Exclusive variants of those (``exclusive'' meaning that proposers obtain revenue conditioned on unique inclusion with no duplicates, otherwise the transaction may still get charged or not but the fee not given to the proposers, e.g., is burnt).
\end{enumerate}
Our results include cases for \emph{base fees}, whether these are charged once or multiple times, which are additive to the above payments to the proposers and are burnt (or equivalently, not provided to the including proposers).
We also examine all mechanisms that are arithmetic interpolations between these corners of the design space. As such, we exhibit a rich mixture of transaction fee mechanisms.\footnote{This mixture covers most commonly-examined cases of MCP TFMs, but is still not fully general; as our technical analysis gets incredibly involved, in this work, we mainly aim to introduce the technical toolkit for looking at equilibria sequences of these mechanisms, along with the most important properties about each of our focused TFMs. We leave to interesting future work using our toolkit to examine more general TFM primitives.}

\paragraph{Model of user strategies.}
More specifically, we are concerned with symmetric equilibria of finite MCP TFM inclusion games (finite number of users $m$ sending transactions).
In this context, broadcasting every transaction to every proposer is potentially highly suboptimal when user bids are endogenous; for example, in PPI, users might want to strategically choose to allocate themselves to (relatively uncongested) proposers.
Hereby, we will use the most natural symmetric model wherein transactions (equivalently, users) choose a number $k$ of proposers (we call them \emph{routes}) that they wish to appear at, chosen uniformly at random.\footnote{Asymmetric equilibria are typically hard to analyze in auctions \citep{auction}, and sometimes imply more coordination that can traditionally be achieved. On the other hand, among all \emph{symmetric} models, our model of route choice is a natural one.}
More precisely, in our model, nature chooses a random subset of $k$ proposers (where the number $k$ is the user's strategy) without replacement to send a user's transaction to.
Users can choose to make their strategy mixed, i.e., choose both bids $b$ and routes $k$ with some probability over each, for example choosing 30\% of the time to send to 1 (randomly-chosen) proposer with a particular bid $b_1$ and 70\% of the time to send to 2 (randomly chosen) proposers with a (generally different) bid $b_2$ to each.
As a result, users' strategies are probability distributions over pairs $(b,k)$.

\paragraph{Our techniques.}
In general, analyzing finite games in such a Bayesian setting is technically complicated, because the exact numbers of users $m$ and proposers $n$ can affect the precise format of the game, including the users' route choices.
We propose, in this work, overcoming this burden (crucially, without restricting the usefulness of the congestion caused by a finite number of transactions and available blockspace) by studying all limiting regimes of exact finite-game Bayesian Nash Equilibria with a fixed number of proposers where the \emph{relative congestion} caused by the finite nature of available transactions and finite total capacity is kept \emph{constant} in that sequence.
One can equivalently view this technique of ours as inducing a structure of sequences of games where all exhibit the same aggregate network congestion effect, the equilibria of which we can characterize.
These limiting sequences are still hard to identify; however, we prove these limits not by directly attempting to solve finite games (even at a large participating number of transactions), but by first examining a continuum game\footnote{Parallels to our continuum game can be drawn to a generalization of the continuum game looked at by \citet{matt_itcs}.} where the effect of the strategic choices of any single transaction is \emph{assumed} to be non-existent in every proposer and every other users' utility. On its own, this directly allows to compute the probabilistic laws of that game and the optimal user strategies. Nonetheless, note that this assumption is in general invalid in finite games, i.e., in no finite game is it the case that a single transaction's deviation does not affect proposer loads and in turn the user's behavior itself.
This is also why there are many more equilibria to this assumed game than limits to actual exact finite game Bayesian Nash Equilibria.
Afterwards, concentration bounds allow us to transform \emph{some} of these ``guessed'' equilibria to actual limits, and furthermore, show that the congestion structure they generate must necessarily lead to the particular continuum-type we found; this part is a meaningfully technically challenging part, as evidenced by the multiplicity of continuum-game equilibria which are not limits of finite games.
Broadly, this latter concentration part involves identifying rare congestion events that become payoff-equivalent in the continuum.

\subsection{Our results}

There are \(n\ge2\) proposers and \(m\) users, each of which generates a transaction (when it is clear from context, we might refer to user and transaction interchangeably).
Each proposer has \emph{capacity} to include up to \(c_m\) transactions.
We define $\rho \defeq c_m / m$ to be the relative congestion of each proposer, and $\beta \defeq n\rho$ to be the total relative congestion of the system as a whole (which can fit up to $n c_m$ transactions).
We assume user values are independently and identically distributed with density \(f\) and cumulative distribution function \(F\).
A user with value \(v\) chooses an action \((b,k)\in\mathbb R_+\times\{0,1,\dots,n\}\).
If \(k>0\), nature selects \(k\) distinct proposers uniformly without replacement and shows them the transaction; if \(k=0\) the transaction is hidden.
We will call $k$ the chosen routing number (or chosen routes) of a user.\footnote{This model corresponds to an \emph{anonymous routing} model, which is fitting in the symmetric equilibria sequences we examine.}

If \(K\) proposers include the same transaction (we will call $K$ the \emph{realized inclusions} or duplicates or copies of a transaction), since the transaction obtains its inclusion value once,\footnote{
We note that this work is all about the game of transaction \emph{inclusion} and not about sequencing or ordering preferences, like, for example, \citet{order_auctions} in the single-proposer setting.}
the user's payoff is $v\bone\{K\ge1\}-p(b,K)-g\cdot G(K)$,
where $p(b,K)$ is the MCP TFM's payment rule to proposers depending on the bid $b$ and realized inclusions $K$,
and $G(K)$ is what we will call the \emph{base fee schedule};
the entire base fee payment $g\cdot G(K)$ is burnt.
Throughout this work, we assume the total payment is given equally to all $K$ including proposers, i.e., each proposer who includes a transaction copy obtains $p(b,K)/K$ (respectively, on the ``exclusive proposer rewards'' model we will examine below, proposers obtain $p(b,K)\cdot\bone\{K=1\}$; the rest is burnt, please see below for more details).

We study symmetric Bayesian equilibria of the finite game, whenever they exist,\footnote{While we cannot guarantee existence for all types of the various games we analyze, our characterizations always hold whenever equilibria exist. We refer the interested reader to \Cref{app:existence}.} and let \(m\to\infty\) with \(n,\rho\) fixed, i.e., we keep the relative congestion of the system fixed as we scale the number of transactions.\footnote{
We make the standard assumption that bids in these \emph{finite} games are atomless. Note that this does not imply that the limits of these sequences of such atomless equilibria, which we will subsequently take, are also necessarily atomless; in fact, as we will see, they have atoms, for example, the pay-per-inclusion MCP TFM with a fixed base fee $g$ exhibits fixed bids of $v^\star - g$ for included transactions.}
Our theorems will unveil multiple symmetric equilibria sequence limits in some mechanisms, with different degrees of duplication. Hence, only examining a single equilibrium or assuming uniqueness of equilibria sequences would be insufficient to obtain a good understanding of the design landscape.

Many quantities will depend on users' values. The first relevant quantity of interest is the \emph{inclusion probability} of a user, i.e., $\Pr[K\ge 1 \mid v]$; higher value generally implies larger inclusion probability. We will compare the inclusion probabilities of different TFMs.
For quantifying the degree of duplication in the system, we call the expected realized inclusions (conditional on inclusion), i.e., $\mu\defeq\E[K\mid K \ge 1]$ the \emph{conditional multiplicity}.
If we additionally condition on value, i.e., $\E[K\mid v, K \ge 1]$ then this gives the value-dependent conditional multiplicity: this is going to be important in our study as different users' types result in quite different duplicate included copies.

Finally, we represent $w\defeq v-g$ as the ``net value'' of a transaction after the lowest amount of base fee eligible to be paid.

\subsubsection{Under congestion}
First, we consider the case where $\beta = n c_m / m < 1$;\footnote{This is a simplification with zero base fees for expository purposes of this introduction. As our main body shows, with a base fee of minimum paid amount $g$, the congested regime is when $\beta < 1 - F(g)$.} this is the congested regime, because the total capacity of all proposers does not suffice to post all interested transactions.\footnote{This is the typical setting considered in much of prior single-leader work \citep{gafni_bayesian,chen_bayesian}.}
On the one hand, PPI incentivizes capacity-constrained users (who can expect to be included if and only if their transaction's value is above an efficient cutoff $v^\star$) to route randomly to \emph{just one} proposer, and we show that transaction duplication (out of those who get included) is asymptotically non-existent.
On the other hand, pay-once incentivizes users to route their txs to \emph{all} proposers, while also increasing the degree of duplication; in particular, we formally prove that even on the asymptotic limit of finite games, there is leftover \emph{inefficiency} (meaning that not all values that could otherwise be included actually end up being included) and the expected duplication cannot match the asymptotically-optimal one of PPI. This gives crucial insight for the real-world behavior around these TFMs.

To ease into our results, we first give an illustrative example in the asymptotic limit:\footnote{We stress that this is only an example, and our theory handles completely generically any value distribution.} consider uniform values $V\sim U[0,1]$, $n=2$ proposers, no base fees, and each proposer offers $\rho=40\%$ capacity. This means that, across both proposers, there is space for at most $\beta=n\rho=80\%$ of transactions. For the pay-once MCP TFM, even for values $v$ that approach the absolute maximum of $1$, their inclusion probability (that they succeed) tops out at 93.6\%, and their duplication conditional on inclusion is expected to be around 1.6 copies. To the contrary, for the PPI mechanism, every value $v>0.2$ has 100\% inclusion probability (efficient) and exactly 1 posted copy.

\paragraph{Pay-once.}
Thus, the economic inefficiency of pay-once in wasting capacity even for high participating types persists despite the large market limit. More precisely, we obtain the stronger result that \emph{every} served finite value $w>t$ retains failure risk of
$ \Pr[K=0\mid w]=\left(\frac tw\right)^{n/(n-1)}>0$, where $t$ is a cutoff value.
As a result, increasing the size of the market alone does not give users certain inclusion under a pay-once MCP TFM; to the contrary, PPI gives certainty above its efficient cutoff.

Prior work \citep{saraf} indicated that the number of copies may be well-controlled for pay-once, and---perhaps surprisingly---that coordination may be achieved even when capacity is divvyed up to increasingly many proposers.
This should be surprising, because in pay-once, all these proposers receive all transactions (as we show), so it is conceivable that they might have an issue choosing non-overlapping subsets among all of them as the same total capacity is divided to an increasingly high number of proposers.
We show that formally examining the equilibria reveals a more nuanced reality, where sometimes coordination cannot be achieved in the pay-once mechanism: either for sufficiently heavy-tailed distributions and/or fixed non-zero relative total congestion $\beta > 0$ (as opposed to $\beta\to 0$), one should expect the duplication in the system to be arbitrarily bad.
For example, formally, we show that holding the total capacity $\beta>0$ fixed and as the number of proposers $n\to\infty$, the expected realized inclusions conditioned on value $\E[K \mid v] \to \log((v-g)/T)\sim O(\log v)$ which can be unbounded for unbounded value distributions (or arbitrarily high, compared to 1 inclusion for PPI), where $T$ is a fixed constant.

Considering now the case when the total relative congestion instead goes to zero ($\beta\to 0$), there is a bifurcation: whereas for sufficiently heavy-tailed distributions duplication can still be maximal (or even unbounded, like for the power law, as the number of proposers grows; we show in \Cref{cor:tails,po-pareto} when this occurs), if the net value distribution (after base fees) is upper-bounded by a value $\bar w$, realized inclusions are asymptotically $1+\Theta(\sqrt{\beta})$, which is indeed close to 1, or to be more accurate they are:
\[ \mu
=
1+\frac13
\sqrt{\frac{2(n-1)}{n\,\bar w f_W(\bar w)}\,\beta}
+o(\sqrt{\beta})
. \]

Regardless of what the belief might be about which of these regimes holds in practice, we fully characterize all the cases of pay-once that could happen.
To show a parametric example of our general calculations, if the value distribution is a Pareto distribution with exponent $\alpha$, we show that, as $\beta\to 0$, the conditional multiplicity goes to
\[ 1+\frac{1}{\alpha+\frac{1}{n-1}} \text{ in closed form.} \]

\paragraph{Pay-per-inclusion.}
As described above, to a first order (in what we call the ``layer 1''), above the efficient value cutoff $v^\star = \max\{g,F^{-1}(1-\beta)\}$, users route to only $k=1$ proposer at equilibrium, bid $b=v^\star-g$, get 1 realized inclusion with 100\% inclusion probability, while users below the cutoff get no copies.
Finite markets reveal what we term a \emph{``layer 2'' asymptotic result:}
close to the efficient cutoff, for every finite $x\in\mathbb{R}$, as $m\to\infty$, the user with value $v=v^\star + h_m\cdot x$ has as optimal strategy to still route to $k=1$ proposer, obtains inclusion probability $\Phi(x)+o(1)$, and bids
$b=v^*-g-h_m\phi(x)/\Phi(x)+o(h_m)$, where $\phi$ and $\Phi$ are the
standard normal density and CDF, and
\[h_m=\frac{n\sqrt{\rho(1-\rho)}}{f(v^\star)\sqrt{m}}.\]
This means that there is a well-specified band of users in finite markets that accept meaningfully non-zero, non-maximal inclusion probability of $\Phi(x)$ whose behavior we can characterize well and in closed form.
Even further below the cutoff, as the market grows, we show that users under PPI \emph{send to all proposers,} rather than any intermediate quantity of proposers (or even mixed strategies). Intuitively, this happens because the asymptotic rate in user utility due to the inclusion probability is of higher order than the rate of duplicate inclusion.

\paragraph{Interpolating TFMs.}
Next, we consider MCP TFMs that arise as linear interpolations between PPI and pay-once: affine TFMs where transactions pay to proposers (equally to each who included them) $b\cdot (1+\eta_b\cdot (K-1))$ when they get included $K\ge 1$ times, where $\eta_b\in [0,1]$: $\eta_b=0$ yields pay-once and $\eta_b=1$ yields PPI.\footnote{Our total user payment accounts in fact for interpolations between base fees as well, i.e.,
$C(b,K)=(b+g)\mathbf{1}_{\{K\ge1\}}+(\eta_b b+\eta_g g)(K-1)_+$,
where $\eta_b, \eta_g\in[0,1]$.}
While the structure of equilibria there cannot be readily analyzed, we derive a few interesting lemmas:
\begin{itemize}
\item As one might expect, as the bid rule approaches pay-per-inclusion ($\eta_b\to 1$), every equilibrium sequence of these affine TFMs has the same efficient cutoff and 1 inclusion per served user as PPI.
\item Given the above lemma, as well as the inefficiency of pay-once, one might expect that there is some inherent monotonicity in the inclusion probability and expected realized inclusions with the coefficient $\eta_b$.
Unfortunately, this is strongly not true: we prove the strong negative result that, for every pair of TFMs given by coefficients $0 < \eta_1 < \eta_2 < 1,$ there exists a distribution $F$ such that, for an equilibrium sequence of both TFMs, all values larger than a constant (even the high values) exhibit lower inclusion probability and higher duplication at $\eta_2$ than the equilibrium sequence of $\eta_1$.
\item Despite the above result, going arbitrarily close to PPI is indeed better than any other affine TFM with coefficient $<1$: formally, if we take any baseline equilibrium sequence of a TFM with coefficient $\eta_b<1$, there exists a coefficient $\eta_b' \ge \eta_b$ (read: a TFM closer to PPI) such that every equilibrium of every affine TFM with any coefficient higher than $\eta_b'$ has lower aggregate degree of duplication and higher aggregate probability of inclusion.\footnote{Upon a first look, a reader might comment that this appears contradictory: it is not, because the latter lower bound is distribution-dependent. This speaks to the interesting structural nature of the equilibria we describe; they are highly non-trivial.}
\end{itemize}

\paragraph{Exclusive proposer rewards.}
We now give a few headline results about three variants of TFMs where a proposer only earns the transaction's bid if they are the transaction's sole includer:
\begin{itemize}
\item Charging users for every copy (akin to PPI) and giving the proposer the bid only when unique no longer ensures low duplication at equilibrium (PPI had no duplicates, this is no longer the case).
In fact, these exclusive rewards can get arbitrarily bad: we manage to exhibit multiple feasible equilibria sequences where duplication is close to the maximum $n$ copies for each included user.
\item Both of the two above observations hold even when users pay only for a unique inclusion and are not charged when duplicates exist.
\item If the user pays once (akin to standard pay-once), all users send to all proposers, and duplication is still present alongside inefficient transaction inclusion.
\end{itemize}
Our results are stable with any base fee schedule (including for the above exclusive TFMs), except wherever pay-once is used as the proposer payment method, where we use base fees also paid once (otherwise the combination of paid-once proposer payments with non--paid-once base fees simulates aspects of PPI payments, as base fee now charges for duplicates, which we believe makes sense to restrict to affine or PPI payments).

\paragraph{Summary.}
Overall, our results reveal that the closer the MCP fee mechanism is to pay-once, the more the burden of optimization is transferred to proposers---who may have little time to coordinate among fast alternating slots---rather than users (or their representatives, such as RPCs) whose incentives are more dominating in the case of TFMs close to pay-per-inclusion.

\subsubsection{Ample blockspace}
Building on the same primitives that we introduced to analyze the congested regime, we examine the ``ample blockspace'' regime, where there is enough \emph{total} capacity (across all proposers) to fit all transactions, but not enough at each individual proposer, i.e., $1 < \beta < n$.\footnote{More precisely, with base fees, this condition is $1 < \frac{\beta}{1-F(g)} < n$. It is clear that duplication is unavoidable in the case of ``even more ample blockspace'' where there is space for the full transaction load on each individual proposer; equilibria there commonly have full conditional multiplicity of $n$ duplicates posted and efficient inclusion. We do not study this latter comparatively simple case.}
We show that some of the results do change here.
Overall, however, it is important to highlight that, even in this regime, pay-once still produces duplicates and imperfect inclusion, whereas PPI does not with non-trivial base fee schedules.

\begin{itemize}
\item PPI leaves surplus capacity unused, and equilibrium sequence bids of transactions tend to zero exponentially with the number of transactions $m$. Also, any base fee charge with $g\cdot [G(r)-1] > 0$ for every $r=2,\cdots,n$, i.e., any weak penalty for duplicate inclusion (such as, but not only, charging the base fee for every inclusion), yields users choosing to route to one proposer and be included once efficiently in the limit.
\item Pay-once fills the full capacity but serves fewer than the efficient proportion of users. Hence, its inefficiency persists despite the ample blockspace.
\item We show a methodology for finding exactly how much duplication exists in the limit, with a single base fee charge. As an example, for PPI under ample blockspace, perhaps as expected, duplicate transactions mirror the available capacity excess to a first order. Pay-once comparatively exhibits higher duplication strictly separated from PPI, i.e., formally $\mu_\text{PPI} < \frac{\beta}{1-F(g)} < \mu_\text{once}$.
\item As an interesting special limiting case, as the capacity of each proposer increases to almost fit the entire transaction load (upper limit of the ``ample blockspace'' regime, i.e., $\beta\to n$), the equilibrium sequences show that: (1) in the limit of PPI, approximately half of the proposers include each user, whereas (2) in the limit of pay-once, approximately all proposers include each user (maximal duplication).
\item At the other boundary, when total capacity is just enough to include all transactions (i.e., $\beta\to 1^+$), substantial pay-once duplication is still the case: the conditional multiplicity of pay-once can be at least $\ge\frac{1}{1-(1-1/n)^n}$ which approaches 1.58 as the number of proposers grows, whereas PPI's instead approaches 1.
\end{itemize}

Our results further imply that, when the base fee is dynamically adjusted such that we are in the ample blockspace regime analyzed herein as sometimes assumed in EIP-1559--like ``posted price'' mechanisms \citep[see, e.g.,][]{tim}, PPI with a non-trivial base fee yields efficient and optimal transaction inclusion, and incentivizes users to only route to one proposer with minimal inclusion bids.

\subsection{Related work}
\label{subsec:litrev}

Closest to our work, part of our results generalizes \citet{matt_itcs}'s continuum of unit-demand buyers bidding in multiple simultaneous first-price auctions (PPI with no base fee, hereby) to finite markets.
Our PPI results also hold with any common reserve $r$ among all proposers, proving convergence of finite markets to their continuum outcome in that setting as well. We refer the interested reader to \Cref{app:common-reserve}.
We show that, as the market approaches the continuum, our PPI mechanism with a common reserve approaches the single-proposer allocation of \citet{LWZ}; however, for values near the cutoff, we uncover a Gaussian inclusion probability behavior that is not obtainable in any continuum model, but only in finite actual markets. At the core of our work is also the study of a much broader class of MCP TFMs to give insight and their equilibrium properties.

\citet{saraf} also study the duplication (which they call throughput) of an overlapping set of TFMs to our work, as well as the cost of bribing concurrent proposers to censor a transaction, assuming an exogenously-specified user bid distribution and users sending to all proposers.
They are exclusively concerned with the proposer--side in a non--full-equilibrium model (the full scope of which is our work).
They also, importantly, assume a fixed bid list that all proposers receive (which they call a ``shared mempool''); as we show, this is crucially not the case in full equilibria with some TFMs (like pay-per-inclusion) where users instead prefer to route to a subset of proposers (behavior which also depends on the chosen TFM), making their bids invisible to other proposers including for their optimization.
More consequentially, our joint equilibrium Bayesian analysis allows us to give insight that a proposer--side model is unable to establish: we can extract the choices of different users and perform comparisons between them (including to reason about the efficiency---in the economic sense---or not of MCP TFMs), leading to, e.g., how duplication and inclusion probability depends on user value.
Separately, we do not engage in surveying the censorship resistance of the mechanisms herein, as it is not the main goal of our work.
The core argument is that we believe that, in order for a censor model to be reasonable to use in practice, it needs to incorporate the censor inside the baseline game of transaction inclusion (as opposed to a separate game than the baseline transaction inclusion). This is not reflected in models like the above, and would require the analysis of extended multi-stage games that are substantially more technically involved than even the current technical tools we invent hereby allow to analyze. As such, we leave the subject of such extended games to very interesting and promising future work.

Preceding this line of work, fruitful analysis of TFMs was initiated by \citet{tim} who studies user and proposer incentives formalizing single-miner transaction fee mechanisms.
\citet{tfm_foundations} prove limits on the joint incentive and collusion guarantees that such mechanisms can provide.
Towards Bayesian models that we study in a multi-proposer setting, \citet{gafni_bayesian} analyze first-price TFMs with a single proposer, and \citet{chen_bayesian} construct single-leader TFMs with positive revenue.
\citet{wu_revenue} then obtain stronger revenue guarantees using cryptographic techniques.
\citet{matt} study a Bayesian proposer who can negotiate with users through an off-chain auction.

In a fixed model of transaction available to each proposer (whereas we allow users to choose their routing endogenously), \citet{leaderless} study a two-stage game with strategic users and multiple block producers, including for a TFM that splits a transaction's payment equal among all proposers, regardless of whether they included it or not.
Mechanisms for setting payments to multiple parties have also been looked at through the censorship lens, addressing different strategic choices than our model.
\citet{censorship_auctions} study the cost of censoring bids in on-chain auctions, and \citet{mcp_bribery} model external bribes as private proposer types.

\section{Model}
\label{sec:model}

\paragraph{Preliminaries.}
There are $m$ users and $n\ge2$ proposers, each with capacity $c_m<m$.
Values $V_i$ are drawn i.i.d. from an atomless distribution $F$.
Every user's pure strategy is a joint choice of a bid $b\ge0$ and a route count $k\in\{0,\ldots,n\}$.
Nature chooses a uniform $k$-subset of proposers and sends the bid to each selected at random proposer.
A user's mixed strategy involves jointly mixing over bids and routes.
Each proposer observes only the transactions submitted to them (with their corresponding bid), but neither the route counts nor other proposers' lists.

\paragraph{Payments and proposer revenue.}
We denote by $K$ the number of proposers that include a transaction.
For functions $A(K), G(K)$, the user's payoff is
$
 v\bone_{\{K\ge1\}}-bA(K)-gG(K).
$
We presume what we will call a ``balanced'' rule where each including proposer receives $bA(K)/K$ and assume that $A(0)=G(0)=0$, $A(1)=G(1)=1$, and $1\le A(r),G(r)\le r$ for $r\ge1$.
The base fee charge $gG(K)$ is burnt.
Separately, when we examine exclusive rewards, each proposer receives $b\bone_{\{K=1\}}$; any remaining user payment, if any, is burnt. The unique-only payment rule \citep[duplicate-penalizing rule in][]{saraf} is the one where the user only pays a uniquely-including proposer, otherwise does not pay.

\begin{center}
\begin{tabular}{lll}
\toprule
Rule & User bid charge, $K\ge1$ & Each includer's revenue \\
\midrule
PPI & $bK$ & $b$ \\
Pay-once & $b$ & $b/K$ \\
Affine & $b[1+\eta_b(K-1)]$ & $b[1+\eta_b(K-1)]/K$ \\
Exclusive & Specified PPI, once, or affine charge & $b\bone_{\{K=1\}}$ \\
Unique-only & $b\bone_{\{K=1\}}$ & $b\bone_{\{K=1\}}$ \\
\bottomrule
\end{tabular}
\end{center}

A proposer, subject to their capacity, maximizes their revenue conditional on the transactions they actually received.
As a result, the proposer's best-response to any user strategies in PPI is to choose the highest actually-received bids, whereas other TFMs induce probabilistic selection rules which rely on the proposers building a belief about how many other proposers are likely to (symmetrically) choose that transaction as well.
For a single user deviation, the proposers' policies as functions of probabilities given the exactly-visible lists are kept fixed (at equilibrium), and so are the other users' strategies. Note that, however, this does not imply that proposer choices remain fixed, because a singular user change might lead to visibility changes in the transaction lists actually received by proposers.

\paragraph{Equilibria and limits.}
We study exact symmetric Bayesian equilibria with atomless finite visible-bid distributions, with jointly independent randomness of all users and proposers.
Bayesian beliefs follow Bayes' rule on-path, and the off-path beliefs should be feasible under random routing.
This also means that we do not hereby study perfect Bayesian equilibria (PBE), but general Bayesian Nash Equilibria.
We let $m\to\infty$, with $n$ fixed, $c_m/m\to\rho\in(0,1)$, and $\beta=n\rho$.
We are able to prove existence and atomlessness in finite games for some types of games, included in \Cref{app:existence}. We conjecture that our structural characterizations can be generalized even when this existence of the particular equilibrium class above is violated, but---in a similar way to \citet{matt_itcs}---our formal characterizations are conditioned to existing equilibria of the type we describe.

\paragraph{Outcomes.}
Let $w=v-g$ be net value; nonpositive net values abstain.
Write $s_m(v)=\Pr(K\ge1\mid v)$ and $\ell_m(v)=\E[K\mid v]$.
Conditional multiplicity is $\mu_m(v)=\ell_m(v)/s_m(v)$ when $s_m(v)>0$.
Across users, denote
\[
 Q_m=\Pr(K\ge1),\quad L_m=\E[K],\quad
 D_m=L_m-Q_m,\quad \mu_m=L_m/Q_m
\]
respectively for: ``coverage,'' ``occupied capacity,'' ``duplicate load,'' and ``aggregate conditional multiplicity.''
Whenever the total capacity is occupied (``fills''), we will have that $L=\beta$ and $\mu=\beta/Q$ in the limit.

\section{Congested blockspace}
\label{sec:congested}

Suppose $0<\beta<\Pr(V>g)$, and define the efficient cutoff $v^\star$ by
$1-F(v^\star)=\beta$.
For PPI and near-PPI results, we assume a positive continuous density around $v^\star$.

\subsection{PPI and pay-once}
\label{subsec:endpoints}

\begin{theorem}[Endpoint outcomes]
\label{thm:endpoints}
Every equilibrium sequence in the stated class has the following limits.
\begin{enumerate}[label=(\roman*),leftmargin=*]
\item Under PPI, with any base fee schedule specified in \Cref{sec:model}, for each fixed $v\ne v^\star$, we have that
\[
 U_m(v)\to(v-v^\star)_+ \qquad \text{and} \qquad
 K\mid v\ \Longrightarrow\ \bone_{\{v>v^\star\}}.
\]
For values above $v^\star$, supported bids tend to $v^\star-g$.

Thus $Q_m,L_m\to\beta$, $D_m\to0$, and $\mu_m\to1$.
\item Under pay-once with the base fee also paid-once, the following statement holds under the more general assumption of $0<\rho<\Pr(V>g)$:
almost every positive net-value user sends their transaction to all proposers, i.e., chooses $k=n$, in any equilibrium of each finite game.
There is, then, a unique $t>0$ satisfying
\[
 \E\!\left[\left(1-(t/(V-g))^{1/(n-1)}\right)_+\bone_{\{V>g\}}\right]=\rho,
\]
for which, and for almost every value $w>0$, it holds that
\[
 K\mid w\ \Longrightarrow\ \operatorname{Bin}(n,p(w)),\qquad
 b_m(w)\to
 \begin{cases}
 w,&w\le t,\\
 \displaystyle\frac{tnp(w)}{1-(1-p(w))^n},&w>t.
 \end{cases}
\,,
\]
where $b_m(v)$ is the function of the equilibrium bid of user of value $v$,
and $p(w) \defeq [1-(t/w)^{1/(n-1)}]_+$ for $w>0$.
\end{enumerate}
\end{theorem}

We defer the above proofs to \Cref{app:finite,app:ppi,app:pay-once}.

Under pay-once, each served finite value $w>t$ has strictly positive failure probability (that it does not get included) of
$(t/w)^{n/(n-1)}>0$ and conditional multiplicity $np(w)/[1-(1-p(w))^n]>1$.
As a result, despite filling the capacity, pay-once's inclusion is inefficient, i.e., $Q<\beta$, whereas PPI is efficient.

\begin{proposition}[Finite PPI cutoff]
\label{prop:gaussian}
Assume bounded value support, a continuous density bounded above and away from zero on $[g,v^\star+\epsilon]$, continuously differentiable near $v^\star$, and
$c_m/m=\rho+o(m^{-1/2})$.
Let $\Phi$ be the standard normal distribution function, and denote
\[
 h_m=\frac{n\sqrt{\rho(1-\rho)}}{f(v^\star)\sqrt m}.
\]
Under PPI, for $v=v^\star+h_mx$, supported actions at every equilibrium eventually use one route and
$s_m(v)=\Phi(x)+o(1)$, uniformly for bounded $x$.
Every fixed $v>v^\star$ also eventually uses one route.
\end{proposition}

The proof of \Cref{prop:gaussian} is given in \Cref{app:ppi}.
There, we also get, assuming $g=0$ and positive continuous density on an interval support $[0,\bar v]$, that each fixed type less than the efficient cutoff, i.e., $0<v<v^\star$, eventually routes to all proposers.

\subsection{Capacity, reliability, and value tails}
\label{subsec:tails}

\begin{corollary}[Pay-once duplication]
\label{cor:tails}
Under the pay-once assumptions of \Cref{thm:endpoints}:
\begin{enumerate}[label=(\roman*),leftmargin=*]
\item A served value with failure probability $\varepsilon$ has expected count
$\ell(v)=n(1-\varepsilon^{1/n})$ and conditional multiplicity
$\mu(v)=n(1-\varepsilon^{1/n})/(1-\varepsilon)$.
\item After the finite-market limit, let $n\to\infty$ with total capacity $\beta$ fixed.
Define $\log_+x \defeq \log(\max\{1,x\})$.
If $\E[\log_+(V-g)]<\infty$, then there is a unique $T>0$ such that
$\E[\log_+((V-g)/T)]=\beta$.
At each fixed positive net value, the inclusion count conditional on value converges in distribution to
$\operatorname{Poisson}(\log_+(w/T))$.
\item At fixed $n$, bounded positive net values give $\mu\to1$ as $\beta\downarrow0$.
If their upper endpoint $\bar w$ has a positive continuous density,
$\mu=1+\Theta(\sqrt\beta)$.
For Pareto net values with exponent $\alpha>0$, at sufficiently small $\beta$, the conditional multiplicity is
\[
 \mu=1+\frac{1}{\alpha+1/(n-1)}.
\]
\end{enumerate}
\end{corollary}

Hence, this result shows that shrinking capacity of higher order than transaction growth eliminates duplication in the limit only when values are bounded. Examining value-dependent duplication instead reveals potentially unbounded conditional multiplicity with heavy tails.
We defer the relevant proofs to \Cref{app:pay-once}.

\subsection{Changing charges and proposer rewards}
\label{subsec:design}

\begin{theorem}[Affine comparisons]
\label{thm:affine}
Consider the affine rule $A(K)=1+\eta_b(K-1)$.
\begin{enumerate}[label=(\roman*),leftmargin=*]
\item If $\eta_{b,m}\to1$ as $m\to\infty$, every equilibrium sequence (for all base schedules) has the PPI allocation of \Cref{thm:endpoints}(i).
\item For every fixed $0<\eta_1<\eta_2<1$, there exist finite atomless equilibria at both coefficients where
limits at $\eta_2$ have lower aggregate coverage and higher conditional multiplicity than the respective ones at $\eta_1$.
\item With once-paid base fees and affine coefficients $\eta_b\ge\eta_0>0$, every limiting outcome satisfies the below:
\[
 Q\ge\beta a^2,\quad D\le\beta(1-a^2),\quad
 \mu\le\min\{n,a^{-2}\},\text{ where}\quad
 a\defeq\frac{1+(n-1)\eta_0}{n}.
\]
Consequently, every fixed baseline equilibrium with positive duplication $D_0>0$ can be strictly improved in both aggregate coverage and multiplicity by all equilibria at sufficiently high coefficients close to 1, i.e., yielding a TFM close to PPI.
\end{enumerate}
\end{theorem}

We defer the proof and constructions of \Cref{thm:affine} to \Cref{app:affine}.

\begin{proposition}[Exclusive rewards]
\label{prop:exclusive}
Fix $n\ge2$, $0<\beta<1$, any admissible base schedule, and any $\zeta>0$.
For any exclusive-rewards rule and every affine user charge (including charging per-inclusion), there exists a value distribution supporting exact finite atomless equilibria with $\mu>n-\zeta$ in the limit.

Specifically, for the pay-once exclusive-reward rule with the base fee also paid-once, where $0<\rho<\Pr(V>g)$, every equilibrium limit satisfies
\[
 Q\le\Pr(V>g)\left[1-\left(1-\frac{\rho}{\Pr(V>g)}\right)^n\right]
 <\min\{\beta,\Pr(V>g)\}.
\]
\end{proposition}

The proofs of \Cref{prop:exclusive} are in \Cref{app:exclusive}.
\Cref{prop:exclusive} essentially yields that exclusively rewarding proposers does not guarantee low duplication, even when users pay for each copy or do not pay at all when duplication exists; to the contrary, there are equilibria sequences with near-maximal duplication.

\section{Ample blockspace}
\label{sec:ample}

Now suppose $\Pr(V>g)<\beta<n\Pr(V>g)$, i.e., that total capacity is sufficient to include every user who could be able to be included, but each proposer still does not possess sufficient space to include everyone. This is what we term the ``ample blockspace'' regime. In \Cref{app:ample}, we give the below characterization:

\begin{theorem}[Service and duplication under ample capacity]
\label{thm:ample}
Every existing equilibrium sequence has the following properties:
\begin{enumerate}[label=(\roman*),leftmargin=*]
\item Under PPI, each fixed positive net value has success tending to one and bid tending to zero.
If $\min\limits_{2\le r\le n}g(G(r)-1)>0$, every such user eventually uses one route and is included exactly once.
Hence $Q,L\to\Pr(V>g)$ and $\mu\to1$.
\item Under a zero or once-paid base fee, PPI has a unique aggregate limiting route law $(\pi_1,\ldots,\pi_n)$, which depends only on $n,\rho,\Pr(V>g)$, and satisfies
\[
 \begin{gathered}
 \pi_j>0,\qquad \sum_{j=1}^n\pi_j=\Pr(V>g),\qquad
 \Pr(k=j),\Pr(K=j)\to\pi_j,\\
 1<\mu=\frac{\sum_jj\pi_j}{\Pr(V>g)}<\frac{\beta}{\Pr(V>g)}.
 \end{gathered}
\]
\item As proven in \Cref{thm:endpoints}(ii), pay-once with a once-paid base fee, in the ample-blockspace regime, fills the total capacity and has
\[
 Q_{\rm once}<\Pr(V>g)=Q_{\rm PPI},\qquad
 \mu_{\rm PPI}<\frac{\beta}{\Pr(V>g)}<\mu_{\rm once}.
\]
\end{enumerate}
\end{theorem}

We view the following as a significant implication of our \Cref{thm:ample}: when the base fee $g$ is such that the ``ample blockspace'' condition $\Pr(V>g)<\beta<n\Pr(V>g)$ is satisfied (e.g., with the dynamically adjusting nature of EIP-1559--like base fee controllers), then the MCP TFM of pay-per-inclusion (with a positive duplicate base fee charge, such as when the base fee is charged per-inclusion) results in users opting to route only to \emph{one} randomly-chosen proposer, and inclusion is efficient with vanishingly small inclusion bids.

\section*{Acknowledgments}
We thank Aadityan Ganesh, Michael Setrin, and Matt Weinberg for helpful discussions.

\section*{AI Methodology}
Codex was used to refine authors' results, which initially started from BNE results about the PPI mechanism as well as partial lemmata about pay-once equilibria, subsequentially supplied to Codex, and to perform autonomous research over affine parameters to identify potential multiplicity of equilibria in the settings of linear combinations of TFMs. Codex was also used to investigate a possible reversion of the authors' conjectured monotonic order of conditional multiplicity and inclusion probability with the affine coefficient, to which factually-checked counterexamples were successfully discovered. Codex models were finally utilized to generate end-to-end the Lean formalizations accompanying this manuscript, available on \url{https://github.com/JasonSome/bayesian-mcp-tfm-lean}; these formalizations correspond to all theorems except proofs of equilibria existence and equilibria constructions. The authors take full responsibility for the entirety of the set of results, the exposition and presentation of economic and technical insight hereby.

\printbibliography

\clearpage
\appendix
\section{Finite incentives and the PPI limit}
\label{app:finite}

For a supported action, denote $S=\Pr(K\ge1)$, $L=\E[K]$, $D=L-S$,
and $w=v-g$. We will start with basic monotonicity properties of inclusion probability and its resulting utility.

\begin{lemma}[Finite incentive envelope]
\label{lem:finite-envelope}
The inclusion probability conditional on value on any supported action is nondecreasing in value.
Indirect utility is convex, nonnegative, and $1$-Lipschitz,
with $U_m(g)=0$ and any supported action's inclusion probability is
in $\partial U_m(v)$. Thus
$U_m(y)-U_m(x)=\int_x^y s_m(v)dv$ for any $g\le x\le y$.
\end{lemma}
\begin{proof}
Myerson's incentive inequalities as they correspond to the inclusion game directly give the desired monotonicity.
Utility is a supremum of affine functions with
slopes in $[0,1]$, including abstention, and $U_m(g)=0$ because inclusion costs at least $g$. The last claim arises from convexity.
\end{proof}

\begin{lemma}[Balanced receipt bounds]
\label{lem:balanced-order}
Let $a_m=\min_r A_m(r)/r>0$, let $\alpha_m(b)$ be the probability of
a rival arriving at a fixed proposer with bid above $b$, and set
$H_m(z)=\Pr\{\operatorname{Bin}(m-1,z)\le c_m-1\}$. Then
\begin{align*}
 &H_m(\alpha_m(a_mb)) \le q_m(b)\le H_m(\alpha_m(b/a_m)),
 \\
 &L(b,k) =kq_m(b),\text{ and}
 \\
 &q_m(b) \le S(b,k)\le nq_m(b)
 .
\end{align*}
Almost every routed equilibrium bid obeys $b\le(v-g)_+$, so
$\alpha_m(b)\le\Pr(V>g+b)$.
\end{lemma}
\begin{proof}
The target's conditional receipt lies in $[a_mb,b]$. Receipt ordering
and independent rivals give the binomial bounds; private routing leaves
each recipient's marginal response unchanged. Linearity and the union
bound give $L,S$. An overbid loses whenever included. If overbids had
positive mass, conditioning all users on such actions reaching one
proposer would force a positive-receipt inclusion at a loss. Routed
zero-bid mass is excluded by visible atomlessness.
\end{proof}

\begin{theorem}[Joint approach to PPI]
\label{thm:near-ppi}
Under Section~\ref{sec:congested}'s scarcity and cutoff-density assumptions,
$a_m\to1$ implies all conclusions of Theorem~\ref{thm:endpoints}(i).
Allowed $G_m$ and the relative rates of $a_m\to1$, $m\to\infty$ are arbitrary.
\end{theorem}
\begin{proof}
For $0<\epsilon<\rho$, let
$u_m(\epsilon)=\inf\{b\ge0:\alpha_m(b)\le\rho-\epsilon\}$,
zero when total visible mass is smaller. The bound $b\le(V-g)_+$ controls
this quantile. A one-route bid $t_m=u_m(\epsilon)/a_m+d$, $d>0$, wins with
probability $1-r_m$, where $r_m=O(m^{-1})$. It beats a supported bid
above $t_m+d$ whenever $r_mw<d$. Thus
$\alpha_m(t_m+d)\le\Pr(V-g\ge d/r_m)\to0$, since values are finite almost surely.
For each $\epsilon'$, eventually
$u_m(\epsilon')\le u_m(\epsilon)/a_m+2d$.
Swap the quantiles and send $d\downarrow0$: all have one subsequential
limit $p$, bid mass above $p+d$ vanishes, and
$\liminf U_m(v)\ge(v-g-p)_+$.
If $p=0$, this and $U_m\le w_+$ force unit success above $g$ by
convexity, contrary to coverage at most $\beta<\Pr(V>g)$.

For $0<d<p$ set $t=p-d/2$. Eventually bids above $t$ have priority
over bids at most $p-d$, and $\alpha_m(t)>\rho-\epsilon$.
If $X$ counts the former at a proposer, their slot shortfall satisfies
$\E(c_m-X)_+/m\le\epsilon+|c_m/m-\rho|+1/(2\sqrt m)$.
Sum over proposers and send $m\to\infty$, $\epsilon\downarrow0$,
then $d\downarrow0$ to obtain
\[
 \E[K\bone_{\{B\le p-d\}}]\to0,\qquad
 e_m:=\E[(p-a_mB)_+K]\to0.
\]
The payment bound gives
$U_m(v)\le(w-p)s_m(v)-pD_m(v)+e_m(v)$, with
$\int e_m(v)dF(v)\to0$. Extract a locally uniform convex utility
limit and a subsequence with $e_m(v)\to0$ almost everywhere.
Subgradient convergence outside a countable set and atomlessness yield
$(w-p)_+\le U(v)\le(w-p)s(v)$ almost everywhere.
Hence success is zero below $g+p$ and one above it.
Integrating the payment bound over bounded value intervals, then using
$D_m(v)\le n$ on the tail, gives $D_m\to0$ and
$Q_m\to\Pr(V>g+p)$. Positive receipts fill capacity except for the
same vanishing shortfall, so $L_m\to\beta$ and
$\Pr(V>g+p)=\beta$. Cutoff density identifies $p=v^\star-g$.

Actual values arbitrarily close to either side of $v^\star$, utility
monotonicity, and the Lipschitz bound extend convergence to hypothetical
values and support gaps: $U_m(v)\to(v-v^\star)_+$ locally uniformly.
Subgradients give success at every fixed $v\ne v^\star$.
For $v>v^\star$, success tends to one, hence $q_m\ge1/n-o(1)$.
A bid below $p-d$ would give type $v^\star$ positive one-route utility;
thus $b\ge p-o(1)$. Now $U_m\le wS-a_mbL$ forces $L\to1$,
and $U_m\le(w-b)S$ gives $b\le p+o(1)$.
Below the cutoff $L\le nS\to0$. Bounded $K$ gives all count and
aggregate limits.
\end{proof}

PPI has $a_m=1$; affine charges have
$a_m=[1+(n-1)\eta_{b,m}]/n$. A diagonal choice of finite markets also
covers successive market-size and coefficient limits.

\subsection{PPI with a common reserve}
\label{app:common-reserve}

Impose a common reserve $r\ge0$ on PPI bids. Each proposer selects the
highest eligible bids, up to capacity, with uniform ties among users.
The following result also allows users to choose named recipients and
a separate bid $b_j\ge r$ for each recipient $j$. Users retain independent
private randomization and a common strategy kernel; visible bids may have atoms.

\begin{proposition}[Common-reserve limit]
\label{prop:common-reserve}
Under Section~\ref{sec:congested}'s scarcity and cutoff-density assumptions,
put $p_r=\max\{r,v^\star-g\}$ and $q_r=1-F(g+p_r)$.
Along every sequence of exact equilibria in either routing model,
\[
 U_m(v)\to(v-g-p_r)_+,
 \qquad s_m(v)\to\bone_{\{v>g+p_r\}}\quad(v\ne g+p_r),
\]
with utility convergence locally uniform, and
\[
 Q_m,L_m\to q_r,\qquad D_m\to0.
\]
Expected total payment at each fixed $v\ne g+p_r$ tends to
$(g+p_r)\bone_{\{v>g+p_r\}}$.
These conclusions allow every base schedule in Section~\ref{sec:model}.
\end{proposition}

\begin{proof}
Write $I_j$ for inclusion by proposer $j$, $\kappa_m=c_m/m$, and
$\alpha_{j,m}(b)=\Pr(B_j>b)$, excluding absent bids.
Almost every submitted coordinate obeys $b_j\le(v-g)_++ng$.
Indeed, deleting a larger coordinate gains $b_j+g-v>0$ if it is the
sole inclusion, and at least $b_j-g(n-2)>0$ if it is a duplicate.
A sampled eligible bid has positive acceptance probability under independent
rivals and uniform ties. In the anonymous model, a supported common bid
above $(v-g)_+$ loses on success. Thus the bid tails are uniformly tight.

Choose $\delta_m\downarrow0$ with $m\delta_m^2\to\infty$, and define
\[
 t_{j,m}=\inf\{b\ge r:\alpha_{j,m}(b)\le\kappa_m-\delta_m\},
 \qquad P_m=\min_j t_{j,m}.
\]
These prices are bounded and $P_m\ge r$. A singleton bid $P_m+d_m$,
where $d_m\downarrow0$, succeeds with probability at least
$1-e^{-2m\delta_m^2}$ at a proposer attaining the minimum.
Counting rival ties as failures gives this bound even with bid atoms.
In the anonymous model all marginals coincide, so a uniform singleton
has the same bound. Consequently,
$\liminf U_m(v)\ge(v-g-p)_+$ along any subsequence with $P_m\to p$.

For $r\le x<P_m$, the number $N_j(x)$ of bids above $x$ is binomial,
with mean greater than $m(\kappa_m-\delta_m)$.
At most $(c_m-N_j(x))_+$ winning bids are below $x$.
The shortfall bound used above, integrated over $0\le x<P_m$, gives
\[
 e_m:=\E\sum_j(P_m-b_j)_+I_j
 \le nP_m\left(\delta_m+\frac1{2\sqrt m}\right)\to0.
\]
Set $e_m(v)=\E[\sum_j(P_m-b_j)_+I_j\mid V=v]$ and
$D_m(v)=\E[(K-1)_+\mid V=v]$. Since $G(K)\ge\bone_{\{K\ge1\}}$,
\[
 U_m(v)\le(v-g-P_m)s_m(v)-P_mD_m(v)+e_m(v).
\]
Lemma~\ref{lem:finite-envelope} applies to bid vectors as well.
Extract a locally uniform utility limit and a subsequence with
$e_m(v)\to0$ for $F$-almost every value. The lower deviation bound and
this inequality give the same envelope argument as
Theorem~\ref{thm:near-ppi}: success tends to
$\bone_{\{v>g+p\}}$ and utility to $(v-g-p)_+$ for $F$-almost every value.
Scarcity forces $p>0$. Integrate the duplicate bound over bounded value
intervals and use $D_m(v)\le n$ on the tail to obtain $D_m\to0$.
Thus $Q_m,L_m\to1-F(g+p)$.

If $p>r$, then eventually $t_{j,m}>r$ for each proposer.
The same shortfall bound at $x=r$ shows that all capacities fill in the
limit, so $1-F(g+p)=\beta$. If $p=r$, capacity gives
$1-F(g+r)\le\beta$. The density near $v^\star$ identifies
$p=\max\{r,v^\star-g\}$ in both cases.
For $p=r$, the bound $U_m(v)\le(v-g-r)_+$ extends the utility limit to
all values. For $p>r$, values on both sides of $v^\star$ and the
convex envelope give the same extension, as in Theorem~\ref{thm:near-ppi}.
Subgradients give fixed-value success away from $g+p_r$.
Expected payment is $v s_m(v)-U_m(v)$, which gives its stated limit.
\end{proof}

\section{PPI routing and the finite cutoff window}
\label{app:ppi}

Assume Proposition~\ref{prop:gaussian}'s distribution conditions. Put
$p^\star=v^\star-g$, $\sigma^2=\rho(1-\rho)$,
$h_m=n\sigma/[f(v^\star)\sqrt m]$, and
$q_m(b)=H_m(\alpha_m(b))$, with $H_m$ from
Lemma~\ref{lem:balanced-order}. For $k$ routes, let $S=\Pr(K\ge1)$,
$D=kq_m-S$, and $H=\E G(K)-S\ge0$. Then
\[
 U_k=(v-g-b)S-bD-gH,\qquad
 k(U_1-U_k)=(v-g)D+gH-(k-1)U_k,
 \qquad U_1=(v-g-b)q_m.
\]

\begin{lemma}[Nonrare actions]
\label{ppi-duplicate}\label{ppi-nonrare}
For each $\epsilon>0$, actions with $q_m\ge\epsilon$, $k\ge2$
have $D\ge d_\epsilon>0$ eventually. Supported actions with
$q_m\ge\epsilon$ have $b\to p^\star$ and eventually $k=1$,
uniformly. Each fixed $v>v^\star$ eventually uses one route.
\end{lemma}
\begin{proof}
Duplication dominates acceptance by two fixed recipients. If their
standardized threshold tends to infinity, both accept with probability
one. Otherwise extract $\alpha_m=\rho+O(m^{-1/2})$.
The bivariate triangular-array CLT has correlation in
$[-\rho/(1-\rho),1]$, with lower endpoint above $-1$ since
$\rho<1/n$. The marginal success bound bounds thresholds below;
Gaussian lower-orthant probabilities stay positive, including at
correlation one. Compactness gives $d_\epsilon$.

Theorem~\ref{thm:near-ppi} gives uniform utility convergence on bounded
support. A nonrare bid below $p^\star-d$ gives type $v^\star$ a
profitable deviation; one above $p^\star+d$ gives utility below its
limit whenever supported. Hence $b=p^\star+o(1)$.
If $k\ge2$, then
$U_m(v)\le(v-v^\star)S-p^\star d_\epsilon+o(1)$, contradicting
the limiting utility. Above the cutoff, $S\to1$ implies
$q_m\ge1/n-o(1)$.
\end{proof}

The atomless visible-bid measure $\nu_m$ becomes Lebesgue measure in
tail coordinate $\alpha_m$; an action contributes $(k/n)dF(v)$.
If all actions below $x$ have success at most $s<1$, the envelope,
density lower bound, and $S\le kH_m(\alpha_m)$ give
\[
 U_m(x)\le C\int_{a_m(s)}^1H_m(a)da,\qquad H_m(a_m(s))=s.
\]
The binomial CLT gives $H_m(\rho+\sigma z/\sqrt m)=\Phi(-z)+o(1)$
on compact $z$-sets; moderate-deviation and Chernoff bounds control the
integral tail. Thus $s_m\to0$ gives $U_m(x_m)=o(h_m)$ for
$x_m=v^\star+O(h_m)$; a fixed lower rank $R$ gives the bound
$Ch_m\int_R^\infty\Phi(-z)dz+o(h_m)$.

\begin{proof}[Proof of Proposition~\ref{prop:gaussian}]
\label{ppi-central-routes}
Put $v=v^\star+h_mx$ and fix $a>\max(0,-x)$. Above a bid $b_m(a)$ of rank
$\rho+\sigma a/\sqrt m$, supported actions are nonrare and one-route.
The bound $b\le v-g$ gives
$\alpha_m(b_m(a))\le[1-F(g+b_m(a))]/n$, so
$b_m(a)\le p^\star-ah_m+o(h_m)$.
This one-route deviation gives utility at least
$h_m\Phi(-a)(x+a)+o(h_m)$. If that rank exceeds total visible mass,
an arbitrarily small positive bid instead has success at least
$\Phi(-a)+o(1)$ and positive utility bounded away from zero.
A rare supported action would, by incentive ordering, make all lower
success vanish; the tail-area bound would give $U_m=o(h_m)$.
The contradiction is uniform for bounded $x$ by fixing one $a$.
Lemma~\ref{ppi-nonrare} gives eventual one-routing.

At differentiability points, incentive ordering sorts one-route bids;
a larger bid in a flat acceptance class is dominated. Any higher bid
is nonrare, so Lemma~\ref{ppi-nonrare} excludes multiple routes before
applying that ordering. Atomlessness therefore gives
$\alpha_m(b_m(v))=[1-F(v)]/n=\rho-\sigma x/\sqrt m+o(m^{-1/2})$.
The binomial CLT yields $s_m(v)=\Phi(x)+o(1)$, and convexity squeezes
all supported actions between nearby differentiability points. This
proves the proposition.
\end{proof}

\begin{corollary}[PPI cutoff prices]
\label{ppi-expansion}
Under Proposition~\ref{prop:gaussian}, uniformly for bounded $x$,
\[
 U_m(v^\star+h_mx)=h_m[x\Phi(x)+\phi(x)]+o(h_m),\qquad
 b_m(v^\star+h_mx)=p^\star-h_m\phi(x)/\Phi(x)+o(h_m).
\]
\end{corollary}
\begin{proof}
The tail-area bound gives
$\lim_{R\to\infty}\limsup_m U_m(v^\star-h_mR)/h_m=0$;
nearby differentiability points and Lipschitz continuity cover exceptions.
Integrate the success profile from $-R$ to $x$, then send $R\to\infty$.
The integral is $x\Phi(x)+\phi(x)$. Since $k=1$,
$b_m=v-g-U_m/s_m$ gives the second formula.
\end{proof}

\begin{proposition}[Broadcasting below the cutoff]
\label{ppi-fixed-routing}\label{ppi-fixed-broadcast}
Suppose $g=0$, $c_m=\rho m$ is an integer, and support is $[0,\bar v]$
with continuous density bounded above and away from zero, continuously
differentiable near $v^\star$. For each $\epsilon>0$, all best responses
of $0<v\le v^\star-\epsilon$ use $n$ routes for all sufficiently
large $m$, uniformly over values and equilibria in the stated class.
\end{proposition}
\begin{proof}
Fix $\bar w<v^\star$, put $N=m-1$, $d=c_m-1$, and
$Q(a)=\Pr\{\operatorname{Bin}(N,a)\le d\}$. Upper one-routing
supplies mass $n(\rho-\eta)$ above bid $\bar w$ for every small
$\eta>0$. Total visible mass thus satisfies
$A\le1-(n-1)(\rho-\eta)\le1-\delta_0$,
$\delta_0=(n-1)\rho/2$. A small positive bid gives utility at least
$vQ(A)>0$, hence every best response has $b<v$ and positive success.

If an action at $v$ succeeds with probability $s$, higher values have
success at least $s$ and marginal at least $s/n$.
Tail-coordinate mass balance gives
$Q^{-1}(s/n)\ge[1-F(v)]/n$.
Put $a_*=[1-F(\bar w)]/n>\rho$ and $\Delta=(a_*-\rho)/2$.
For $v\le\bar w$, $q\le s\le nQ(a_*)$.
Since $-Q'/Q\ge(Na-d)/[a(1-a)]$,
$Q(a_*)/Q(\rho+\Delta)\le e^{-m\Delta^2}$ eventually.
Thus every supported bid has $\alpha(b)\ge\rho+\Delta$.

Let $\theta=\delta_0/(f_{\max}\bar v)$, $h=\theta v/m$,
and choose a best response of $y=v-h$ with success $s_y$.
If $s_y<s/n$, the rank band $s_y<Q(\alpha)<s/n$ can contain only
types in $(y,v)$, hence visible mass at most $f_{\max}h$.
The bound $-Q'/Q\le N/(1-a)\le m/\delta_0$ gives
$\log[(s/n)/s_y]\le v/\bar v\le1$.
In either case $s_y\ge s/(en)$, and imitation yields
\[
 U(v)\ge\theta vs/(enm),\qquad
 v-b\ge U(v)/(kq)\ge\theta v/(en^2m).
\]
This rent bound remains uniform near zero.

For two recipients, rival higher-bid indicators have marginals $\alpha$,
and the upper one-route population gives
$t=\Pr(X_1=0,X_2=1)\ge\rho-\eta$.
Choose $\eta<\min(\rho/2,\rho\Delta/[2(1-\rho)])$.
Conditional on first load $j\le d$, second hits among its nonarrivals
are binomial with mean at least
\[
 (m-c_m)\frac{\rho-\eta}{1-\rho-\Delta}
 =(\rho+\kappa)m,\qquad
 \kappa=\frac{\rho\Delta-(1-\rho)\eta}{1-\rho-\Delta}>0.
\]
Other arrivals only increase that load. Chernoff therefore gives
$q_2/q_1\le e^{-cm}$ uniformly. Couple routes by a random proposer
ordering: adding a contact to $k<n$ increases success by at least
$q_1-kq_2$ and expected count by $q_1$. Its payoff gain is at least
$vq_1[\theta/(en^2m)-ne^{-cm}]>0$. Taking
$\bar w=v^\star-\epsilon$ proves the assertion.
\end{proof}

For a positive per-inclusion base, the proof applies on compact intervals
inside $(g,v^\star)$: replace $v$ in the rent bound by $v-g$.
The gain is at least $q_1[\theta(v-g)/(en^2m)-vne^{-cm}]>0$.

\section{Pay-once: finite routing, limiting allocation, and entry}
\label{app:pay-once}

Users pay bid and base once on success; each includer receives $b/K$.
Retain Section~\ref{sec:model}'s information and equilibrium conventions.
Write $w=v-g$, $W=(V-g)_+$, and select abstention at $v=g$.
An atomless positive value law means atomlessness conditional on $V>g$.

\subsection{Finite routing and the proposer allocation}

\begin{lemma}[Participation and broadcasting]
\label{po-broadcast}
Suppose $n\ge2$ and $1\le c<m$. If the positive value law is atomless,
almost every positive net-value type bids $0<b<w$, obtains positive
utility, and routes to all $n$ proposers. More generally, every
positive-bid action with positive surplus uses all routes, up to null sets.
\end{lemma}

\begin{proof}

Adding a contact at the same bid preserves existing contacts' information
and weakly increases success. Suppose a positive-probability set $A$ instead
uses a fixed $k<n$, with $b,w-b\ge a>0$. Such sets cover any violation.
Condition all users on $A$ and all opponents' routes on the target's
size-$k$ subset. This event has probability
$\Pr(A)^m\binom nk^{-(m-1)}$. Its common proposer marginals satisfy
$\sum_i p_i\le c$, so average failure is at least $(1-c/m)^k>0$.
An added recipient sees only the target and includes its positive bid.
The resulting strict average gain contradicts equilibrium.

Positive-value abstention mass permits a profitable small bid when all
opponents abstain. Positive-value zero-bid mass is also impossible:
restrict to $w\ge a>0$ and one route count, and use the same common-route
event. A small positive replacement preserves old success because a zero
bid can enter only with fewer than $c$ positive rivals. It also converts
a positive average failure into success. This covers $k=n$.
If overbidding had positive mass, conditioning all users on an overbidding
set and one route count would force some positive bid to be included at
a loss. Finally, split any positive mass of zero-utility positive values
into separated lower and upper value groups. Some lower-group action
succeeds when all users belong to that group; an upper type can imitate
its action distribution at positive surplus. Thus $0<b<w$ and utility
is positive almost surely, and the first argument gives broadcasting.
\end{proof}

Define
\[
 S(p)=1-(1-p)^n,\qquad
 H(p)=\frac{S(p)}{np}
     =\int_0^1(1-p+pu)^{n-1}\,du,\qquad H(0)=1.
\]
Here $H$ decreases from $1$ to $1/n$, with $H'<0$ and $H''\ge0$.

\begin{lemma}[Finite proposer kernel]
\label{po-kernel}
For a common list of $M>c$ positive bids, the symmetric proposer game
has one inclusion-marginal vector. For a slot value $L>0$, it is
\[
 P(b,L)=
 \begin{cases}
 0,&b\le L,\\
 H^{-1}(L/b),&L<b<nL,\\
 1,&b\ge nL,
 \end{cases}
 \qquad p_i=P(b_i,L),\qquad \sum_i p_i=c.
\]
An interior coordinate makes $L$ unique. If every coordinate is zero or one,
a whole interval of prices can give the same vector.
Conditional on the list, $K_i\sim\operatorname{Bin}(n,p_i)$.
\end{lemma}

\begin{proof}

Against marginal $p_i$, including user $i$ earns $b_iH(p_i)$.
Maximize the strictly concave potential
$\sum_i b_i\int_0^{p_i}H(u)\,du$ subject to $0\le p_i\le1$ and
$\sum_i p_i=c$. Its maximizer is unique.
Its first-order conditions give the kernel and score threshold $L$.
To implement exact capacity, place intervals of lengths $p_i$
consecutively on $[0,c)$, draw $U$ uniformly on $[0,1)$, and select
intervals containing $U,U+1,\ldots,U+c-1$. Randomize their order for
anonymity. The selected set has the prescribed marginals and contains
all scores above $L$, none below it, and enough equal scores to fill
capacity; hence it is a best response. Independent proposer seeds give
the binomial law. For $M\le c$, include every positive bid.
\end{proof}

\subsection{All limits with atomless finite bids}

\begin{theorem}[Pay-once allocation and bids]
\label{po-limit}
The conclusions of Theorem~\ref{thm:endpoints}(ii) hold whenever
$c_m/m\to\rho\in(0,\Pr(V>g))$ and positive net values are atomless
and finite almost surely, along every sequence of exact symmetric
equilibria with atomless positive visible bids. Write $t$ for its
capacity root, $p_*(w)$ for its limiting marginal, $b_*(w)$ for its
limiting bid, and $s_*(w)=S(p_*(w))$, with $p_*(0)=0$.
For a sampled user,
\[
 \Pr(K=j)\longrightarrow
 \E\binom nj p_*(W)^j[1-p_*(W)]^{n-j},\qquad
 \mu_m\longrightarrow\frac{n\rho}{\E S(p_*(W))}>1.
\]
\end{theorem}

\begin{proof}

First assume $W>0$ almost surely. By Lemmas~\ref{po-broadcast}--\ref{po-kernel},
all proposers see the same list and use the kernel $P$.
Extend that kernel to arbitrary focal bids and write
$Q_m(b)=\E S(p_i(b,B_{-i}))$. This analytical extension agrees with
actual play on a common full-measure set under the equilibrium bid law
$G_m$, by Fubini. All incentive comparisons below use that set.
The convex utility envelope, positive success, and own-bid monotonicity
of $P$ imply pure, nondecreasing bids $B_m(w)$ almost surely:
at a differentiability point all best replies have the same success,
so the payoff identity fixes the bid. Atomless values avoid the
countably many exceptional points. Two-sided accumulation in atomless
$G_m$ gives
\[
 Q_m(B_m(w))=(w-B_m(w))Q'_m(B_m(w)).
 \tag{PO.1}\label{po-e9}
\]
Let $I=\{i:0<p_i<1\}$. Define $d_i=1/(B_i[-H'(p_i)])$ and $\omega_i=d_i/\sum_{j\in I}d_j$.
With $\phi(p)=S'(p)H(p)/[-H'(p)]$, implicit differentiation gives
\[
 \partial_{B_i}p_i=\frac{H(p_i)}{B_i[-H'(p_i)]}(1-\omega_i),\qquad
 B_i\partial_{B_i}S(p_i)=\phi(p_i)(1-\omega_i)\mathbf1_I(i).
 \tag{PO.2}\label{po-e10}
\]
Clipping derivatives are zero off $I$. Exceptional clipping lists are
null: conditioning on the other coordinates fixes a boundary bid at one
value; an all-zero-or-one boundary requires two bids with ratio $n$.
The local derivative bound $C_n/b$ and Fubini justify differentiating
$Q_m$, even for singular atomless $G_m$. Integer capacity requires at
least two interior coordinates whenever $I\ne\varnothing$. Their bids
lie in $(\widehat L,n\widehat L)$, so their $d_i$ ratios are bounded.
Thus, for a constant $\kappa_n>0$,
\[
 \sum_i\omega_i\mathbf1_I(i)\le1,\qquad
 1-\omega_i\ge\kappa_n\quad(i\in I).
 \tag{PO.3}\label{po-e20}
\]

\emph{Positive capacity price.}
Since $0<B_m(W)<W$, joint value-bid laws are tight. Extract a limit with
bid marginal $G$ and demand $D_G(L)=\E_G P(B,L)$.
The bounded monotone kernel and the empirical CDF bound give locally
uniform demand convergence for positive $L$; a focal bid changes demand
by at most $1/m$. Choosing $M$ with $\Pr(W>M)<\rho/2$ bounds empirical
prices above with probability tending to one.
A positive bid group with marginal tending to one is impossible:
on its bounded-value, positive-bid part,
$Q'_m(b)\le(C_n/b)\E[(1-p_i)^{n-1}\mathbf1_I(i)\mid B_i=b]\to0$.
Integrating \eqref{po-e9} gives a positive left side and a vanishing right
side. This excludes a nontrivial root interval of $D_G(L)=\rho$,
which would require mass $\rho$ at bids at least $nL$ with marginal one.
It also excludes any positive-bid mass at a zero limiting price.

If all bids tend to zero, finite IC makes interim success tend to one
common constant across positive values. For $h(w)=w/(1+w)$,
the empirical covariance of $h(W_i)$ and $S(p_i)$ is nonnegative and
has expectation tending to zero. The pairwise covariance formula,
applied between two separated interior value quantiles and then with
the outer quantile masses tending to zero, forces
$m^{-1}\sum_i|p_i-\rho|\to0$ in mean. Here $S$ has a continuous inverse
and $m^{-1}\sum_i p_i=\rho_m$. Multiply \eqref{po-e9} by $B_i$,
truncate to $W_i\le M$, and use \eqref{po-e10}:
\[
 \E[W_i\mathbf1_{\{W_i\le M\}}\phi(p_i)(1-\omega_i)\mathbf1_I(i)]
 =\E[B_i\mathbf1_{\{W_i\le M\}}
       \{S(p_i)+\phi(p_i)(1-\omega_i)\mathbf1_I(i)\}].
\]
The right side tends to zero. By \eqref{po-e20}, the weight correction
on the left is at most $M\|\phi\|_\infty/m$, so its limit is
$\E[W\mathbf1_{\{W\le M\}}]\phi(\rho)>0$ for suitable $M$.
Hence empirical prices converge to a unique positive root $L$, without
saturated positive bid mass.

\emph{Rare success.}
For iid triangular rows $Y_{m,j}\in[0,1]$ converging in law to $Y$, if
$\operatorname{ess\,inf}Y<\rho<\E Y$, then
\[
 m^{-1}\log\Pr\{\textstyle\sum_jY_{m,j}\le r_m\}
 \to-I_Y(\rho),\quad
 I_Y(\rho)=\sup_{u\le0}\{u\rho-\log\E e^{uY}\},\quad r_m/m\to\rho.
\]
Chernoff gives the upper bound. For the lower bound, tilt to mean
$z<\rho$ above the lower support endpoint; bounded variance makes the
tilted mean concentrate near $z$. Reverse the tilt, then send $z\uparrow\rho$.
Bounded summands give convergence of the moment generating functions
and derivatives, so the argument applies to changing row laws.

Suppose $x_m\to x\in(0,\infty)\cap\supp G$ and
$D_G(xH(s))>\rho$ for each $s>0$. Allow $G$ to have mass at infinity,
with $P(\infty,L)=1$. The exact event identity is
\[
 \{p_i\ge s\}=
 \left\{\sum_{j\ne i}P(B_j,x_mH(s))\le c_m-s\right\}.
 \tag{PO.4}\label{po-tail-event}
\]
For $0<\delta<\min(\epsilon,\rho)$, the limiting summands
$Y_\delta=P(B,xH(\delta))$ and $Y_\epsilon=P(B,xH(\epsilon))$
are ordered strictly on a positive-mass neighborhood of $B=x$.
Moreover $\operatorname{ess\,inf}Y_\delta\le\delta<\rho<\E Y_\delta$.
At its negative optimizing tilt $u$, therefore,
$\log\E e^{uY_\epsilon}<\log\E e^{uY_\delta}$.
The preceding tail estimate and Chernoff give
$\Pr(p_i\ge\epsilon)/Q_m(x_m)\to0$ exponentially.
Equations~\eqref{po-e10}--\eqref{po-e20} then give
\[
 \liminf_m\frac{Q'_m(x_m)}{Q_m(x_m)}
 \ge\frac{\kappa_n\min_{[0,\epsilon]}\phi}{xS(\epsilon)}.
\]
Send $\epsilon\downarrow0$: $Q'_m/Q_m\to\infty$.
For a limiting supported bid $0<x\le L$, uniqueness of the root gives
the required demand inequality, so \eqref{po-e9} forces $w-x\to0$.

Positive types cannot create an additional bid atom at zero.
Otherwise choose a positive-mass value block $[a,d]$, $a>0$, whose
bids vanish. Since $G(B>0)>\rho$, choose $u>0$ with $G(B>u)>\rho$.
No fixed-length window of $\log B_m(W)$ within this block can retain
positive mass. If one did, scale its left endpoint to one and extract
a rival law on $[0,\infty]$. A regular supported action has a finite
positive scaled limit $x$; the mass above $u$ becomes mass greater than
$\rho$ at infinity. The preceding ratio estimate diverges, whereas
scaled IC gives $Q'_{\mathrm{scaled}}/Q=a_m/(w-B_m(w))\to0$,
where $a_m\to0$ is the bid scale.
Consequently bids at distinct supported block quantiles have ratios
tending to zero or infinity. Relative to an interior block value $w$,
the rival law tends to $F_W(w)\delta_0+[1-F_W(w)]\delta_\infty$.
The event identity and the triangular tail estimate give
\[
 m^{-1}\log Q_m(B_m(w))\to
 -D_{\mathrm{KL}}(\rho\Vert1-F_W(w)).
\]
For two interior probability quantiles $x<y$, these rates are strictly
ordered because $1-F_W(x)>1-F_W(y)>\rho$. But finite IC gives
$1\le Q_m(B_m(y))/Q_m(B_m(x))
\le[x-B_m(x)]/[x-B_m(y)]\to1$, a contradiction.

\emph{Identification.}
A fixed interior bid block supplies order-$m$ interior coordinates,
so the focal $\omega_i\to0$ on compact positive bid intervals.
For $b>L$, pass \eqref{po-e9} to the limit and use $b=L/H(p)$ and
$H(p)+pH'(p)=(1-p)^{n-1}$ to obtain $w(1-p)^{n-1}=L$.
The rare-success argument gives $b=w\le L$ for excluded positive types
and excludes zero limiting bids. Capacity identifies $L=t$ uniquely:
the defining demand decreases strictly through $\rho$, from one to zero.
The conditional binomial law and bounded convergence prove all inclusion
statements, with strict multiplicity since $np>S(p)$ on $0<p<1$.

With nonparticipants, use zero dummy bids. Positive users exceed capacity
except with exponentially small probability. The derivative formulas
ignore dummy coordinates. If all positive bids vanished, apply the
covariance argument within the positive sample: its limiting marginal
would be $\rho/\Pr(V>g)$, so the same truncated positive contradiction
holds. The vanishing-bid block uses $W\ge a>0$ and thus leaves the
nonparticipation atom intact; its relative upper tail is $\Pr(V>g+w)$.
All other steps apply unchanged, with integrands zero at $W=0$.
\end{proof}

\subsection{The Gaussian participation window}

Retain Theorem~\ref{po-limit}'s assumptions and require positive value
probability in every open subinterval of a neighborhood of $t$.
Put $a=n-1$, $p_L(w)=[1-(L/w)^{1/a}]_+$ (zero at $w=0$),
$D(L)=\E p_L(W)$, and $D(t_m)=\rho_m$. Define
\[
 \begin{gathered}
 A_* =\E\frac{\mathbf1_{\{W>t\}}}{b_*(W)[-H'(p_*(W))]},\quad
 \sigma_{\mathrm{po}}^2=\frac{\Var(p_*(W))}{A_*^2},\\
 h_m=\frac{\sigma_{\mathrm{po}}}{\sqrt m},\quad C_* =\frac{2n}{at},\quad
 T(y)=y\Phi(y)+\varphi(y).
 \end{gathered}
\]
Here $\varphi,\Phi$ are the standard normal density and CDF.
The constants $A_*,\sigma_{\mathrm{po}}$ are positive and finite:
the integrands are bounded and local full support makes $p_*(W)$ nonconstant.

\begin{theorem}[Pay-once entry window]\label{po-gaussian}
For $w=t_m+h_mx$, uniformly on compact $x$-sets,
\[
 \begin{aligned}
 b_m(w)&=t_m+h_my(x)+o(h_m),&
 \Pr(K\ge1\mid w)&=C_*h_mT(y(x))+o(h_m),\\
 \E[K\mid w]&=C_*h_mT(y(x))+o(h_m),&
 U_m(w)&=C_*h_m^2[x-y(x)]T(y(x))+o(h_m^2),
 \end{aligned}
\]
where $y(x)$ uniquely solves $x=2y+\varphi(y)/\Phi(y)$.
Also $(\widehat L_m-t_m)/h_m\Rightarrow N(0,1)$, including conditional
on a regular supported focal bid in the window. Bid, success, and utility
limits extend to every best reply at hypothetical central types;
exact broadcasting retains its almost-everywhere meaning.
If $\rho_m-\rho=o(m^{-1/2})$, replace $t_m$ by $t$.
\end{theorem}

The following estimates isolate the finite work needed for the theorem.
Write $\delta_m=m^{-1/2}$ and let
$\E_{G_m}P(B,\bar t_m)=\rho_m$.
Set $J(p)=nH(p)/[-H'(p)]$; then \eqref{po-e10} becomes
$bQ'_m(b)=\E[J(p)(1-p)^a(1-\omega)\mathbf1_{\{0<p<1\}}]$.

\begin{lemma}[Price, tail, and centering estimates]
\label{po-g-price}\label{po-g-hazard}\label{po-g-center}
The root $\bar t_m$ is unique eventually and tends to $t$.
Uniformly over focal bids, a fixed compact positive price interval
and order-$m$ interior rival block fail with probability $O(e^{-cm})$.
For the clipped price and every fixed $r\ge1$,
\[
 \begin{gathered}
 \E|\widehat L_m-\bar t_m|^r=O(\delta_m^r),\qquad
 \E(\widehat L_m-\bar t_m)=o(\delta_m),\qquad
 \sup_{b>0}\|p_i-P(b,\bar t_m)\|_{L^r}=O(\delta_m),\\
 \sqrt m(\widehat L_m-\bar t_m)\Rightarrow N(0,\sigma_{\mathrm{po}}^2)
 \quad\text{for the unclipped price}.
 \end{gathered}
\]
A focal bid changes the regular root by $O(m^{-1})$.
For some $\epsilon,C>0$, supported differentiable bids
$\bar t_m-\epsilon\le b\le\bar t_m$ satisfy
\[
 \frac{Q_m(b)}{Q'_m(b)}\le\frac C{\sqrt m+m(\bar t_m-b)}.
 \tag{PO.5}\label{po-g-hazard-bound}
\]
Finally, $a_m(w)=\E[p_i\mid W_i=w]$ satisfies
$|a_m(w)-p_{\bar t_m}(w)|\le C\delta_m$ almost everywhere, with
$o(\delta_m)$ at each fixed regular $w\ne t$; hence
$\bar t_m-t_m=o(\delta_m)$.
\end{lemma}

\begin{proof}
\emph{Prices.}
By Theorem~\ref{po-limit}, the limiting bid CDF is continuous at the
clipping thresholds $t,nt$. The bounded kernel derivative
$\partial_LP=\mathbf1_{\{L<b<nL\}}/[bH'(P)]$ therefore gives
population demand slope tending to $-A_*$, by discarding and then
shrinking neighborhoods of those thresholds. A fixed interior bid block
supplies a slope bounded away from zero, in the sample except with
exponentially small probability. Hoeffding's inequality gives
\[
 \Pr(|\widehat L_m-\bar t_m|>u)
 \le Ce^{-cmu^2}+Ce^{-cm},\qquad C/m\le u\le\epsilon,
 \tag{PO.6}\label{po-g-subgaussian}
\]
uniformly over a focal demand contribution of at most $1/m$.
The bounded triangular-array CLT and local empirical slope convergence
give the root CLT. Integrating this tail bound gives all clipped moments;
the centered limit and uniform integrability give the signed mean.
The kernel derivative then gives the allocation moments.

\emph{Supported lower tails.}
Let $d=\bar t_m-b$, $N=m-1$, and
$F_s=\Pr(p_i\ge s)$. By \eqref{po-tail-event}, it is the probability
that $\sum_{j=1}^N P(B_j,bH(s))\le c_m-s$ for $s>0$; $F_0=\Pr(p_i>0)$ uses a strict inequality.
For $d\ge R\delta_m$, with fixed large $R$, define
$Y_s=P(B,bH(s))$, $r_s=(c_m-s)/N$, and
$K_s(u)=\log\E e^{u(Y_s-r_s)}$.
The fixed interior block gives
$\E Y_s-r_s\asymp d+s$, tilted variance bounded above and below,
and $\partial_sY_s\ge c>0$ on that block. The negative mean-zero
tilt has $|u_s|\asymp d+s$. Its rate $I_s=-K_s(u_s)$ obeys
$I'_s=|u_s|(\E_{u_s}\partial_sY_s+1/N)\ge c(d+s)$ almost everywhere.
Thus $I_s-I_0\ge c(ds+s^2)$.

After standardizing the sum, reverse the tilt. The prefactor is
$\E_{u_s}[e^{v_sZ_s}\mathbf1_{\{Z_s\le0\}}]$, where
$v_s=|u_s|\sqrt{N\Var_{u_s}Y_s}$.
Berry--Esseen and bounded variation give error $O(N^{-1/2})$ from
$e^{v_s^2/2}\Phi(-v_s)\asymp(1+v_s)^{-1}$.
Choosing $\epsilon$ small bounds its relative error below one;
the strict indicator gives the same lower bound for $F_0$.
Consequently
\[
 F_s/F_0\le C\frac d{d+s}e^{-cm(ds+s^2)}\quad(0<s\le\epsilon).
\]
For larger $s$, use $F_s\le F_\epsilon$. Integration yields
$\E p_i/F_0\le C/(md)$, whereas \eqref{po-e10}--\eqref{po-e20}
give $Q'_m(b)\ge cF_0$ and $Q_m(b)\le n\E p_i$.
For $d\le R\delta_m$, the price CLT gives $Q'_m\ge c_R>0$ and
$Q_m=O_R(\delta_m)$. Together these prove \eqref{po-g-hazard-bound}.

\emph{Centering, including unbounded values.}
Put $L=\bar t_m$, $b=B_m(w)$, and $p_0=P(b,L)$.
A fixed regular type below $t$, supplied by local full support, puts
all nearby and larger types above $L-\epsilon$ by bid monotonicity.
Lower types have exponentially small marginal by \eqref{po-g-subgaussian}.
For $L-\epsilon\le b\le L$, \eqref{po-g-hazard-bound} and finite IC
give $w-b=O(\delta_m)$, so both $a_m(w)$ and $p_L(w)$ are $O(\delta_m)$.
For $L\le b\le L+\epsilon$, $Q'_m\ge c>0$ and
$w-b\le C[(b-L)+\delta_m]$. Kernel smoothness off entry gives
\[
 |Q_m-S(p_0)|\le C\delta_m,\qquad
 |Q'_m-\partial_bS(P(b,L))|
 \le C\delta_m+Ce^{-cm(b-L)^2}.
\]
Multiplying the second error by $w-b$ makes it $O(\delta_m)$.
The identity $H+pH'=(1-p)^a$ then turns finite IC into
$|p_0-p_L(w)|\le C\delta_m$.

For $L+\epsilon\le b\le B$, with fixed large $B$, write
$R=\E[J(p)(1-p)^a(1-\omega)\mathbf1_{\{0<p<1\}}]$.
Finite IC is exactly
\[
 (L/w)^{1/a}=\left(\frac L{b[Q_m+R]}\right)^{1/a}R^{1/a}.
\]
Here $Q_m$ is bounded below. Allocation moments, Minkowski's inequality,
and $\omega=O(m^{-1})$ on the regular event give, with $q_0=1-p_0$,
$R^{1/a}=J(p_0)^{1/a}q_0+O(\delta_m)$ and
$Q_m+R=S(p_0)+J(p_0)q_0^a+O(\delta_m)$.
The deterministic right side equals $q_0$, including saturation.
Thus the allocation error is $O(\delta_m)$ uniformly in $w$.
Choose $B$ above $n$ times the regular price bound. For $b>B$,
$Q_m=1+O(e^{-cm})$ and $R=O(e^{-cm})$, so the same identity gives
$(L/w)^{1/a}=O(e^{-cm/a})$ and proves the remaining uniform bound.

At each fixed $w<t$, acceptance is exponentially small. At fixed
finite $w>t$, the bid stays inside both clipping boundaries;
Taylor expansion, the signed price mean, the second moment, and the
$O(m^{-1})$ own-price correction improve the error to $o(\delta_m)$.
It is zero at $W=0$. Dominated convergence using the uniform bound gives
$D(\bar t_m)=\rho_m+o(\delta_m)$, since expected filled capacity differs
from $c_m$ exponentially rarely. Finally
$D'(L)=-[\Pr(W>L)-D(L)]/(aL)<0$ continuously near $t$,
which proves $\bar t_m-t_m=o(\delta_m)$.
\end{proof}

\begin{proof}[Proof of Theorem~\ref{po-gaussian}]
The entry expansion
$P(b,L)=2(b-L)_+/(at)+O((b-L)^2+|L-t||b-L|)$ and the price estimates
give, uniformly for bounded $y$,
\[
 Q_m(\bar t_m+h_my)/h_m\to C_*T(y),\qquad
 Q'_m(\bar t_m+h_my)\to C_*\Phi(y).
\]
For central equilibrium types, monotonicity and local full support give
$b_m\to t$, and $b_m<w_m$. A bid farther than order $h_m$ below
$\bar t_m$ would have
$w_m-b_m=Q_m/Q'_m=o(\bar t_m-b_m)$ by the hazard bound, a contradiction.
Thus every scaled bid has a bounded subsequence; finite IC identifies
its limit by $T(y)=(x-y)\Phi(y)$.
Writing $M(y)=\varphi(y)/\Phi(y)$,
$[2y+M(y)]'=1+\Var(Z\mid Z\le y)\in(1,2)$, and its range is
$\mathbb R$. This proves uniqueness and compact-uniform convergence.
Success follows from the displayed limit, expected count from
$np-S(p)=O(p^2)$, and utility from $(w-b)Q_m(b)$.
The centering estimate gives the primitive price CLT.
The convex envelope squeezes every hypothetical best reply between
nearby regular types; $b=w-U_m/\Pr(K>0)$ then gives its bid limit.
The derivative of $D$ permits replacement of $t_m$ by $t$ under the
stated capacity rounding condition.
\end{proof}

\begin{corollary}[Copies in the central window]\label{po-gaussian-copies}
Put $\alpha_*=2/(at)$, $Z\sim N(0,1)$, and
$R_2(y)=(y^2+1)\Phi(y)+y\varphi(y)$. At regular central types,
\[
 \frac{p_i}{h_m}\Rightarrow\alpha_*[y(x)-Z]_+,\qquad
 \frac{\Pr(K=j\mid w)}{h_m^j}\to
 \binom nj\alpha_*^j\E[y(x)-Z]_+^j\quad(1\le j\le n).
\]
Both $\E[K\mid K\ge1,w]-1$ and $\Pr(K\ge2\mid K\ge1,w)$ equal
$\frac{h_m}{t}R_2(y(x))/T(y(x))+o(h_m)$.
At $x=0$, $y(0)=-0.61200318\ldots$ and $R_2(y(0))/T(y(0))=1.02197525\ldots$.
\end{corollary}
\begin{proof}
The entry expansion and all fixed-order price moments give the first
limit with moment convergence. Expand the conditional binomial
probabilities, bounding the remainder by the next moment, and divide
the $j=2$ term by leading success. Its coefficient is
$\binom n2\alpha_*^2/(n\alpha_*)=1/t$; higher terms are $o(h_m)$.
\end{proof}

\subsection{Capacity, reliability, and value tails}

We prove Corollary~\ref{cor:tails} and give its constants and boundary
refinement. Throughout, first take $m\to\infty$ at fixed $n,\beta=n\rho$;
then vary $n$ or $\beta$. Write $Q=\E S(p_*(W))$ and $\mu=\beta/Q$.

\begin{proof}[Proof of Corollary~\ref{cor:tails}(i)]
\label{po-reliability}
Substitute $p_*=1-\varepsilon^{1/n}$ into the binomial mean and divide
by success $1-\varepsilon$. The same substitution gives
\[
 \Pr(K\ge2\mid K\ge1,w)
 =1-\frac{n(1-\varepsilon^{1/n})\varepsilon^{(n-1)/n}}
               {1-\varepsilon}.
\]
Both expected copies and conditional multiplicity increase with served
value: $p_*(w)$ increases, and strict concavity of $S$ makes $np/S(p)$
increase. If $W\le M<\infty$, every served type retains failure
$(t/w)^{n/(n-1)}\ge(t/M)^{n/(n-1)}>0$.
\end{proof}

\begin{proof}[Proof of Corollary~\ref{cor:tails}(ii)]
\label{po-poisson}
Put $L_T(w)=[\log(w/T)]_+$, with $L_T(0)=0$.
For $L\ge0$, the expression $n[1-e^{-L/(n-1)}]$ tends to $L$
and is at most $2L$.
The log-moment assumption therefore gives demand convergence to
$\E L_T(W)$, uniformly on compact positive cutoff intervals.
This function is continuous and strictly decreasing at its positive
root, with limits infinity and zero as $T\downarrow0$ and $T\uparrow\infty$.
Bracketing gives $t(n)\to T$, where $\E L_T(W)=\beta$.
Thus $np_*(w)\to L_T(w)$ and $p_*(w)\to0$ at each fixed finite value;
the binomial law gives the stated Poisson limit. Bounded convergence
and capacity also give
\[
 Q\to\E[1-e^{-L_T(W)}],\qquad
 \mu\to\frac{\beta}{\E[1-e^{-L_T(W)}]}.
\]
If $W>T$ almost surely, the root simplifies to
$T=\exp(\E\log W-\beta)$.
\end{proof}

\begin{proof}[Proof of Corollary~\ref{cor:tails}(iii), with endpoint constants]
\label{po-bounded}\label{po-pareto}
For finite positive upper endpoint $M$, capacity gives $t\uparrow M$
as $\beta\downarrow0$. Hence $\sup_w p_*(w)\to0$, and
$S(p)=np+O(p^2)$ gives $Q/\beta\to1$.
For a rate, suppose $\Pr(W>M-u)\sim c_0u^\gamma$, $c_0,\gamma>0$,
and put $d=M-t$. Uniformly on the active interval,
$p_*(w)=(w-t)[1+o(1)]/[(n-1)M]$. Integration by parts against
$M-W$, followed by $u=dz$, gives, for each integer $j\ge1$,
\[
 \E p_*(W)^j\sim
 \frac{c_0\Gamma(\gamma+1)\Gamma(j+1)}{\Gamma(\gamma+j+1)}
 \frac{d^{\gamma+j}}{[(n-1)M]^j}.
\]
The case $j=1$ and capacity give $d\sim d_\beta$, where
\[
 d_\beta=\left(\frac{(\gamma+1)(n-1)M\beta}{nc_0}\right)^{1/(\gamma+1)},
 \qquad t=M-d_\beta+o(d_\beta).
\]
Expanding $Q=n\E p_*-\binom n2\E p_*^2+O(\E p_*^3)$ then yields
\[
 \mu-1\sim\frac{(n-1)\E p_*^2}{2\E p_*}
          \sim\frac{d_\beta}{(\gamma+2)M}.
\]
At a positive continuous endpoint density, take $\gamma=1$ and
$c_0=f_W(M)$ to get
\[
 \mu-1=\frac13\sqrt{\frac{2(n-1)\beta}{nMf_W(M)}}+o(\sqrt\beta).
\]
Under Theorem~\ref{po-gaussian}'s local-support condition, the same
moments give $\Var(p_*(W))\sim\E p_*^2$. Combining this with
$A_*\sim2c_0d^\gamma/[(n-1)M]$ yields
$\sigma_{\mathrm{po}}^2\sim d_\beta^{2-\gamma}/[2c_0(\gamma+1)(\gamma+2)]$.

For the Pareto law $\Pr(W>w)=(w_0/w)^\alpha$, $w\ge w_0>0$,
direct integration at $t>w_0$ gives
\[
 \E p_* =\frac{\Pr(W>t)}{1+\alpha(n-1)},\qquad
 Q=\frac{n\Pr(W>t)}{n+\alpha(n-1)}.
\]
Capacity thus gives
$t=w_0\{n/([1+\alpha(n-1)]\beta)\}^{1/\alpha}>w_0$
when $\rho<1/[1+\alpha(n-1)]$. Taking $\mu=\beta/Q$ proves the
constant multiplicity in Corollary~\ref{cor:tails}(iii).
\end{proof}

\begin{corollary}[Duplication when total capacity reaches demand]
\label{po-jensen}
If $W>0$ almost surely, each exact finite equilibrium satisfies
\[
 Q_m\le1-(1-\rho_m)^n,\qquad
 \mu_m\ge\frac{n\rho_m}{1-(1-\rho_m)^n}.
\]
Hence $\liminf_{\beta\to1^+}\mu(\beta)\ge[1-(1-1/n)^n]^{-1}$,
which tends to $(1-e^{-1})^{-1}$ as $n\to\infty$.
Each fixed atomless positive value law has a strict boundary gap.
\end{corollary}
\begin{proof}
Lemma~\ref{po-broadcast} fills capacity, and Lemma~\ref{po-kernel}
gives $Q_m=\E S(p_i)$, $\E p_i=\rho_m$.
Strict Jensen gives both bounds, with equality exactly when
$p_i=\rho_m$ almost surely over the sampled user and bid list.
The limiting capacity root and outcome integrals are continuous for
$\beta\in(0,n)$. At $\beta=1$, equality would require $p_*(W)=1/n$
almost surely, hence constant $W$. This excludes equality for atomless laws.
For $W\sim\operatorname{Unif}[0,1]$, $n=2$, integrating
$p_*(w)=(1-t/w)_+$ and success gives at $\beta=1$
\[
 1-t+t\log t=\tfrac12,\qquad Q=(1-t)^2,\qquad
 t=.1866823088\ldots,\qquad \mu=1.5117485537\ldots>\tfrac43.
\]
\end{proof}

\section{Finite equilibrium existence}
\label{app:existence}

For pay-once, let $n,m\ge2$, $1\le c<m$, and use $S,H$ from
Lemma~\ref{po-kernel}. The bounded-law result permits bid atoms.

\begin{lemma}[Continuation properties]
\label{ex-potential}
The common-list continuation of Lemma~\ref{po-kernel} is continuous
in positive bids and nondecreasing in each own bid. If
$b_i/n>\max_{j\ne i}b_j$, then $p_i=1$.
\end{lemma}
\begin{proof}
Continuity follows from the unique potential maximizer. Adding its two
optimality inequalities gives $(b_i'-b_i)\int_{p_i}^{p_i'}H\ge0$.
The final assertion follows by transferring marginal capacity to $i$.
Use these continuations at unused lists under broadcast/abstention beliefs.
\end{proof}

\begin{proposition}[Bounded values]
\label{ex-bounded-pay-once}
If $W=(V-g)_+\le\bar w<\infty$ almost surely, balanced pay-once has an
exact symmetric finite equilibrium. Types $V\le g$ abstain; almost every
positive type broadcasts and bids $0<b<W$, with positive utility.
An atomless positive net-value law gives pure nondecreasing bids almost
surely, but their distribution can have atoms.
\end{proposition}
\begin{proof}
If all values are zero, abstain. Otherwise restrict positive types to
$\{\varnothing\}\sqcup[\epsilon,\bar w]$, with isolated abstention
and broadcast active bids. Lemma~\ref{ex-potential}, rejecting zero bids,
gives continuous payoffs $(w-b)S(p_i)$ on each participation pattern.
Joint laws with fixed value marginal are compact and convex; best responses
have nonempty convex values and closed graph. Kakutani--Fan--Glicksberg
gives a fixed point. Zero types abstain. Positive mass of abstaining types
$W>\epsilon$ would profit by bidding $\epsilon$ when all rivals abstain.

Let $q=P_\epsilon(W\ge a,0<B\le t)$, where $\epsilon\le t<\bar w/(2n)$.
Raising its bids to $2nt$ weakly increases success and ensures inclusion
against class rivals. When all users belong to the class, average success
is at most $S(c/m)$ by concavity. Thus
\[
 aq^m[1-S(c/m)]\le2ntq,
 \qquad q^{m-1}\le\frac{2nt}{a[1-S(c/m)]}\quad(q>0).
\]
In a weak limit as $\epsilon\downarrow0$, this bound (with threshold
$a/2$ on open sets) excludes active zero bids of positive types.
Positive abstention mass is likewise impossible. At the remaining profiles
the kernel is continuous. Flooring a competing law's positive bids at
$\epsilon$ passes its inequality by bounded convergence; zero bids give
abstention payoff, and bids above $\bar w$ cannot gain.

Write $Q(b)=\E S(p_i(b,B_{-i}))$. It is continuous for $b>0$ and
nondecreasing. Almost every used bid has $Q(b)>0$ by the event that all rivals abstain
or bid no more. Bids above value lose money; a small decrease improves a
bid equal to value. Thus $0<b<w$ and utility is positive. The convex function
$u(w)=\max_b(w-b)Q(b)$ has, at each differentiability point, a unique
optimal bid: its success is $u'(w)>0$ and $b=w-u(w)/u'(w)$.
Its exceptional set is countable. Atomless values therefore give pure
bids; two-type incentive inequalities and monotonicity of $Q$ order them.
A hidden change to $k$ routes preserves each contacted list and changes
success to $1-(1-p_i)^k$. This cannot help for $b\le w$; otherwise
utility is nonpositive. Broadcast beliefs complete unused lists.
\end{proof}

\begin{proposition}[A fixed atomless family]
\label{ex-narrow-pay-once}
Fix $n\ge2$, $w_0>0$, and $0<\rho_-<\rho_+<1$. There are
$\delta_*>0,C<\infty$ such that every iid net-value law on
$[w_0-\delta,w_0+\delta]$, $0\le\delta\le\delta_*<w_0$, and every
$\rho_m=c/m\in[\rho_-,\rho_+]$ admit an exact broadcast equilibrium with
\[
 \|b_m-B_m\|_\infty\le C\delta,
 \qquad B_m=\frac{w_0(1-\rho_m)^{n-1}}
 {H(\rho_m)-\rho_mH'(\rho_m)/(m-1)}.
\]
The bid function extends to a continuously differentiable function with
positive lower and finite upper derivative bounds independent of $m,c$
and the value law. Thus a fixed atomless law gives atomless finite bids.
\end{proposition}
\begin{proof}
Against constant bid $B_m$, marginal $p$ costs
$b(p)=B_mH(q(p))/H(p)$, where $q(p)=(c-p)/(m-1)$; clip at endpoint
marginals. The payoff $w_0S(p)-nB_mpH(q(p))$ is strictly concave, since
\[
 [pH(q(p))]''=-\frac{2H'(q(p))}{m-1}
       +\frac{pH''(q(p))}{(m-1)^2}>0.
\]
Its derivative vanishes at $p=\rho_m$. This proves global optimality,
with $0<B_m<w_0$ and a strict preference for all routes.

The $B_m$ lie in a compact subset of $(0,w_0)$. Nearby marginals are
uniformly interior. The clearing equation
$m^{-1}\sum_iH^{-1}(\lambda/b_i)=\rho_m$ has multiplier derivative
bounded away from zero and, by its normalization, uniform bid derivatives
through order three. For $W=w_0+\delta X$, $X\sim\nu$ on $[-1,1]$, define
\[
 \mathcal F_\delta(f)(x)=
 \E[(w_0+\delta x-f(x))S'(p)p_b-S(p)].
\]
At $f=B_m$, put $k_m=-(1-1/m)H(\rho_m)/H'(\rho_m)>0$; then
$p_b=k_m/B_m$. With $\mathsf Eh=\int h\,d\nu$,
\[
 \begin{aligned}
 D\mathcal F_0&=a_m(I-\mathsf E)+e_m\mathsf E,\\
 a_m&=\big[w_0S''(\rho_m)-nB_m[pH(q(p))]''_{p=\rho_m}\big]
                         (k_m/B_m)^2,\qquad
 e_m=-w_0S'(\rho_m)k_m/B_m^2.
 \end{aligned}
\]
Own and common bid changes give $a_m,e_m$, respectively. Both are
uniformly negative, so the inverse on $L^\infty(\nu)$ is
$a_m^{-1}(I-\mathsf E)+e_m^{-1}\mathsf E$.
The uniform derivative bounds make
$h\mapsto h-(D\mathcal F_0)^{-1}\mathcal F_\delta(B_m+h)$ a contraction
on a uniform small ball, with a root of norm $O(\delta)$ and negative
local payoff curvature.

For global bids, $R_m=H(q(p))/H(p)$ has derivative bounded above and
away from zero by $1/n\le H\le1$ and $1/n\le-H'\le(n-1)/2$.
Rivals in $B_m\pm\epsilon$ sandwich the target response between the
two constant-rival responses; the clipped inverse gives uniform payoff
continuity. Strict concavity then preserves a uniform gap outside the
local bid neighborhood for small $\delta_*$. Local curvature checks
that neighborhood, and above-value bids cannot gain. For fixed rivals,
$\partial_w\mathcal F=\E[S'(p)p_b]>0$ and
$\partial_b\mathcal F<0$, both uniformly bounded away from zero.
Implicit differentiation gives the claimed derivative bounds. Interior
marginals and $b<w$ make every added route strictly useful;
Lemma~\ref{ex-potential} completes proposer incentives.
\end{proof}

\begin{proposition}[Ample capacity and costly duplicates]
\label{ex-ample-singleton}
Suppose the positive net-value law is atomless and bounded by $H_0$,
$c_m/m\to\rho$, $n\rho>\Pr(V>g)$, and $1\le c_m<m$.
Let $A(1)=G(1)=1$, $1\le A(r),G(r)\le r$, and
$d=\min_{2\le r\le n}g(G(r)-1)>0$.
For all large $m$ an exact symmetric equilibrium has atomless bids,
one route for positive types, and abstention otherwise. Its bids are
\[
 q_m(w)=\Pr\!\left\{\operatorname{Bin}\left(m-1,
             \frac{\Pr(V-g>w)}n\right)\le c_m-1\right\},
 \qquad b_m(w)=w-\frac{\int_0^wq_m(x)\,dx}{q_m(w)}.
\]
Duplicates vanish, aggregate success tends to $\Pr(V>g)$, and bids
tend uniformly to zero.
\end{proposition}
\begin{proof}
At all local lists believe each visible user chose one route, and include
the highest bids, filling spare capacity even with zero bids. Receipts
equal bids, so this is optimal. Set $q_{0,m}=q_m(0)\to1$ by strict
capacity surplus. Writing $U(w)=\int_0^wq_m$, reporting value $x$ gives
$U(x)+(w-x)q_m(x)\le U(w)$. This checks all one-route bids and gives
utility at least $wq_{0,m}$. Moreover,
$db_m=(\int_0^wq_m)/q_m^2\,dq_m$: since $c_m<m$, $q_m$ strictly
increases across positive value mass. Thus bids are positive and atomless.
Any bid at each contacted proposer wins with probability at least
$q_{0,m}$, since fewer than $c_m$ positive rivals forces inclusion.
Two fixed contacts both include with probability at least $2q_{0,m}-1$.
A deviation to $k\ge2$ routes therefore earns at most
$w-d(2q_{0,m}-1)<wq_{0,m}$ once
$d(2q_{0,m}-1)>H_0(1-q_{0,m})$. Nonpositive types cannot gain.
Finally $K\le1$, $q_m\ge q_{0,m}$, and
$0\le b_m\le H_0(1-q_{0,m})$ give the limits.
\end{proof}

\section{Affine charge comparisons}
\label{app:affine}

\subsection{Uniform bounds and fixed-baseline comparisons}
\label{aff-bounds}

\begin{proof}[Proof of Theorem~\ref{thm:affine}(i) and (iii)]
Part (i) follows from Theorem~\ref{thm:near-ppi}, since
$\min_rA(r)/r=[1+(n-1)\eta_b]/n\to1$.
For (iii), set $a=[1+(n-1)\eta_0]/n$, $w=v-g$.
Receipts lie in $[ab,b]$. For the visible-bid tail $\alpha_m$ from
Lemma~\ref{lem:balanced-order}, extract
$u_m(\epsilon_j):=\inf\{b:\alpha_m(b)\le\rho-\epsilon_j\}
\to p_j\downarrow p$ as $\epsilon_j\downarrow0$.
The bound $b\le w$ controls these quantiles. A one-route bid above
$u_m(\epsilon_j)/a$ gives $U(v)\ge(w-P)_+$, $P=p/a$.
If $p=0$, the envelope forces unit success above $g$, contrary to scarcity.
Thus $P>0$. Bids above $t<p$ fill capacity except for an $o(m)$
binomial shortfall, giving $L_m\to\beta$.

Let $T_{m,M}$ be payment from bids at most $M$, and $N_{j,m}(t)$
count local bids above $t$. Except for $(c_m-N_{j,m}(t))_+$ slots,
selected receipts exceed $at$. Locally measurable truncation and $b\le w$ give
\[
 T_{m,M}\ge at\left[L_m-n\Pr(V-g>M)
                   -m^{-1}\sum_j\E(c_m-N_{j,m}(t))_+\right].
\]
Since $T_{m,M}\le nM$, extract limits at bid-law continuity points
$M$, then send $M\to\infty$, $\epsilon_j\downarrow0$, $t\uparrow p$.
Monotone convergence yields $T\ge ap\beta=a^2P\beta$.
The payment identity gives $T(v)\le PS(v)$: use $U\ge0$
for $w\le P$, and $wS-w+P=PS-(w-P)(1-S)\le PS$ otherwise.
Exhausting bounded value intervals gives $T\le PQ$, hence
$Q\ge\beta a^2$, $D\le\beta(1-a^2)$ and $\mu\le\min\{n,a^{-2}\}$.
The extraction was arbitrary. For $Q_0=\beta-D_0<\beta$, choose
$\eta_0$ above the baseline coefficient and
$(n\sqrt{Q_0/\beta}-1)/(n-1)$. The resulting strict gap
$\beta a^2>Q_0$ gives all comparisons, including eventually finite ones.
\end{proof}

\begin{proof}[Pointwise comparisons following Theorem~\ref{thm:affine}]
\label{aff-fixed-value}
Failure would give coefficients $\eta_j\uparrow1$ and limit outcomes
retaining a strict failure or multiplicity gap at a fixed $v>v^\star$. Choose from each defining sequence a finite
market approximating that value's outcome. The resulting diagonal
contradicts Theorem~\ref{thm:near-ppi}. The theorem gives $K\mid v\Rightarrow1$; bounded $K$ then gives
conditional multiplicity tending to one. The argument also handles
any fixed finite collection of values.
\end{proof}

\subsection{Shared finite constructions}
\label{aff-hybrid}

Here $n=2$, $g=0$, $0<\eta<1$, $\xi=(1-\eta)/2$, and
$\kappa=1-\xi$. Singleton and duplicate receipts are $b$ and $\kappa b$.
Atomless low, active, and upper value groups have masses $s,a,h$ and
supports $(0,d]$, $[v_-,v_+]$, and $[v_H,\infty)$, respectively,
where $s+a+h=1$, $v_H>v_+$, and $h/2<\rho<a+h/2$.
Low and active users broadcast; upper users choose one fair route.

\begin{lemma}[Low-group existence]
\label{aff-body}
Suppose higher bids have minimum receipt above $d$. Put
\[
 I=\mathcal D(\rho\Vert1-h/2),\qquad
 J=\sup_{t<0}\{2\rho t-\log(ae^{2t}+he^t+s)\},
\]
where $\mathcal D(x\Vert y)=x\log(x/y)+(1-x)\log((1-x)/(1-y))$.
If $J>2I$, all large markets have an exact symmetric low-group equilibrium
with positive atomless bids. Every positive low value strictly prefers
two routes to any one-route best response. Higher strategies are fixed inputs to this low-user game.
\end{lemma}
\begin{proof}
For counts $L,A,H_1,H_2$, residual capacities are
$r_j=\min\{L,(c-A-H_j)_+\}$. At equal $r_j$, use independent
exact-capacity lotteries with marginals $[(1-\lambda/b_i)/\xi]_0^1$.
At unequal $r_j$, sort bids downwards and maximize the exact potential
\[
 \max_{\max(0,R-L)\le z\le\min(r_1,r_2)}
 \left\{\sum_{i\le R-z}b_i+\kappa\sum_{i\le z}b_i\right\},
 \qquad R=r_1+r_2.
\]
Duplicate the top $z$ and divide the other selected bids to fill capacities;
unequal counts distinguish roles. These are Nash continuations.
Let $\ell_m,h_m$ be the binomial lower-tail and boundary probabilities
at $c-1$, with success parameter $1-h/2$ and $m-1,m-2$ trials.
Let $d_m,e_m$ be the probabilities of underflow at both proposers with
$m-1,m-2$ rivals. Then positive bids have marginal success at least
$\ell_m$, duplication at most $d_m$, and
\[
 \ell_m,h_m=e^{-m(I+o(1))},\qquad d_m,e_m\le e^{-m(J-o(1))},
 \qquad\varepsilon_m:=\frac{8\eta(m-1)e_m}{\ell_m(h_m-e_m)}<\frac12
\]
for large $m$, by binomial probabilities and the joint Chernoff bound.
Also $d_m\le\ell_m/4$ and $e_m<h_m$.

Restrict broadcasts to bids $vt$, $t\in[\epsilon,1/(1+\epsilon)]$,
with multiplicative noise $1+\epsilon Z$, $Z$ uniform. Continuous payoffs
give a compact distributional fixed point. The bid $v/2$ ensures
$U(v)\ge v\ell_m/4$, hence $v-b\ge b\ell_m/4$ on support.
In a zero-noise limit, cross a cluster of mass $\mu_b$ at $b>0$.
Given one rival in it, the remaining higher-bid count has parameter
in $[a+h/2,1-h/2]$; since $a+h/2>\rho$, its boundary probability
is at least $h_m$. Removing both-underflow states leaves probability
$h_m-e_m$ of one low slot at one recipient and none at the other.
Exchangeability gives gain at least $bs\mu_b\ell_m(h_m-e_m)/8$.
A harmful jump adds a duplicate; potential increments
$\kappa b_{z+1}-b_{R-z}$ require a rival at $b$ or $\kappa b$.
Its loss is at most $\eta b(m-1)s(\mu_b+\mu_{\kappa b})e_m$;
the equal-capacity rule is continuous. Hence
$\mu_b\le\varepsilon_m(\mu_b+\mu_{\kappa b})$, whose iteration
excludes all atoms. A positive-value zero cluster has the same first-copy
gain and vanishing payment loss. All limiting discontinuities are therefore
null, so noisy inequalities pass to exact bid optimality.

A one-route optimum earns $U_1\ge v\ell_m/2$; adding the hidden second
route gains $U_1-[v-(1-\eta)b]\Pr(K=2)
\ge v(\ell_m/2-d_m)>0$. The low game depends only on higher counts.
\end{proof}

\begin{lemma}[Upper bids and boundary matching]
\label{aff-upper-envelope}
Let $1\le c<m$ and let an atomless upper class $v\ge t$ choose one
fair route with priority over lower bids. Define its rank success $q(v)$ and envelope by
\[
 \begin{aligned}
 q(v)&=\Pr\{\operatorname{Bin}(m-1,\Pr(V>v)/2)\le c-1\},\\
 U_H(v)&=(t-R)q(t)+\int_t^v q(x)\,dx,
 &b_H(v)&=v-U_H(v)/q(v),\qquad 0<R<t.
 \end{aligned}
\]
These atomless bids, with floor $R$, check all upper one-route bids.
If lower contracts have success at most $\bar s<q(t)$ and maximum
utility $(t-R)q(t)$ at $t$, they are better below $t$ and worse above it.
Put $E=(t-R)(1-q(t))+\int_t^\infty(1-q(x))\,dx$, extending $q=1$
past a bounded upper support. Then
\[
 U_H(v)\ge v-R-E,\qquad R\le b_H(v)\le R+E/q(t).
\]
If every hidden upper contact has acceptance at least $1-\delta$,
$\eta R(1-2\delta)>E$ excludes all hidden two-route upper bids.
\end{lemma}
\begin{proof}
Reporting $x$ gives $U_H(x)+(v-x)q(x)\le U_H(v)$, also checking gaps
and unused upper bids. Since $db_H=U_Hq^{-2}\,dq$ and $q$ strictly
increases across positive value mass, bids are atomless. The utility
identity gives the error bound; monotonicity gives the bid bounds.
The slope gap gives both boundary comparisons. A hidden bid $R\le b\le v$
earns at most $v-R-\eta R(1-2\delta)$; bids above value give no positive
utility. The error condition puts this below upper utility, or below the
upper floor contract for $v<t$, hence below the lower optimum.
\end{proof}

\begin{lemma}[Finite hybrid implementation]
\label{aff-implementation}
Retain the three groups and $J>2I$. Suppose their supports are bounded.
Fix $z\in[0,1)$, $e=1-z$, $r=(\rho-h/2)/a$, and
$\bar q=(r-z/2)/e\in(0,1)$. For $\lambda>0$ set
\[
 \begin{aligned}
 u_{2,z}(v,q)&=z\left(v-\frac\lambda{1-\xi q}\right)
                +e[v(2q-q^2)-2\lambda q],\\
 u_{1,z}(v,q)&=(z/2+eq)\left(v-\frac\lambda{1-\xi q}\right),
 &U_{j,z}(v)&=\max_{q\in[0,1]}u_{j,z}(v,q),\\
 M_1(v)&=\max_{q\in[0,1]}q\left(v-\frac\lambda{1-\xi q}\right),
 &M_2(v)&=(v-\lambda)^2/v\quad(v>\lambda).
 \end{aligned}
\]
Suppose the maximizing $q(v;\lambda)$ for $u_{2,z}$ is interior and
$\E_Aq(V;\lambda)=\bar q$. Put $h_0=1-\xi\bar q$ and
$T_0=(1-\bar q)h_0/[h_0^2+z\xi/(2e)]$.
Assume a bid box $[L,U]$ and upper floor $R$ satisfy
\[
 \kappa U<h_0L<h_0U<L,\quad \kappa L>d,\quad
 L<T_0v_-<T_0v_+<U<v_-,\quad U<R.
\]
On active values require $U_{2,z}>\max\{U_{1,z},M_1,v-R\}$,
and also $U_{2,z}>M_2$ if $z>0$. On upper values require
$v-R>\max\{U_{2,z},U_{1,z},M_1,M_2\}$.
Then exact symmetric finite equilibria with atomless bids exist for all
large markets. Their limiting laws are zero inclusions in the low group,
$z\delta_1+e\operatorname{Bin}(2,q(v;\lambda))$ in the active group,
and one inclusion in the upper group.
\end{lemma}
\begin{proof}
Construct low bids by Lemma~\ref{aff-body}. Given their count $L_B\ge1$,
$Z=1-(1-G_B(\min_i b_{B,i}))^{L_B}$ is common uniform randomness,
independent of active values given counts; split its binary digits for
multiple draws. Use it only where low users are rejected, so seed users'
payoffs are unchanged. Common counts $L_B,N$ reveal the other upper count
$H_{-j}=m-L_B-N-H_j$.

Put $d_j=\min\{N,(c-H_j)_+\}$, $\Delta=|d_1-d_2|$,
$d_* =\min d_j$, and $\ell=\lfloor zN/2\rfloor$.
Near expected counts, at unequal $d_j$, assign $\ell+\Delta$ exclusives
to the larger-capacity proposer and $\ell$ to the other. On the complement
of their union $E$, use independent exact-capacity lotteries with
\[
 q_i=(1-\Lambda/b_i)/\xi,\qquad
 \Lambda=\frac{N-|E|-\xi(d_*-\ell)}{\sum_{i\notin E}b_i^{-1}}.
\]
The receipt guards put mixed receipt $\Lambda$ between each exclusive
owner's $b_i$ and nonowner's $\kappa b_i$, proving best responses.
At equal capacities use the continuous clipped rule; elsewhere use the
pure rule of Lemma~\ref{aff-body}. Equal counts have probability
$O(m^{-1/2})$, other exceptions $e^{-\Omega(m)}$, also given one or
two categories.

Choose $E$ by favoring a target proportionally to bid rank, filling
uniformly, then dividing into the required sizes. For $E\ne\varnothing$,
a target's exclusive probability is
\[
 \frac{|E|-1}{N-1}+\frac{N-|E|}{N-1}
     \frac{\operatorname{rank}(b_i)}{N(N+1)/2}=z+O(m^{-1/2}).
\]
Sampling concentration and bounded reciprocal bids give uniform interim
convergence, for every box-supported law $G$, to the stated menu with
$\lambda_G=h_0/\E_G[B^{-1}]$. Private noise gives a compact law fixed
point. The derivative
$\partial_bu_{2,z}=-z+2e\lambda_G[v(1-q)-\lambda_G]/(\xi b^2)$
has inward endpoint signs: worst multipliers $h_0L,h_0U$ reduce the
conditions to $L<T_0v_-$ and $T_0v_+<U$.

Crossing a tie increases favored probability. Couple fillings to swap
that target from mixed to exclusive; equal bids preserve the multiplier.
The gain $(v-b)(1-q)^2+\eta bq^2$ is uniformly positive, so an atom
$\mu_b$ gives gain $c\mu_b/m^2$. Receipt gaps exclude scaled ties;
the rare pure rule's only negative jump costs $Cm e^{-cm}\mu_b$.
Equal-count rules are continuous. Thus zero-noise limits are atomless,
and their best-response inequalities hold exactly.
Strict concavity, $\partial_{qq}u_{2,z}=-2ev-2z\lambda\xi^2/(1-\xi q)^3$,
gives the first-order condition
\[
 \lambda/v=\Phi(q):=\frac{1-q}{1+z\xi/[2e(1-\xi q)^2]}.
\]
The strictly decreasing $\Phi$ identifies a unique clearing multiplier
in $[v_-\Phi(\bar q),v_+\Phi(\bar q)]$, hence the active limit.
Lemma~\ref{aff-upper-envelope}, with $t=v_H$ and floor $R$, supplies
atomless upper bids; their failure and error $E$ are exponentially small.

For unused bids below $R$, believe the target broadcasts. Outside the
box, omit it from exclusives and use the clipped game: its multiplier
change is $O(1/m)$, priorities persist, and its menu is bounded by
$M_1,M_2$. Low bids use the low game and have vanishing success.
Bids at least $R$ have one-route beliefs and upper priority; multiple
exceptional lists use the broadcast potential after that priority.
These locally optimal continuations and the strict menu gaps exclude
all joint deviations. For $z=0$, $M_2=U_{2,0}$; concavity and endpoint
guards check outside-box bids. Lemmas~\ref{aff-upper-envelope} and
\ref{aff-body} check hidden upper and low route deviations.
\end{proof}

\subsection{Multiplicity and interior reversals}

\begin{example}[Two exact finite equilibrium sequences]
\label{ex:multiple}
Let $n=2$, $g=0$, $\rho=.495$, $\eta=.1$, and
\[
 \begin{aligned}
 F={}&.015\operatorname{Unif}[0,.01]+.335\operatorname{Unif}[.9495,.9505]\\
 &+.33\operatorname{Unif}[1.0495,1.0505]+.32\operatorname{Unif}[1.1495,1.1505].
 \end{aligned}
\]
The efficient cutoff $1/150$ has positive density. Making either the
lowest positive band or the lowest two positive bands active gives
\[
 \begin{array}{c|ccc}
 a&Q&D&\mu\\\hline
 .335&.9037313357\ldots&.0862686642\ldots&1.095458308\ldots\\
 .665&.8208315197\ldots&.1691684802\ldots&1.206094035\ldots.
 \end{array}
\]
Both are limits of exact finite equilibria at the same primitives.
\end{example}
\begin{proof}
Use Lemma~\ref{aff-implementation} with $z=0$ and
$(a,h,R)=(.335,.65,.71)$ or $(.665,.32,.762)$.
Here $r=(\rho-h/2)/a$, $\lambda=(1-r)/\E_A[V^{-1}]$,
$p(v)=1-\lambda/v$, and $b(v)=\lambda/(1-\xi p(v))$;
the multipliers are $.467910404556163\ldots$ and
$.494813937923547\ldots$.
The largest exposures are below $.507723,.528974$. On this range,
the one-route optimizer $(1-\sqrt{\lambda/v})/\xi$ is interior;
comparison with $U_{2,0}=vp^2$ reduces to
$\xi>(1+\sqrt{1-p})^{-2}<.352$, satisfied by $\xi=.45$.
The upper-price guard is
$2\lambda-\lambda^2/v_+<R<
\min\{2\lambda-\lambda^2/v_H,v_H-M_1(v_H)\}$.
Its active/upper, upper/active, and upper/one-route margins exceed
$(.00452,.01720,.00396)$ and $(.00544,.01463,.000898)$, respectively.
Take $L=v_-(1-r)/(1-\xi r)-.001$ and
$U=v_+(1-r)/(1-\xi r)+.001$; substitution gives all receipt and inward
guards. The rate gaps are $J-2I>.11120,.03617$.
These bounds follow with outward rounding from the positive root
$z_0=e^t<1$ of
$a(1-\rho)z_0^2+h(1/2-\rho)z_0-\rho s=0$ and the logarithm bound
\[
 \left|\log x-2\sum_{j<N}\frac{y^{2j+1}}{2j+1}\right|
 \le\frac{2|y|^{2N+1}}{(2N+1)(1-y^2)},\qquad y=\frac{x-1}{x+1}.
\]
Finally, on a uniform band $[l,u]$,
$\E p(V)^2=1-2\lambda\log(u/l)/(u-l)+\lambda^2/(lu)$.
Weighting active bands gives $D$, then $Q=.99-D$ and $\mu=.99/Q$.
\end{proof}

\begin{proof}[Proof of Theorem~\ref{thm:affine}(ii)]
Proposition~\ref{aff-tail-result} gives both aggregate comparisons and the
pointwise comparisons throughout an unbounded upper tail, for one common
law at every fixed interior coefficient pair.
\end{proof}

\subsection{A reversal with uniform values}
\label{aff-uniform}

\begin{proposition}[Uniform values]
\label{aff-uniform-result}
Let $F=\operatorname{Unif}[0,1]$, $n=2$, $g=0$, and
$c_m=\lfloor .4995m\rfloor$. For each of
$(\eta_1,x_1)=(.1,.75)$ and $(\eta_2,x_2)=(.18,.7)$, exact symmetric
finite equilibria with atomless visible bids exist for all sufficiently
large $m$. Set
\[
 \delta=.001,\qquad t_i=\frac{\delta}{2x_i(1-\log x_i)-1},\qquad
 \lambda_i=x_it_i,\qquad R_i=2\lambda_i-\lambda_i^2/t_i.
\]
Types below $t_i$ broadcast; types above $t_i$ use one route. For each
fixed value away from $0,\lambda_i,t_i$,
\[
 K\mid v\Longrightarrow
 \begin{cases}
 0,&v<\lambda_i,\\
 \operatorname{Bin}(2,1-\lambda_i/v),&\lambda_i<v<t_i,\\
 1,&v>t_i.
 \end{cases}
 \qquad
 D_i=t_i(1-x_i^2+2x_i\log x_i),\quad Q_i=1-\delta-D_i.
\]
Load tends to $.999$, with $D_2>D_1$ and $Q_2<Q_1$.
At $v=.00109$, success falls from one to $.49009354\ldots$ and
conditional multiplicity rises from one to $1.16680848\ldots$.
\end{proposition}

Write $\xi=(1-\eta)/2$, $\kappa=1-\xi$, $\rho=.4995$, and omit
the row index. The intended lower-user limit is
\[
 p(v)=(1-\lambda/v)_+,\qquad
 B(v)=\frac{\lambda v}{\kappa v+\xi\lambda}\quad(v>\lambda).
 \tag{U.1}\label{aff-u-product}
\]
Both rows satisfy $\xi>1/(1+\sqrt x)^2$ and $R<\lambda/\kappa$.
These give the strict product route and upper-contract inequalities of
Lemma~\ref{aff-implementation}. We first construct finite equilibria
with restricted bids, then remove the restrictions and check routing,
including types whose success vanishes.

\subsubsection{A finite construction and removal of bid restrictions}
\label{aff-u-global}\label{aff-u-atoms}\label{aff-u-floor}

\paragraph{Restricted game.}
Lower values $v\le t$ broadcast and initially bid in
$[\ell(v),\min(v,U)]$, where
\[
 \begin{gathered}
 \chi=.69,\quad v_0=.005t,\quad
 s=\lambda+.0025t,\quad z_c=\lambda+.005t,\quad U=B(t)+.005t,\\
 \ell(v)=\chi(v-v_0)_+
       \left[\frac{v-s}{z_c-s}\right]_0^1.
 \end{gathered}
 \tag{U.2}\label{aff-u-floor-def}
\]
Upper users take one fair route and bid above a floor $R_m>U$, to be
specified below. The parameters obey
$\ell(v)\le\chi(t-v_0)<\lambda$, $\chi(z_c-v_0)>\kappa U$,
and $1/2<\chi<1/(1+\eta)$, with $\chi t<U$.
Thus $\Pr(V\le t,B\le\kappa U)\le z_c$ for every restricted law.

For a conditional lower bid law $G$, put $r=(t-\delta)/(2t)$ and let
$\Lambda_G$ solve $\int p_{\Lambda_G}(b)dG(b)=r$, where
$p_z(b)=[(1-z/b)/\xi]_0^1$. Uniformly over the compact set of
restricted joint value/bid laws,
\[
 .54t<\Lambda_G\le(1-\xi r)U<U,\qquad \Lambda_G>\kappa U.
 \tag{U.3}\label{aff-u-root-guard}
\]
For the lower bound evaluate demand at $\Lambda_0=.54t$ using only
full-floor values. With $l=\max(z_c,v_0+\Lambda_0/\chi)$, its mean is
at least $[t-l-(\Lambda_0/\chi)\log((t-v_0)/(l-v_0))]/(\xi t)$.
The two lower bounds are $.0543982$ and $.0597053$, whereas the
required $r$ values are below $.0342385$ and $.0503276$.
The upper bound follows from $b\le U$. Only bids bounded away from
zero contribute near this root; their positive mass and reciprocal-bid
bounds make the root unique, continuous in $G$, and uniformly stable.

\paragraph{Proposer policies.}
Upper priority and the common lower list reveal both upper counts
$H_1,H_2$ as in Lemma~\ref{aff-implementation}. Let
$d_j=(c_m-H_j)_+\wedge N$, $d=\min(d_1,d_2)$,
$\Delta=|d_1-d_2|$, and $T=d_1+d_2$.
Choose a continuous weight $w_G(b)=w(p_{\Lambda_G}(b))$, where $w$
is zero through $r/4$, one from $r/2$, and linear between.
Its mean is at least $r/(2-r)$. Positive-weight bids exceed
$\Lambda_G$ by a fixed margin. Choose $a>0$ so they exceed
$\Lambda_G+3a$ and $\Lambda_G-a>\max(\kappa U,.52t)$.

On guarded count states, sample $\Delta$ lower users successively
without replacement with weights $w_G$. Assign them exclusively to the
larger residual capacity; use the symmetric clipped equilibrium of
capacity $d$ on the remaining list. Define
$H(z)=\sum_i p_z(b_i)$ and $W=\sum_iw_G(b_i)$. The guards require
\[
 H(\Lambda_G-a)>d+\Delta+\epsilon m,\quad
 H(\Lambda_G+a)<d-\epsilon m,\quad W>\Delta+\epsilon m,
 \tag{U.4}\label{aff-u-guards}
\]
for sufficiently small fixed $\epsilon>0$, together with counts close
to their means. Removing any sampled set changes demand by at most
$\Delta$; its multiplier remains between $\Lambda_G-a$ and
$\Lambda_G+a$. An exclusive user's first-copy receipt exceeds this
multiplier, and its second-copy receipt is below it. Hence each sampled
allocation is a proposer Nash equilibrium.

We also need a guard which rejects low bids under the pure fallback.
Set $d_0=.001t$, $\psi_z(b)=(b-z)_+/(d_0+(b-z)_+)$, and choose
$z_G$ by $\int\psi_{z_G}dG=3r$. This continuous root satisfies
$z_G>.52t$: the floor forces more than $.21239$ of lower types above
$.54t$, whose contribution to $\psi_{.52t}$ exceeds $.20227>3r$.
Require $\sum_i\psi_{z_G}(b_i)>T+\epsilon m$. More than $T$ bids
then exceed $z_G$, so the pure potential maximizer rejects every bid
at most $z_G$.

Use a continuous weight $\alpha$, equal to one on inner guards and
zero before any outer guard fails, to mix this allocation with the
pure potential maximizer. Choose its demand cutoffs monotone in their
slacks. For equal residual capacities use the clipped rule directly.
Bounded-variable concentration makes the inner event's complement
$O(e^{-cm})$, uniformly over restricted laws; $\E\Delta=O(\sqrt m)$.
All sampling denominators on the mixing support exceed $\epsilon m$.

Use Lemma~\ref{aff-implementation}'s minimum-bid seed, now from bids
at most $z_s=.5t$ and conditional on their count. For diffuse inputs,
it is uniform and independent of nonseed bids and upper routes given categories.
The probability of a seed-free list is exponentially small, since $b\le v$
gives seed mass at least $z_s$; use the pure rule on such lists. On seeded
states, both rules reject seed users, so manipulating the seed or crossing
its cutoff preserves their payoff.

\paragraph{Noisy fixed points and their limits.}
Represent an action by $q\in[0,1]$ and submit
$b=\ell(v)+[\min(v,U)-\ell(v)][(1-\varepsilon)q+\varepsilon Z]$,
with $Z$ uniform on $[0,1]$. Positive noise makes ordinary and scaled
ties and seed-threshold crossings null. The root, sampling, guards, and
seed transform are continuous under weak convergence of these laws.
Bounded convergence gives continuous interim payoffs; the compact
fixed-point argument of Lemma~\ref{aff-implementation} gives a noisy
restricted equilibrium.

Uniformly in restricted laws and focal bids, its two-route payoff
converges, as $m\to\infty$, to
$u_G(v,b)=v(2p-p^2)-2bp(1-\xi p)$, with $p=p_{\Lambda_G}(b)$.
Indeed, guards fail exponentially rarely; weighted sampling includes a
fixed nonseed target with probability $O(\E\Delta/m)$; and the common
multiplier concentrates by the uniform slope bound. Seed bids have zero
exposure on the inner event, which also controls bids approaching zero.

Every limit with $m\to\infty$, $\varepsilon\to0$ thus maximizes this
restricted product payoff. Its unconstrained optimum has exposure
$(1-\Lambda/v)_+$ and bid $\Lambda v/(\kappa v+\xi\Lambda)$.
If $\Lambda<\lambda$, the cap is inactive and the floor can only
increase exposure; mean demand exceeds $r$. If $\Lambda>\lambda$,
the floor lies below entry and the cap can only reduce exposure;
demand is below $r$. Thus $\Lambda=\lambda$, and the active graph
is \eqref{aff-u-product}, with maximum $U-.005t$. Inactive bids obey
$b\le v\le\lambda$; atomless values and the active graph exclude an
atom at $\lambda$, while $\kappa U<\lambda$ excludes the other clipping
boundary. The active interval supplies a positive reciprocal-bid slope.
Uniform payoff convergence excludes a fixed cap strip for all
$m\ge m_0$, $\varepsilon\le\varepsilon_0$: otherwise supported
type/action pairs along $m\to\infty$, $\varepsilon\to0$ would limit
to a product maximizer there. We can therefore send noise to zero at
fixed large $m$.

\paragraph{Finite atomlessness.}
Fix such an $m$ and send $\varepsilon\downarrow0$. Put
\[
 \begin{aligned}
 L_m&=\Pr\{\operatorname{Bin}(m-1,(1+t)/2)\le c_m-1\},\\
 N_m&=m-2c_m-2,\\
 H_m&=\Pr\{\operatorname{Bin}(m-2,t)=N_m\}
           2^{-2c_m}\binom{2c_m}{c_m-N_m-1},\\
 E_m&=\Pr\{\operatorname{Bin}(m-2,z_c)\ge m-2c_m\}.
 \end{aligned}
 \tag{U.5}\label{aff-u-atom-events}
\]
The event defining $L_m$ accepts any positive focal lower bid at one
proposer and overloads the other with upper users. The feasible bid
$\max(v/2,\ell(v))\le\chi v$ gives utility at least
$(1-\chi)vL_m$; retain half in the noisy game. Thus a supported bid
has $v-b\ge(1-\chi)bL_m/2$.

For a first-copy crossing, designate a rival in the same bid cluster,
require exactly $N_m$ other lower users, and allocate the $2c_m$ upper
users so one proposer has $c_m-1$ bids above the cluster. The other is
upper-overloaded. For every lower ranking this event has probability
at least $H_m$; exchangeability gives an average gain of at least half
a first copy by crossing the cluster. A harmful pure-selector jump adds
a second copy, at rival bid $b$ or $\kappa b$. Such a duplicate leaves
at least $m-2c_m+1$ other lower users unserved, all with bids at most
$\kappa b$. Conditional on the designated rival, its probability is
at most $E_m$.

Consequently the limiting conditional lower atom masses satisfy
\[
 \mu_b\le e_m(\mu_b+\mu_{\kappa b}),\qquad
 e_m=\frac{4\eta(m-1)E_m}{(1-\chi)L_mH_m}<\tfrac12.
 \tag{U.6}\label{aff-u-atom-contraction}
\]
To verify the last inequality, let $d(a\Vert b)$ be binary relative
entropy. The rates of $L_m,H_m,E_m$ are respectively
\[
 I_0=d(\rho\Vert(1+t)/2),\quad
 I_1=d(\delta\Vert t)+(1-\delta)
 d\!\left(\tfrac12-\tfrac{\delta}{1-\delta}\middle\Vert\tfrac12\right),
 \quad J=d(\delta\Vert z_c).
\]
The two margins $J-I_0-I_1$ exceed $1.3890\cdot10^{-5}$ and
$1.7325\cdot10^{-5}$; also $J>2I_0$.
Iteration of \eqref{aff-u-atom-contraction} excludes positive atoms.
Crossing a zero cluster gives a positive success gain for a positive-value
subcluster at vanishing payment, excluding an atom at zero.

To cover a possible limiting seed atom, first integrate out the seed
conditional on its count. Seed bids contribute zero to guarded demand
and weights and are rejected by both rules; the remaining mixture is
continuous. Thus only the pure selector has negative jumps.
Atomlessness then permits bounded convergence of the noisy incentive
inequalities, giving an exact finite restricted equilibrium.

\paragraph{Removing floor and cap.}
Let $b_*=\chi(t-v_0)<\lambda$. On an inner typical event, clipped demand
at a fixed $\lambda_-\in(b_*,\lambda)$ exceeds capacity by a linear
margin. Removing $O(\sqrt m)$ exclusive users and changing one focal
bid cannot erase it. All $b\le b_*$ are ineligible for exclusive
assignment and are rejected there, uniformly. Their success tends to
zero. Positively floored types have $v>s$ and a feasible product payoff
at least $(s-\lambda)^2/s>0$. No below-floor bid, with either route
count, can improve their payoff.
For an exceptional $U<b<R_m$, use the clipped completion and make
that bid ineligible for exclusive assignment. Its multiplier impact is
$O(1/m)$. Values $v\ge U$ have a strict product payoff gap to such
bids; values $v<U$ would overbid. Thus both bid restrictions are inactive. We next check routing and upper bids.

\subsubsection{Low bids and endogenous rank regularity}
\label{aff-u-low}\label{aff-u-rank}\label{aff-u-regularity}

At the identified active graph, $z_G>\lambda+.01t$. For the first row,
values $v/t\ge.85$ give mean $\psi_{.76t}>.145>3r$; for the second,
$v/t\ge.8$ gives mean $\psi_{.71t}>.193>3r$. On a prefix through
$\lambda+e$, for fixed small $e>0$, the focal is ineligible for exclusive
assignment, the pure rule rejects it wherever $\alpha>0$, and only
the lower-root guard can depend on its bid. This dependence is
nondecreasing. Thus success $S(b)$ and duplication $D(b)$ are
nondecreasing, and the canonical part adds to the full pure kernel.
Equal success at two used bids would make the larger bid strictly more
costly. The two-type incentive constraints therefore sort used bids.
Writing $M(b)=\Pr(V\le t,B\le b)$, we obtain $M(B(v))=v$ for
almost every remote value $v<\lambda-e$.

\paragraph{Low-bid duplication.}
For $b\le .501t$, the unequal-capacity canonical block rejects the
focal user. Pure duplication is bounded by $e^{-mJ+o(m)}$ from
\eqref{aff-u-atom-events}. In the equal-count canonical block, acceptance
requires $\sum_{i=1}^{m-1}Y_i<c_m$, where $Y_i=1/2$ for an upper
rival and $Y_i=p_{.501t}(\ell(V_i))$ for a lower rival.
Chernoff bounds give a rate $J_C>I_0$. One reproducible bound uses
$a=\max(x+.005,.005+.501/.69)$ and
$f(y)=(1-.501/[.69(y-.005)])/\xi$ for $y\ge a$:
\[
 \E e^{-\theta Y}\le(1-t)e^{-\theta/2}
       +t\left[a+\int_a^1e^{-\theta f(y)}dy\right].
\]
Left Riemann sums for the decreasing integrand and alternating exponential
bounds at $\theta=.27,.23$ give $J_C-I_0>5.3735\cdot10^{-6}$ and
$3.8930\cdot10^{-6}$. Thus $D(b)=o(L_m)$ uniformly on this interval.
A one-route optimizer for any lower value has utility at least $vL_m/2$
by bidding $v/2$. If its bid is at most $.501t$, local information gives
\[
 U_2(v,b)-U_1(v,b)
 =U_1(v,b)-[(v-b)+\eta b]D(b)
 \ge v[L_m/2-D(b)]>0.
 \tag{U.7}\label{aff-u-route-identity}
\]
This also excludes abstention and handles values approaching zero.

\paragraph{The pure rank kernel.}
Put $u=1-t$, $P(w)=(1+t)/2-w$, and
$Q_m(w)=\Pr\{\operatorname{Bin}(m-1,P(w))\le c_m-1\}$.
Choose $M_*$ slightly above $\lambda$ with
$M_*<\min(\delta,\delta u/(1-\delta))$ and
$d(\delta\Vert M_*)>I_0$. Uniformly for $M(b)\le M_*$, the
unequal-capacity pure marginal satisfies
\[
 q_P(b)=Q_m(M(b))[1+O(e^{-cm})].
 \tag{U.8}\label{aff-u-pure-rank}
\]
Indeed let $X$ count own upper rivals and lower bids above $b$, and $Y$
count other upper rivals. The event $X<c_m$, $Y\ge c_m$ gives exactly
one focal copy. Conditional on $X=k<c_m$,
$Y\sim\operatorname{Bin}(m-1-k,(u/2)/(u/2+M(b)))$ has mean above
$c_m$ by a linear margin, proving the claimed relative approximation.
Any remaining pure inclusion, with both upper counts below capacity,
requires at least $m-2c_m$ lower bids at most $b$. Its rate is at least
$d(\delta\Vert M_*)>I_0$, and it is exponentially smaller than $Q_m$.

\paragraph{Stieltjes bounds.}
First-copy rank crossings and the same conditional low-bid count give
Stieltjes inequalities
\[
 dS\ge c mQ_m(M)dM,\qquad
 dD_P\le C mE_m[dM(b)+dM(\kappa b)].
 \tag{U.9}\label{aff-u-stieltjes}
\]
For $b\le .5t$, the canonical derivative satisfies
$D_C'(b)\le Ce^{-mJ_C+o(m)}/b$, since statewise
$p'(b)\le1/(\xi b)$. In this range
$S\asymp Q_m(M)$ and $S+\eta D=O(Q_m(M))$.
The envelope and $(\log Q_m)'=O(m)$ imply
$M-b\ge c\min(M,1/m)\ge c\min(b,1/m)$.

Let $\mathcal S$ be the closed used support. The continuous
bounded-variation function
$U(M(b))-M(b)S(b)+b[S(b)+\eta D(b)]$ is nonnegative and zero on
$\mathcal S$. Its derivative measure vanishes there; this follows by
decomposing the complementary open intervals, whose endpoint values
are both zero. The Stieltjes chain rule and the value envelope yield
\[
 (M-b)dS=(S+\eta D)\mathbf1_{\mathcal S}\,db+\eta b\,dD
 \quad\hbox{on }\mathcal S.
 \tag{U.10}\label{aff-u-ic-measure}
\]
Using \eqref{aff-u-stieltjes} and the rent bound, absorb the same-level
duplicate term to get
$dM(b)\le C(1+1/(mb))db+\epsilon_m dM(\kappa b)$, where
$\epsilon_m\le CmE_m/L_m\to0$.
Iterating on any Borel set $A$ leaves at most $t\epsilon_m^j$ and uses
$\int_{\kappa^j A}(1+1/(mx))dx=\int_A(\kappa^j+1/(mb))db$.
Hence $M$ is locally absolutely continuous on $(0,.5t]$, with
$M'(b)\le C[1/(1-\epsilon_m\kappa)+1/((1-\epsilon_m)mb)]$.

Since $\kappa\lambda<.5t$, the scaled term near the crossover is now
absolutely continuous and bounded by $CmE_m\,db$. The remaining
same-level term is absorbed by \eqref{aff-u-ic-measure}; the singleton
rent bound and $J>2I_0$ suffice. Canonical kernels and guard weights are
locally absolutely continuous there. Thus $M$ is locally absolutely
continuous on positive crossover intervals too. Since $b>\kappa U$
there, pure first-copy crossings cannot occur at a rival bid $b/\kappa$.
Consequently $dS_P=2Q_m'(M)[1+O(e^{-cm})]dM$; pure-duplicate
corrections remain exponentially small after differentiation.

\subsubsection{Entry moments and the exact route comparison}
\label{aff-u-impact}\label{aff-u-moments}\label{aff-u-entry}

Freeze a focal-independent balanced list and its capacity $d$. Let
$H(z)=\sum_i p_z(B_i)$, $K(z)=\sum_{\kappa B_i<z<B_i}1/B_i$,
and let $L$ be the upper endpoint of $H(L)=d$. For an interior focal
marginal, $H(\Lambda)+p=d$ and $\Lambda=b(1-\xi p)$ give
\[
 p=\frac{(b-L)_+}{\xi(b+1/\overline K)},\qquad
 p'=\frac{1-\xi p}{\xi(b+1/K(\Lambda))},
 \tag{U.11}\label{aff-u-impact-formula}
\]
where $\overline K$ averages $K$ between $L$ and $\Lambda$.
If $K\ge cm$ and $b,L$ stay in a positive compact interval, these are
relative $1+O(1/m)$ approximations to
$p_0=(b-L)_+/(\xi b)$ and $p_0'=L\mathbf1_{L<b}/(\xi b^2)$.
For $A=\E(b-L)_+$, $B=\E(b-L)_+^2$, and $P=\Pr(L<b)$,
\[
 q_0^2>(q_0+\eta bq_0')r_0
 \quad\Longleftrightarrow\quad
 \frac{A^2}{PB}>\frac\eta\xi+\frac{2A}{bP},
 \qquad q_0=\E p_0,\quad r_0=\E p_0^2.
 \tag{U.12}\label{aff-u-moment-condition}
\]
Choose a rival-only good event on which the inner guards hold for every
focal bid in one fixed entry interval. One focal demand term is bounded,
so the linear guard margins give complement probability $O(e^{-cm})$.
The focal is ineligible for exclusive assignment there, and the rival
sample and seed are focal-independent. We prove
$\liminf A^2/(PB)\ge1/2$ and $A/(bP)\to0$ for bids approaching
$\lambda$, with moments restricted to this event. Since both coefficients
are below $1/5$, this yields a strict route margin.

\paragraph{Balanced deficit.}
For each rival set $X_i=1/2$ if upper and $X_i=p_b(B_i)$ if lower;
let $Y_i$ be a fair sign if upper and zero otherwise. With
$W=c_m-\sum_iX_i$ and $Y=\sum_iY_i$, the exact post-balancing deficit is
\[
 D=W-\sum_{i\in E}h_i,\qquad h_i=1/2-p_b(B_i),\qquad |E|=|Y|.
 \tag{U.13}\label{aff-u-deficit}
\]
Eligible $h_i$ lie in a fixed positive interval $[a,A]$ near entry.
Thus $D>0$ requires $W>0$, $|Y|<W/a$.
The empirical demand slope divided by $m$ converges to $-a_0<0$;
only reciprocal bids bounded away from zero enter it. Concentration and
removal of $o(m)$ users therefore give
$(b-L)_+=D_+/(ma_0)[1+o(1)]$ on the relevant states. A fixed root or
derivative error has exponentially small probability.

For weighted sampling, with $W_e=\sum_iw_i\ge cm$ and
$\bar h_w=\sum_iw_ih_i/W_e$, couple successive draws to iid draws
with probabilities $w_i/W_e$, redrawing repeated indices. Through $d$
draws, redraw count has mean $O(d^2/m)$ and second moment
$O(d^2/m+d^4/m^2)$. Hence the selected sum $S_d$ satisfies
$\E S_d=d\bar h_w+O(d^2/m)$ and
$\E(S_d-d\bar h_w)^2=O(d+d^4/m^2)$.
These bounds also hold for uniform sampling. The seed has zero clipped
demand near entry, so conditioning on demand or tilting it leaves the
conditional seed transform uniform.

\paragraph{Moderate tails.}
Tilt the iid $X_i$ by $e^{-\theta_mX_i}$ so that its sum has mean
$c_m$. Below entry by more than the central scale,
$\theta_m\to0$ and $\theta_m\sqrt m\to\infty$.
Under the tilted law, upper probability $u_m\to u>0$, variance
$\sigma_m^2\to\sigma^2>0$, and the weighted mean $\bar h_w\to h_0>0$.
Conditional on the upper count $H$ and lower bids, upper signs remain
fair. For $W\to\infty$, $W=o(\sqrt m)$, Stirling's formula and the
parity-two Riemann sum give, for $k=0,1,2$,
\[
 \E[D_+^k\mid H,\text{lower bids}]
 =\sqrt{\frac2{\pi H}}\,
       \frac{W_+^{k+1}}{(k+1)\bar h_w}[1+o(1)].
 \tag{U.14}\label{aff-u-parity-moments}
\]
Here $k=0$ means $\Pr(D>0\mid\cdot)$. The sampling error summed over
$d=O(W)$ is $O(W^{k+1/2})$ for $k\ge1$, by the preceding second
moment and the Lipschitz bound for $(W-S_d)_+^k$. For $k=0$, discard
$O(\sqrt W)$ indices around $d=W/\bar h_w$ and sum the Chebyshev
bounds $CW/j^2$ outside them. Both errors are lower order.

Scalar Berry--Esseen and integration by parts give
\[
 \E_\theta[e^{-\theta W}W_+^j]
 =\frac{j!}{\sigma\sqrt{2\pi m}\theta^{j+1}}[1+o(1)],
 \qquad
 \E_\theta[e^{-\theta W}D_+^k]
 \sim\frac{k!}{\pi h_0\sigma\sqrt u\,m\theta^{k+2}}.
 \tag{U.15}\label{aff-u-moderate-moments}
\]
The Berry--Esseen error is $O(m^{-1/2}\theta^{-j})$ because the
exponential-polynomial function has variation $O(\theta^{-j})$.
After $s=\theta W$, the Gaussian integral gives the first expression;
\eqref{aff-u-parity-moments} gives the second. Restrict first to
$\varepsilon/\theta<W<M/\theta$; the conditional binomial bound
$C(W+1)/\sqrt m$ and exponential-polynomial tails remove this
restriction. Atypical upper counts cost $e^{-c\sqrt m}$ and are negligible.
This proof allows lattice or singular lower laws. It also proves the same
formula for $D=W-a_m|Y|$ with $a_m\to a>0$, by taking deterministic
$h_i=a_m$.

Untilting supplies a common factor $e^{-o(m)}$. It cancels in the
moment ratio, which tends to $1/2$. The conditional mean deficit is
asymptotic to $1/\theta_m$, so the conditional price shortfall tends
to zero after division by $ma_0$.

\paragraph{Central entry and entry from above.}
If the mean deficit divided by $\sqrt m$ tends to $z\in\mathbb R$,
the bivariate CLT and the sampling bound yield
$D/\sqrt m\Rightarrow z-\sigma N_X-h_0\sqrt u|N_Y|$, with independent
standard normals. The density of
$Z=\sigma N_X+h_0\sqrt u|N_Y|$ is
$2\phi(s/\omega)\Phi(h_0\sqrt u\,s/(\sigma\omega))/\omega$,
where $\omega^2=\sigma^2+h_0^2u$. It is log-concave. The conditional
positive shortfall $z-Z$ has increasing hazard, so its survival obeys
$S(x+y)\le S(x)S(y)$. Integration gives
$\E[(z-Z)^2\mid Z<z]\le2\E[z-Z\mid Z<z]^2$.
Bounded summands give uniform integrability of the first two moments.
Above the central scale, the deficit concentrates around its positive
mean and the ratio tends to one. These three cases prove the asserted
entry moment bounds, with all relevant moments $e^{-o(m)}$ up to powers
of $m$.

\paragraph{Comparison at an optimizer.}
Equations \eqref{aff-u-impact-formula}--\eqref{aff-u-moment-condition}
give $q_G^2-(q_G+\eta bq_G')r_G\ge c q_G^2$ for good-state marginals,
with fixed $c>0$. At an interior maximizer of
$u_1^G=(v-b)q_G$, its first-order condition gives
$u_2^G\ge(1+c)\max u_1^G$. Bad states change each optimized payoff
by at most $\epsilon_m\le C\Pr(G^c)$; thus
\[
 u_2(v,b_*)-\max_bu_1(v,b)\ge c\max_bu_1^G(v,b)-2\epsilon_m.
 \tag{U.16}\label{aff-u-optimizer}
\]
For values approaching $\lambda$, a good-state optimizer approaches
entry and is interior: bids bounded below entry yield $O(e^{-cm})$
utility, whereas feasible bids approaching entry below value yield
$e^{-o(m)}$. The right side of \eqref{aff-u-optimizer} is positive.
Good canonical kernels are differentiable: on each clipping face,
thresholds are strictly increasing functions of at least one diffuse
rival bid (a reciprocal-sum root or a scaled bid). The finitely many
breakpoints therefore have zero probability, and $p'\le1/(\xi b)$
permits differentiation under expectation.

\subsubsection{Sub-entry kernels and certified rate separation}
\label{aff-u-rates}

On a compact sub-entry interval $[.5t,\lambda)$, tilt $X_b$ to mean
$c_m/(m-1)$. The active graph alone gives tilt bounded away from zero,
positive variance, and a continuous nonlattice component. Upper signs
remain fair with $\operatorname{Cov}_\theta(X_b,Y)=0$.
For fixed bounded $w$, integer $y$, and $h>0$, uniformly on such compacts,
\[
 \Pr_\theta\{W\in[w,w+h],\textstyle\sum Y_i=y\}
 =\frac{h}{2\pi(m-1)\sigma_X\sigma_Y}[1+o(1)].
 \tag{U.17}\label{aff-u-local-limit}
\]
To prove this interval limit, bound the
interval indicator above and below by functions with Fourier support
$[-T,T]$ and integrals $h+O(1/T)$. Fourier-invert the lattice coordinate
on $[-\pi,\pi]$. Weak convergence and the active continuous component
give a uniform characteristic-function gap away from $(0,0)$; the upper
$\pm1$/lower-zero mixture excludes other lattice periods. Bounded third
moments give the Gaussian expansion near zero. First take $m\to\infty$,
then $T\to\infty$. The same argument gives an interval upper bound
$C(h+\varepsilon)/m$, uniform in location, and is unchanged after fixing
finitely many lower marks. It therefore permits integration of fixed
polynomials against $e^{-\theta W}$ by truncation and a geometric tail
bound.

Let $C_m(b)$ and $R_m(b)$ denote the canonical marginal and duplicate
probability, including equal counts. For unequal counts which can accept
the focal bid, choose the receipt-guard width so that
$b>(1-\xi/2)U$. Then every sampled $h_i=1/2-p_b(B_i)$ is bounded
positively. (The normalized bounds $(1-\xi/2)U/t$ are below $.65881$
and $.63853$.) Fixed-size weighted samples converge to iid samples
from the tilted weight-biased law. Their contribution for imbalance $y$
contains $\E_\theta e^{-\theta\sum_Eh_i}$; the sum over imbalances
converges geometrically. The equal-count term has coefficient one,
independent of guards, and bounds the combined coefficient away from
zero. At a guard boundary the unequal-count weight can vanish; within
$O(1/m)$ of a receipt boundary it is $O(1/m)$ and contributes lower
order. The focal derivative of a normalized guard is also $O(1/m)$.
These observations make the following estimates uniform at transitions.

The active interior strip gives $cm\le\overline K\le Cm$ on tilted
typical states. Its lower bound persists when a shortfall moves the root
down; the upper bound uses reciprocal bids at least $\kappa b$.
Saturation requires $W\ge c'm$ and incurs extra exponential cost.
Using the exact identity $p=D_+/(1+b\overline K)$ in
\eqref{aff-u-local-limit} gives
\[
 C_m(b)\asymp m^{-2}e^{-(m-1)I_m(b)},\qquad R_m/C_m=O(1/m).
 \tag{U.18}\label{aff-u-coarse}
\]
Tilted empirical failures cost $e^{-cm}$ and, on $D>0$, their untilting
weight is at most one. The active strip also gives a uniform strict
rate decrease over every fixed bid increment: consequently
$C_m(b+\epsilon)/C_m(b)\ge e^{c_\epsilon m}$.

For bids below $.5t$, the rank envelope already gives
$B_m(v)=v-\int_0^vQ_m(w)dw/Q_m(v)+o(1/m)=v-O(1/m)$ on positive
compact value intervals. If a remote used bid $b\ge.5t$ had
$v-b\ge\epsilon_0>0$, the envelope and monotonicity would give
$U_m(v)\le C Q_m(v)/m+vS_C(b)+o(Q_m(v))$.
Comparing with $(v-b)S(b)-\eta bD(b)$ and
\eqref{aff-u-coarse} forces $S_C(b)\ge c_{\epsilon_0}Q_m(v)$.
Raising $b$ by $\epsilon_0/4$, still below $v$ and below entry,
then multiplies canonical success exponentially, with duplicate ratio
$O(1/m)$ and negligible pure losses. This contradicts optimality.
Compactness makes the argument uniform; hence $B_m(v)\to v$ on every
compact subset of $(0,\lambda)$. The full limiting lower bid law is
now identified, and its CDF is continuous at entry.

Under a fixed sub-entry tilt, $K/m\to\nu(b)>0$. Integrating
$(W-s)_+$ and $(W-s)_+^2$ against $e^{-\theta W}$ gives respectively
$e^{-\theta s}/\theta^2$ and $2e^{-\theta s}/\theta^3$.
The common imbalance coefficient cancels. Averaging the exact $p'$ and
$2pp'$ from \eqref{aff-u-impact-formula}, with $a_m=C_m'/C_m$, yields
\[
 a_m=mA(b)[1+o(1)],\quad A(b)=\theta(b)\nu(b)/\xi,
 \quad \frac{R_m}{C_m}=\frac2{\xi b a_m}[1+o(1)],\quad
 \frac{R_m'}{C_m}=\frac2{\xi b}[1+o(1)].
 \tag{U.19}\label{aff-u-fixed-moments}
\]
A canonical one-route optimizer therefore
loses to a two-route deviation by a fraction tending to $1-2\eta/\xi>0$.
Equation~\eqref{aff-u-optimizer} transfers this comparison when canonical
utility dominates the exceptional pure contribution.

\paragraph{Rate certificate.}
Normalize $z=b/t$ and set
$B_0(y)=y$ for $y\le x$, $B_0(y)=xy/(\kappa y+\xi x)$ for $y>x$,
and $p_z(y)=[(1-z/B_0(y))/\xi]_+$. Since $\kappa B_0(1)<1/2$,
the marginals satisfy $p_z(y)<1$ for $z\ge1/2$. Define
\[
 \begin{aligned}
 Z(\theta,z)&=(1-t)e^{-\theta/2}+t\int_0^1e^{-\theta p_z(y)}dy,\\
 F_{\rm tilt}(\theta,z)&=\E[(X_z-\rho)e^{-\theta X_z}],\\
 I_C(z)&=\sup_{\theta\ge0}[-\theta\rho-\log Z(\theta,z)],\qquad
 I_P(z)=d(\rho\Vert(1+t)/2-zt).
 \end{aligned}
 \tag{U.20}\label{aff-u-rates-def}
\]
The following outward bounds locate one crossover and certify derivative
separation. Derivatives $A=-I'_{C,b}$ and $d=-I'_{P,b}$ use physical bid units.
\[
\begin{array}{c|cc}
 \eta&.1&.18\\\hline
 (z_{\rm lo},z_{\rm hi})&(.739,.742)&(.691,.694)\\
 F_{\rm tilt}(.02,z_{\rm hi})>&1.3661\cdot10^{-6}&1.9361\cdot10^{-6}\\
 F_{\rm tilt}(1.4,.5)<&-1.4812\cdot10^{-5}&-1.9121\cdot10^{-5}\\
 \text{test tilt at }z_{\rm lo}&.0359&.0431\\
 I_C(z_{\rm lo})>&1.5133776\cdot10^{-7}&2.1413452\cdot10^{-7}\\
 I_P(z_{\rm lo})<&1.1851732\cdot10^{-7}&1.6545428\cdot10^{-7}\\
 \text{tilt bracket at }z_{\rm hi}&(.02584,.02585)&(.02842,.02844)\\
 I_C(z_{\rm hi})<&7.86282\cdot10^{-8}&9.35290\cdot10^{-8}\\
 I_P(x)>&1.0729786\cdot10^{-7}&1.5414125\cdot10^{-7}\\
 A/d>&1.6211&2.2606
\end{array}
\tag{U.21}\label{aff-u-certificate}
\]
For $.5\le z\le z_{\rm hi}$, $\partial_zF_{\rm tilt}<0$ since its
integrand has the sign of $\partial_zp_z$ and
$1-\theta(p_z-\rho)\ge.2993$ for $\theta\le1.4$.
The tilt is unique and lies in $(.02,1.4)$. Thus
$A\ge .02(1-x)e^{-1.4}/(\xi B_0(1))$, whereas
$d=[P(z)-\rho]/[P(z)(1-P(z))]\le.002$.
This proves the last row and strict decrease of $I_C-I_P$.
The endpoint bounds locate its unique zero. For $z\ge z_{\rm hi}$,
$I_C(z)\le I_C(z_{\rm hi})<I_P(x)\le I_P(z)$, excluding another zero.

The certificate uses exact rational decimal inputs. Integrate $[0,z]$
exactly and use composite midpoint quadrature with $5000$ panels on
$[z,x]$ and $[x,1]$. For $f=e^{-\theta p_z}$, $f''\ge0$ and its
upper bounds on the two pieces are
$(\theta^2/\xi^2+2\theta/\xi)/z^2$ and
$(\theta^2+2\theta)/x^2$. For
$g=(p_z-\rho)e^{-\theta p_z}$, bounds on $|g''|$ are
$[(\theta^2+2\theta)/\xi^2+2(1+\theta)/\xi]/z^2$ and
$(\theta^2+4\theta+2)/x^2$.
Multiplying these by interval length times panel width squared divided
by $24$ bounds each remainder, one-sided for $f$. Exponential Taylor
bounds and the identity
$\log q=2\sum_{j\ge0}r^{2j+1}/(2j+1)$, $r=(q-1)/(q+1)$,
with its geometric tail bound give rational enclosures. The largest
$F_{\rm tilt}$ error is below $4.30\cdot10^{-12}$; endpoint $Z$ errors
are below $9\cdot10^{-15}$. At a tilt bracket midpoint with half-width
$h$, add $h^2/8$ to the objective since its second derivative is minus
a variance bounded by $1/4$. The allowances are $3.125\cdot10^{-12}$
and $1.25\cdot10^{-11}$. These bounds also verify the earlier strict
entropy and floor inequalities; the low-bid transform uses the stated
monotone Riemann bound.

\subsubsection{The crossover and completion of the equilibrium}
\label{aff-u-crossover}\label{aff-u-boundary}

To check routing where the pure and canonical rates cross, write $w=M_m(b)$, $\Delta_v=w-b$, $q=Q_m(w)+C_m(b)$,
$x_C=C_m(b)/q$, and $a=a_m(b)$. Suppressing the exponentially small
pure-duplicate terms justified in \eqref{aff-u-stieltjes}, the exact
incentive identity is
\[
 \Delta_v[2Q_m'(w)M_m'(b)+2C_m'(b)-R_m'(b)]
 =2q-(1-\eta)R_m(b)+\eta bR_m'(b)=:N.
 \tag{U.22}\label{aff-u-crossover-ic}
\]
Define the barrier $T_m=N/(2a q-R_m')$. Its denominator is positive,
and \eqref{aff-u-fixed-moments} gives, uniformly for $0\le x_C\le1$,
\[
 T_m=\frac{1+(\eta/\xi)x_C}{mA(b)}+o(1/m).
 \tag{U.23}\label{aff-u-barrier}
\]
Identity convergence and
\eqref{aff-u-certificate} give
$r_m=a/(Q_m'(w)/Q_m(w))>1+\gamma$ on the certified interval, for
fixed $\gamma>0$. At $\Delta_v=T_m$,
\eqref{aff-u-crossover-ic} gives $M_m'=r_m$ and $x_C'=0$.
Below the barrier by a fixed multiple of $1/m$, it gives
$M_m'>1+\gamma/2$ and $x_C'<0$ directly from \eqref{aff-u-crossover-ic}.

The initial pure-dominated region has
$\Delta_v=(Q_m'/Q_m)^{-1}[1+o(1)]>T_m$.
A first used-support excursion from $\Delta_v-T_m=-\epsilon/(2m)$
to $-\epsilon/m$ cannot occur: on that band $\Delta_v'\ge\gamma/2$
and $x_C'<0$. Since $\Delta_v,T_m=O(1/m)$, its length is $O(1/m)$;
continuity of $A$ and \eqref{aff-u-barrier} make $T_m$ increase by
at most $o(1/m)$, while $\Delta_v$ increases. Thus
$\Delta_v\ge T_m-o(1/m)$ on used support.

On unused bid gaps, $M_m$ and pure first-copy success are constant. At a right endpoint, the common boundary
type's left difference quotients give
$\Delta_v\ge N/(2C_m'-R_m')+o(1/m)\ge T_m+o(1/m)$.
Canonical kernels are continuously differentiable; the remaining pure
variation is Lipschitz with exponentially small coefficient by scaled-bid
regularity. Each reentry therefore resets the barrier on its valid side.
The Stieltjes identity holds almost everywhere on the residual used set,
which suffices for the integrated excursion argument.

Substitution in \eqref{aff-u-crossover-ic} now gives
$q_m'(b)\le a_m(b)[1+o(1)]q_m(b)$, also in integrated form across
gaps. At any one-route optimizer, comparison with a bid lower by
$h=\tau/m$, followed by $m\to\infty$ and $\tau\downarrow0$, yields
$v-b\ge a_m(b)^{-1}[1-o(1)]$. This covers nondifferentiable and gap
endpoints. Combining with \eqref{aff-u-route-identity} and
\eqref{aff-u-fixed-moments},
\[
 U_2(v,b)-U_1(v,b)
 \ge \frac{q_m(b)}{a_m(b)}[1-2\eta/\xi-o(1)]>0.
 \tag{U.24}\label{aff-u-crossover-route}
\]
The term $(v-b)D$ is lower order since $D/q=O(1/m)$.

All lower one-route optimizers are covered: \eqref{aff-u-route-identity}
through $.501t$, the crossover argument from $.5t$, and
\eqref{aff-u-fixed-moments}--\eqref{aff-u-optimizer} where canonical
success dominates. Optimizers approaching entry use the moment bounds;
fixed active values use the strict product route inequality.

Finally set
\[
 q_{H,m}(v)=\Pr\{\operatorname{Bin}(m-1,(1-v)/2)\le c_m-1\},
 \qquad R_m=t-U_{A,m}(t)/q_{H,m}(t).
\]
The upper envelope of Lemma~\ref{aff-upper-envelope} matches lower
utility at $t$ and has $R_m\to R$. Since $(1-t)/2<\rho$, its success
tends uniformly and exponentially to one. All lower contracts, including
unused bids below $R_m$, have success bounded by $\bar s<1$: their
limiting marginal is at most $(1-\lambda/R)/\xi<1$, exclusive assignment
has probability $O(m^{-1/2})$, and atypical counts are exponentially rare.
Lemma~\ref{aff-upper-envelope} now checks both sides of the boundary
and hidden upper route deviations: its failure and envelope error are
exponentially small, while $\eta R>0$. Assign best responses at null
types to complete the exact finite equilibria.

Integrating $p(v)^2$ on $[\lambda,t]$ gives $D$ in the proposition.
The two values are $.000006416257\ldots$ and $.000011847600\ldots$;
$t_1<.00109<t_2$ gives the stated pointwise comparison from
\eqref{aff-u-product}. Bounded inclusion counts pass these strict aggregate
comparisons to all sufficiently large common finite market sizes.

\subsection{A reversal throughout a fixed unbounded tail}
\label{aff-tail}

\begin{proposition}[Every interior coefficient pair]
\label{aff-tail-result}
For every $0<\eta_1<\eta_2<1$, there is one atomless unbounded value
law $F$, with $n=2$, $g=0$, and $c_m=\lfloor.495m\rfloor$, admitting
exact finite atomless equilibrium sequences at both coefficients.
Selected limits satisfy, for every fixed $v\ge M$,
\[
 S_1(v)=L_1(v)=\mu_1(v)=1,
 \qquad S_2(v)<1<L_2(v),\quad \mu_2(v)>1.
\]
They also satisfy $Q_2<Q_1$ and $\mu_2>\mu_1$.
The same law has positive density near its efficient cutoff and throughout
$[M,\infty)$, with gaps between its lower value bands.
\end{proposition}
\begin{proof}
\par\smallskip\noindent\emph{Parameters and menus.}
\label{aff-tail-menu}
Write $\xi=(1-\eta_2)/2$, $\kappa=1-\xi$. For sufficiently small
$\theta>0$, set
\[
 \delta=\theta^2,\quad h=.32,\quad s=.01+\delta,\quad
 a=.67-\delta,\quad v_A=11\xi/(10\theta),\quad \rho=.495.
\]
Choose small $d_0,w,\zeta>0$ and large $M>v_A+w$ below. The common law is
\[
 \begin{aligned}
 F={}&s\operatorname{Unif}[0,d_0]+h\operatorname{Unif}[2-w,2+w]\\
 &+a[(1-\zeta)\operatorname{Unif}[v_A-w,v_A+w]
                         +\zeta\operatorname{Pareto}(M,10)].
 \end{aligned}
\]
The Pareto survival is $(M/v)^{10}$. All parameters are fixed before $m$
grows. At $\eta_2$, the value-two group chooses one fair route. A common
regime $A$ of probability $1-\theta$ serves it; regime $R$ of probability
$\theta$ rejects it. The last group broadcasts. Put
\[
 \begin{aligned}
 q_A^0&=19/20,&q_R^0&=9/10,&\lambda_r^0&=1-\xi q_r^0,\\
 B&=1+\xi/50,&R&=1-37\xi/40,&\Delta&=B-1,\\
 e_A^0&=10\delta/(9a),&e_R^0&=5(h+\delta)/(4a),&
 e_r(b)&=e_r^0((B-b)/\Delta)^2.
 \end{aligned}
\]
For $1\le b\le B$, a target is exclusive with probability $1-e_r(b)$
and otherwise has independent marginals $q_r(b)=(1-\lambda_r/b)/\xi$.
With $(\omega_A,\omega_R)=(1-\theta,\theta)$, its menu is
\[
 S=1-\sum_r\omega_re_r(1-q_r)^2,\quad
 D=\sum_r\omega_re_rq_r^2,\quad T=b(S+\eta_2D),\quad L=S+D.
\]
An active bid law $G$ clears capacity at
\[
 \lambda_r=\frac{(1-\xi/2)\E_Ge_r-(\xi/2)\delta_r}{\E_G(e_r/b)},
 \qquad \delta_A=\delta/a,\quad\delta_R=(h+\delta)/a.
\]
At constant bid one this gives $\lambda_r^0$. Since
$\delta_r/e_r^0=2q_r^0-1$, bulk bids in $[1,1+u]$ and conditional
tail mass at most $2\zeta$ make the moment map uniformly close to
$\lambda^0$ as $u,\zeta\downarrow0$. Thus an arbitrarily small closed
rectangle about $\lambda^0$ is invariant.

At its center, $q_A\in[.95,.97]$, $q_R\in[.9,.92]$, and
\[
 \kappa B<R<1,\qquad \kappa B<\lambda_A^0<R<\lambda_R^0.
\]
For fixed $\xi$, as $\theta\downarrow0$, $e_A^0=O(\theta^2)$ and
$e_R^0\to40/67$. Hence $T'=1+O(\theta)$ and $T''=O(\theta)$
uniformly on $[1,B]$. Each failure term is a positive multiple of
$[(B-b)(\lambda_r/b-\kappa)]^2$. The bracket is positive, decreasing,
and strictly convex before $B$; its square has positive second derivative
through $B$. Thus $S'>0$ before $B$, $-S''\ge c_\xi\theta$, and
$T''S'-T'S''>0$ for small $\theta$. Also
\[
 \limsup_{\theta\downarrow0}v_AS'(1)
 \le\frac{11}{10}\frac{40}{67}\frac65=\frac{264}{335}<.8.
\]
Indeed the $R$ contribution to $S'(1)$ is
$\theta e_R^0(1+\lambda_R^0/5)/\xi$; the $A$ contribution times
$v_A$ vanishes. Consequently the bulk optimizes at one, while $T'/S'$
increases to infinity and gives each sufficiently large finite value a
unique optimizer $b(v)<B$. Choose $M$ above all floor thresholds in a
small multiplier rectangle. Continuity preserves these strict conditions
and the bulk derivative for small positive value widths.

Active utility is at least $v-B$. Its gap over partial one-route utility
at $v_A$ is at least
\[
 \theta v_A+(1-\theta)R-B=31\xi/200-\theta R>0;
\]
the gap increases with value. An active one-route bid has success at most
$1/2+O(\theta)$. Canonical one-route cost, for these large values, is
at least $\lambda_A^0/\kappa>B$. Canonical two-route cost is at least
$2\lambda_A^0-(\lambda_A^0)^2/v>B$, since
$2\lambda_A^0-B=1-48\xi/25\ge1/25$.
These give uniform active deviation margins for $v\ge v_A-w$.

At partial value two, the intended cost is $R+\theta(2-R)<1$.
Bids at least one give utility at most one. A canonical one-route bid
also gives at most one: $(2-1/q)(1-\xi q)$ increases to
$\kappa<\lambda_A^0$ on $0<q\le1$. For two routes,
\[
 2\lambda_A^0-(\lambda_A^0)^2/2-R
 =\tfrac12-\xi/40-361\xi^2/800\ge1199/3200.
\]
Broadcasting a partial bid adds a positive duplicate charge in $A$.
Choose the multiplier rectangle, value widths, and tail mass small enough
to preserve all these margins.

\par\smallskip\noindent\emph{The lower coefficient.}
At $\eta_1$, make the value-two band active and the entire last group
upper. Put $\xi_1=(1-\eta_1)/2$, $q_1=3/5$,
$e_1=5\delta/h$, and $z_1=1-e_1$. Then
$h(z_1+2e_1q_1)+a=.99$. At active value two, Lemma~\ref{aff-implementation}
gives
\[
 \lambda_1=\frac{2(1-q_1)}{1+z_1\xi_1/[2e_1(1-\xi_1q_1)^2]}.
\]
Write $H=1-\xi_1q_1$ and $C=H^{-1}+\xi_1(1-q_1)/(2H^2)$.
The active cost divided by $\lambda_1$ tends to $C>H^{-1}$, and
\[
 \frac1{1-\xi_1}-C
 =\frac{\xi_1(1-q_1)(2H-1+\xi_1)}{2(1-\xi_1)H^2}>0,
 \qquad 2-C\ge .18/H^2>0.
\]
Choose a floor $R_1$ strictly between the active cost and
$\lambda_1\min\{(1-\xi_1)^{-1},2\}$, with a fixed relative margin.
The hybrid first-order condition gives its normalized cost at value $v$ as
\[
 e_1(1+q)+z_1\left[\frac1{1-\xi_1q}
              +\frac{\xi_1(1-q)}{2(1-\xi_1q)^2}\right].
\]
For $v\ge v_A-w$, $\lambda_1/(e_1v)=O(v_A^{-1})$ forces $q\to1$.
This cost tends to $(1-\xi_1)^{-1}$, while the canonical two-route cost
ratio tends to two. The exclusive one-route menu also loses. Thus the
upper group prefers its singleton floor. Narrow bands preserve the
receipt, inward, and user guards of Lemma~\ref{aff-implementation}.
Lemma~\ref{aff-upper-envelope} covers the unbounded upper class: its
rank failure is exponentially small through $M$, and above $M$ is bounded
by $e^{-cm}(v/M)^{-10c_m}$. Its integrated and value-weighted failure
are therefore exponentially small.

\par\smallskip\noindent\emph{Finite continuations.}
\label{aff-tail-fallback}
At $\delta=0$, the low-group triples $(a,h,s)$ are
$(.32,.67,.01)$ and $(.67,.32,.01)$. Substituting $e^t=.16,.12$,
respectively, in Lemma~\ref{aff-body} gives $J-2I>.1397,.1101$.
The logarithm bound in Example~\ref{ex:multiple} certifies these strict
inequalities. They persist for small $\delta$ and a vanishing private
upper mass. Choose $d_0$ below every high receipt. That lemma supplies
the low equilibrium and the common minimum seed.

At $\eta_2$, privately route values above $t_m=m^{1/5}$. Their mass is
$h_m=a\zeta(M/t_m)^{10}=O(m^{-2})$; their conditional share among
private arrivals is $\pi_m=h_m/(h+h_m)$. Active bids lie in
$[1,U_m]$, with $U_m<B$ specified below, and partial bids in $[R,R+u_P]$.
The private upper floor exceeds $U_m$. Given common counts, private
compositions are independent and $m\pi_m=O(m^{-1})$.

The atypical rule uses $\kappa B+\xi Bm\pi_m<R$. If one proposer
has fewer than $c_m$ active/private arrivals, it accepts them all; the
other ranks private bids before active duplicate receipts, then low bids.
Otherwise, at unequal private totals, if all first copies fit jointly,
include all private users and use the active potential to cover the active
list and duplicate its highest bids. If first copies exceed total capacity
but the active list fits, assign all active targets exclusive owners,
with owner counts $a_1+a_2=N$ and
$\max(0,c_m-H_j)\le a_j\le c_m$; serve private uppers, owners, then
partials. If $N>2c_m$, assign only the highest $2c_m$ targets, $c_m$
to each proposer. An owner accepts with probability at least $1-m\pi_m$,
so nonowner receipts are at most $\kappa B+\xi Bm\pi_m<R$.
These inequalities check best responses at every private composition.
At equal totals, maximize the strictly concave common potential
\[
 \Psi(q)=\sum_i b_i(q_i-\xi q_i^2/2)+\mathcal R(q),
\]
where $\mathcal R$ is the concave private receipt frontier. Its unique
marginals are continuous; variational inequalities give own-bid
Lipschitz constant $1/\xi$. The low game handles residual low slots.

\par\smallskip\noindent\emph{Typical lists and the fixed point.}
\label{aff-tail-typical}
Use the seed only where low users are rejected. Independently mark each
active target mixed with probability $e_r(b_i)$, and force a rank-favored
target exclusive as in Lemma~\ref{aff-implementation}. Balance exclusive
owners against conditional mean private inclusions, namely $H_j$ in $A$
and $\pi_mH_j$ in $R$. For mixed set $I$, $M_I=|I|$,
$W_I=\sum_Ib_i^{-1}$, owner count $\ell_j$, and actual included private
count $P_j$, use independent exact-capacity lotteries with
\[
 d_j=c_m-P_j-\ell_j,\qquad
 \Lambda_j=(M_I-\xi d_j)/W_I,\qquad
 q_{ji}=(1-\Lambda_j/b_i)/\xi.
\]
At the central bid vector the mixed masses are $10\delta/9$ in $A$ and
$5(h+\delta)/4$ in $R$. The $A$ marginal is $.95$. In $R$, allowing
all private users to be upper lowers the marginal from $.9$ to
$1/2+2\delta/[5(h+\delta)]$. Thus all private compositions retain
interior marginals. In $A$, owners and private users occupy
$(.99-\delta-10\delta/9)m/2<c_m$ slots per proposer; $R$ has fewer
owners. The strict receipt guards hold on fixed neighborhoods of these
means. Independent private lists give mixed receipts
$b_i(1-\xi\E q_{-j,i})=\E\Lambda_{-j}$, between owner and nonowner
receipts and on the required side of the partial price. Mix continuously
with the atypical rule outside a smaller neighborhood.

Bounded-sum concentration, $W_I=\Theta(m)$, and owner balancing give
uniform allocation error $O(m^{-1/2})$ from the stated moment menu.
Equal private totals have probability $O(m^{-1/2})$; other exceptions
have exponentially small probability. For an unused common bid, exclude
it from exclusives. If $L_j=(|I|-\xi\E d_j)/W_I$ before inserting it,
its receipt advantage is
\[
 b-L_{-j}-\xi(b+1/W_I)p_{-j}.
\]
Reject below both prices, use
$p_j=(b-L_j)/[\xi(b+1/W_I)]$ above both before saturation, and assign
exclusively between the two prices. Balancing gives $|L_1-L_2|=O(m^{-1})$.
The intermediate case has probability $O(m^{-1/2})$ uniformly in $b$:
after multiplication by $W_I$, its test is a bounded interval for the
independent mixed flags with weights $b/b_i-(1-\xi/2)$. Near either
price a positive bulk fraction has weights bounded away from zero,
since $\lambda_r^0-(1-\xi/2)=\xi(1/2-q_r^0)\ne0$.
Their probabilities are bounded away from zero and one. Represent each
as a fair coin with positive probability. A fixed-length interval meets
each inclusion chain in a bounded number of sets, so the central
binomial coefficient bounds its probability by $O(m^{-1/2})$.
Forcing one flag changes only that interval's constant length.
Thus unused contracts converge uniformly to the canonical menus.

\label{aff-tail-fixedpoint}
Use the compact noisy fixed point of Lemma~\ref{aff-implementation}
jointly for active and partial bids, with bulk box $[1,1+u]$ and tail
box $[1,U_m]$. At equal private totals the continuous frontier above
also varies continuously with private laws; noise removes private ties.
Provisional upper bids may be replaced later, since their priority makes
lower allocations depend only on their counts.
At an active atom, a favored-target exchange preserves other allocations
when both exchanged bids coincide. If both marks would be mixed, its gain is
\[
 (v-b)(1-q_1)(1-q_2)+\eta_2bq_1q_2>0.
\]
Both mixed probabilities are at least $ct_m^{-2}$ up to the moving cap.
The gain $c\mu_b/(m^2t_m^4)$ dominates the only harmful atom jump,
$C(t_m+2B)me^{-cm}\mu_b$, from unequal-count atypical states.
Equal-count rules are continuous. Private atoms are excluded by the
positive-probability private-overload rank event. The bulk derivative
and constant limiting partial success exclude their upper bid caps.
The next step excludes the moving cap. Thus the zero-noise limits give
exact atomless best responses at every sufficiently large $m$.

\par\smallskip\noindent\emph{The moving boundary and comparisons.}
\label{aff-tail-boundary}
Uniformly on the multiplier rectangle, with
$A=\sum_r\omega_re_r^0(1-q_r(B))^2/\Delta^2>0$,
\[
 \begin{aligned}
 S(b)&=1-A(B-b)^2+O((B-b)^3),\\
 T(b)&=B-(B-b)+O((B-b)^2),\\
 b(v)&=B-(2Av)^{-1}+O(v^{-2}),\\
 v-\max_b(vS-T)&=B-(4Av)^{-1}+O(v^{-2}).
 \end{aligned}
\]
Shrink the rectangle so $A_{\max}<2A_{\min}$ and choose
$1/(4A_{\min})<k<1/(2A_{\max})$. Then $U_m=B-k/t_m$ lies above
the optimizer at $t_m$ and below its boundary-matched singleton floor.
The utility error $O(t_m/\sqrt m)=o(t_m^{-1})$ excludes a strip below
the active cap. All lower contracts have success at most
$1-c/t_m^2$: active failure has this bound, partial success tends to
$1-\theta$, and exceptional canonical marginals stay below one.
The allocation error $O(m^{-1/2})=o(t_m^{-2})$ preserves this gap.

For $v\ge t_m$, upper rank failure satisfies
\[
 1-q_H(v)\le D_m(v/t_m)^{-10c_m},\qquad
 D_m=\left[\frac{e(m-1)h_m}{2c_m}\right]^{c_m}
       =e^{-\Omega(m\log m)}.
\]
Its integrated and value-weighted bounds are $O(t_mD_m)$.
Apply Lemma~\ref{aff-upper-envelope} with
$R_m=t_m-U_{A,m}(t_m)/q_H(t_m)$, which lies above $U_m$.
It checks boundary matching and all upper bid and route deviations;
its envelope error is smaller than $\eta_2R_m(1-2D_m)$.
The earlier strict gaps check active and partial deviations. Low users
use Lemma~\ref{aff-body}. This completes both finite constructions.

Extract a convergent multiplier subsequence. Every fixed $v\ge M$
eventually lies below $t_m$ and has its unique interior optimizer.
There $e_r>0$ and $q_r>1/2$, so $S_2(v)<1$, $L_2(v)>1$, and
$\mu_2(v)>1$. The $\eta_1$ upper class has one certain limiting inclusion.
Finally, its partial-group failure is $O(e_1)=O(\theta^2)$, whereas
at $\eta_2$ that group loses mass $h\theta$. Both low groups have
vanishing service. Hence $Q_2<Q_1$ for small $\theta$, after the
narrow-band perturbations. Both markets fill capacity, so
$\mu_2=.99/Q_2>.99/Q_1=\mu_1$.
\end{proof}

\section{Exclusive proposer rewards}
\label{app:exclusive}

We prove Proposition~\ref{prop:exclusive}, using its tariffs and the base schedules of Section~\ref{sec:model}.

\subsection{A common finite construction}
\label{app:exclusive-kernel}

Put $r=1/(n-1)$, $q=1-p$, $S_k(p)=1-q^k$, $Z_k(p)=kpq^{k-1}$, and
$H_{G,k}(p)=\E G(\operatorname{Bin}(k,p))$.
The expected bid charges are $T_{\mathsf U,k}=Z_k$ and $T_{\eta,k}=(1-\eta)S_k+\eta kp$.
Omitted subscripts mean $k=n$.

\begin{lemma}[Exclusive-reward capacity lottery]
\label{lem:exclusive-kernel}
For $N>c$ positive bids, define the unique root
\[
 p_i=[1-(\lambda/b_i)^r]_+,\qquad \sum_i p_i=c.
\]
Independent feasible lotteries with these marginals are best responses under a broadcast posterior.
For $N\le c$, include all positives; reject zeros.
Against broadcasting or abstaining rivals, $(b,k)$ gives $K\mid\mathbf B_{-i}\sim\operatorname{Bin}(k,p_i)$.
\end{lemma}
\begin{proof}
Demand decreases continuously from $N$ to zero.
The marginal vector lies in the convex hull of $c$-subsets; take a measurable decomposition and average within ties.
Independent seeds give receipts $b_i(1-p_i)^{n-1}$, equal to $\lambda$ on selected coordinates and at most $\lambda$ elsewhere.
Every selected subset is optimal; for $N\le c$, all receipts are zero.
Each deviator recipient sees the same list, proving the binomial law.
The posterior broadcasts visible messages and makes omitted users abstain, feasibly and with correct on-path beliefs.
\end{proof}

\begin{lemma}[Finite atomless broadcast equilibria]
\label{lem:exclusive-narrow}
Fix $0<\rho<h\le1$, $g\ge0$, and an allowed base schedule.
For suitable $H>0,\epsilon>0$, let mass $h$ have any atomless law on $[H(1-\epsilon),H(1+\epsilon)]$; hold others at abstention.
For all large finite markets, the same primitives admit exact symmetric broadcast equilibria under $\mathsf U$ and every affine tariff, with pure monotone bidding and positive atomless bids.
Their limits satisfy
\[
 v=\lambda f_a(p(v))+g h_G(p(v)),\qquad
 h\E[p(V)\mid\text{active}]=\rho,\qquad b(v)=\lambda[1-p(v)]^{-1/r},
\]
where $h_G=H_G'/S'$ and
\[
 f_{\mathsf U}=q^{-1/r},\quad
 f_0=\frac{n-1+q^n}{nq^{2n-1}},\quad
 f_1=\frac{n-1-(n-2)q}{q^{2n-1}},\quad
 f_\eta=(1-\eta)f_0+\eta f_1.
\]
The solution is unique locally, with $K\mid v\Rightarrow\operatorname{Bin}(n,p(v))$ at almost every active value.
\end{lemma}
\begin{proof}
First fix $g=0$ and $\lambda>0$.
For success $s\in(0,1)$ on $k$ routes, put $A=-\log(1-s)$ and $y=A/k$.
The required bid is $\lambda e^{(n-1)y}$, with expected costs
\[
 C_{\mathsf U,k}=\lambda Ae^{-A}\int_0^1e^{(n-t)y}\,dt,\quad
 C_{0,k}=\lambda s e^{(n-1)y},\quad
 C_{1,k}=\lambda A\int_0^1e^{(n-1-t)y}\,dt.
\]
Costs increase strictly in $y$, so more routes reduce cost at fixed success.
The $n$-route marginal cost $\lambda f_a(p)$ increases strictly from $\lambda$ to infinity.
Each $v>\lambda$ thus has one positive optimum, with uniform gaps over lower-route actions on compact exposure intervals and over all tariffs.
Base charges perturb payoffs by at most $ng$.

Normalize $H=1$, put $p_0=\rho/h$, and set $\lambda_0=1/f_a(p_0)$ and $b_0=\lambda_0(1-p_0)^{-1/r}$.
Choose $I_a=[b_0(1-\delta),b_0(1+\delta)]$.
On an all-active list,
\[
 \lambda^r=\frac{N-c}{\sum_jb_j^{-r}},\quad
 \omega_i=\frac{b_i^{-r}}{\sum_jb_j^{-r}},\quad
 \partial_{b_i}p_i=\frac{r(1-p_i)}{b_i}(1-\omega_i).
\]
The reference payoff derivative has an $O(m^{-1})$ correction when $N$ is near $hm$.
Direct differentiation gives the reference best-bid elasticity
$\partial\log B_a/\partial\log\lambda=1-(n-1)f_a/(qf_a')\in[0,2/3]$; it is zero for $\mathsf U$.
For mixtures, $qf_a'/f_a$ lies between its endpoint ratios.
Thus rivals in $I_a$ put reference best bids within radius $3\delta/4$ for small $\delta$.
Narrow values and small $g/H$ preserve inward finite expected derivatives at both endpoints.

Uniformly over a focal bid, the rival clearing equation is
$\lambda^r=(N_H-c+p)/\sum_{j\ne i}b_j^{-r}$, with $0\le p\le1$.
Focal price effects vanish uniformly on typical counts, so the reference gaps exclude outside bids and lower-route deviations.
Affine bids above value are unprofitable.
For $\mathsf U$, cost diverges as $b\to\infty$ when $k<n$; for $k=n$, $bZ_n=np\lambda$ gives endpoint utility strictly below the optimum.

Choose $I_a=[\underline b,\overline b]$ so narrow that $(\overline b/\underline b)^r(1-c_m/m)<1$.
Coordinates are active for $N>c_m$; otherwise acceptance is constant.
Payoffs and bid derivatives are uniformly bounded, so exponentially unlikely atypical counts preserve both guards.
Scaling $H$ handles any fixed $g$.

Joint value-bid laws with the prescribed value marginal form a compact convex set.
Continuous payoffs give a nonempty, convex, compact best-response correspondence with closed graph; Kakutani gives a symmetric fixed point.
The guards remove both action restrictions.
Writing expected payoff as $vA_m(b)-C_m(b)$, smoothness and positive overload probability give $A_m'(b)>0$.
An interior bid atom would require a positive mass of values to satisfy $vA_m'(b)=C_m'(b)$, impossible under atomless values.
Convex indirect utility has only countably many nondifferentiability points; elsewhere strict increase of $A_m$ gives one optimal bid, ordered by value.

The triangular-array law of large numbers for bounded inverse bids and the vanishing own-price term give the limiting equations.
Locally $\lambda f_a'+gh_G'>0$, so exposure increases with value and decreases with price; capacity determines one limit.
Monotonicity and Lemma~\ref{lem:exclusive-kernel} give almost-everywhere convergence and the inclusion law.
Compact tariff dependence makes the choices uniform.
\end{proof}

With a once-paid base, this construction under $\mathsf U$ gives $b(v)=v-g$ and $p(v)=1-(\lambda/(v-g))^r$ in the limit.

\subsection{Near-maximal duplication under genuine scarcity}
\label{app:exclusive-scarce}

\begin{lemma}[Local marker continuation]
\label{lem:exclusive-markers}
Suppose $m-nc\ge n-1$.
A list has one bid in $(0,t)$, distinct high bids in a prescribed subset of $[t,\infty)$, and $N\le nc$ positives.
Its zero-message count $r\in\{0,\ldots,n-1\}$ can define a role for a common optimal policy under feasible off-path beliefs; role zero rejects the low bid.
\end{lemma}
\begin{proof}
Partition positives into $n$ blocks of size at most $c$.
Block zero takes the highest $\min(c,N-1)$ high bids; another takes the low bid.
Role $r$ includes block $r$ and rejects zeros.
Believe all positives broadcast; $m-N\ge n-1$ lets observed zero identities extend to $n-1$ distinct markers.
Marker $\ell$ routes to roles $\ell,\ldots,n-1$, a legal uniform-routing realization.

Conditional on the observed role, choose other roles and marker identities uniformly; observed markers receive indices at most $r$.
These relabeling-invariant beliefs give each role its observed zero count.
Assigned positives earn their full bid; unassigned positives earn zero because another role includes them.
Thus the common policy maximizes receipts, with each role determined by its local observations.
The posterior correlates feasible off-path actions under Section~\ref{sec:model}'s convention.
\end{proof}

\begin{proof}[Scarce construction for Proposition~\ref{prop:exclusive}]
\label{thm:exclusive-scarce}
Choose $p_0\in(1/n,1)$ near one and $h=\rho/p_0\in(\rho,\beta)$.
Give mass $h$ the narrow high law of Lemma~\ref{lem:exclusive-narrow}, and the remainder law $\operatorname{Unif}(g,g+L)$.
High users broadcast bids in $I=[\underline b,\overline b]$; low users abstain.
Because $(1-p_0)^{n-1}<(1-1/n)^{n-1}$, narrowing $I$ permits
$\lambda_{\rm typ}<t<\underline b(1-1/n)^{n-1}$.
Use marker policies at eligible lists with one bid below $t$ and other bids in $I$; use the canonical lottery elsewhere.
Its broadcast posterior is correct on path and preserves fallback classification across proposers.

A deviator with $0<b<t$ preserves the zero-message count, hence sees rejecting role zero when $N\le nc$.
Otherwise there are $N_H\ge nc$ high rivals, whose canonical price is at least
$\underline b(1-c/N_H)^{n-1}\ge\underline b(1-1/n)^{n-1}>t$.
Thus all bids below $t$, including zero, are rejected for every route count.

For $b\ge t$, recipients use the canonical lottery.
The rival clearing equation gives $0<\lambda_-\le\lambda\le\lambda_+<t$ on a count event with exponentially small complement, uniformly in $b$.
Hence success is at least $p_{\min}=1-(\lambda_+/t)^r>0$.
Affine tariffs cost at least $t$ per success.
For $\mathsf U$, this cost is $R_k=bkpq^{k-1}/(1-q^k)$.
The mean $b/R_k=k^{-1}\sum_{j<k}q^{-j}$ increases with $k$, so
\[
 R_k\ge R_n=\frac{n\lambda}{\sum_{j=0}^{n-1}(\lambda/b)^{jr}}
 \ge R_*:=\frac{n\lambda_-}{\sum_{j=0}^{n-1}(\lambda_-/t)^{jr}}>0.
\]
The expression increases in $b$ and $\lambda$: its price derivative leaves coefficients $1-jr\ge0$ and a positive constant.
Choose $L<\min(t,R_*)$.
A low user's payoff on the good event is at most $-p_{\min}(\min(t,R_*)-L)$ and elsewhere at most $L$.
Additional base charges are nonnegative, so every deviation is unprofitable for large $m$.

High users retain their positive-rent canonical optimum; only below-$t$ deviations change, to rejection.
Lemma~\ref{lem:exclusive-narrow} gives finite atomlessness and common value primitives.
Since $h>\rho$, capacity fills and $L_m\to\beta$; also
$\E[p\mid\text{high}]=p_0$.
Concavity gives $Q\le hS_n(p_0)$ and $\mu\ge np_0/S_n(p_0)\to n$ as $p_0\uparrow1$.
Finally $\Pr(V>g)=1>\beta$, and $v^\star=g+L(1-\beta)/(1-h)<g+L$ lies inside the low component.
\end{proof}

\subsection{Efficient and duplicative equilibria at the same primitives}
\label{app:exclusive-coexistence}

Fix $\eta_b>0$, $c_m=\lfloor\rho m\rfloor$, and $\chi=(1-1/n)^{n-1}$.
Put $d_* =\min_{2\le j\le n}g(G(j)-1)$ and
$\theta_* =\tfrac12[\tfrac14+(2\pi)^{-1}\arcsin(-\rho/(1-\rho))]>0$.
Choose $H>a>0$, $\ell<a\chi$, and $H-a<(\eta_ba+d_*)\theta_*$.
The atomless net-value law has mass $1-\beta$ in $(0,\ell)$ and mass $\beta$ in $[H,H_{\max}]$, with essential infimum $H$ but possible gaps.

High users choose one route, with
\[
 \begin{gathered}
 q_m(w)=\Pr\{\operatorname{Bin}(m-1,\Pr(W>w)/n)\le c_m-1\},\\
 U_m(w)=(H-a)q_m(H)+\int_H^wq_m(t)\,dt,
 \qquad b_m(w)=w-U_m(w)/q_m(w).
 \end{gathered}
\]
Low users abstain.
On lists with positives at least $a$, the singleton-route posterior makes highest-bid selection optimal.
Use markers for eligible lists with one bid below $a$ and the canonical broadcast continuation elsewhere; reject zeros.

The envelope verifies one-route deviations above $a$.
Since $db_m=U_mq_m^{-2}\,dq_m$, bids increase across positive-probability intervals and are atomless.
The central limit theorem gives $q_m(H)\to1/2$, hence a uniform positive rent.
Below-$a$ deviations succeed only on overloaded marker lists, with probability at most
$n\Pr\{\operatorname{Bin}(m-1,\rho)\ge nc_m\}=e^{-\Omega(m)}$.
Their utility therefore falls below that rent.

At $b=a$, higher-bid rival counts at two recipients have categorical probabilities $\rho,\rho,1-2\rho$ and correlation $-\rho/(1-\rho)$.
The bivariate central limit theorem gives joint acceptance tending to $2\theta_*$.
Acceptance increases with $b$, so every $b\ge a$, $k\ge2$ has $\Pr(K\ge2)\ge\theta_*$ for large $m$, and
$u_k(H,b)\le H-a-(\eta_ba+d_*)\theta_*=-\varepsilon<0$.
The value slope is at most one, so $u_k(w,b)\le w-H-\varepsilon$.
By bounded convergence, $R_m:=\int_H^{H_{\max}}(1-q_m(t))dt\to0$, whereas
$U_m(w)\ge w-H+(H-a)q_m(H)-R_m$.

Low bids below $a\chi$ are rejected by role zero or by the canonical price bound $\lambda\ge a\chi$ on larger lists.
Higher bids cost more per success than low net values.
Thus low users abstain, $D_m=0$, and $Q_m\to\beta$.

For coexistence, it suffices to treat $0<\zeta<n-1$.
Choose $\rho<h<\beta/(n-\zeta)<\beta$ and the fixed net-value law
\[
 (1-\beta)\operatorname{Unif}(0,\ell)
 +(\beta-h)\operatorname{Unif}(H,H+\Delta)
 +h\operatorname{Unif}(T,T+\delta).
\]
The efficient construction serves both upper components.
For a second equilibrium, only the highest component broadcasts under the scarce continuation of Section~\ref{app:exclusive-scarce}.
Scaling $T$ and narrowing $\delta$ makes its threshold exceed $H+\Delta$, while large $T/g$ preserves the guards.
Both lower components abstain because lower bids are rejected and higher bids cost more than their values.
The broadcast equilibrium fills capacity with $Q_{\rm dup}\le h$, hence $\mu_{\rm dup}\ge\beta/h>n-\zeta$.
Both sequences share fixed primitives; the efficient cutoff lies in a support gap.

\subsection{Once-paid charges force broadcast}
\label{app:exclusive-once}

\begin{lemma}[Finite routing under exclusive once-paid charges]
\label{lem:exclusive-once-routing}
If bid and base are paid once on success and $1\le c<m$, almost every positive-bid participant broadcasts in every finite atomless equilibrium.
\end{lemma}
\begin{proof}
Suppose a positive-bid, $k<n$ component has probability $s>0$ and atomless law $H$.
If all users choose it and route to a fixed $k$-subset $T$, recipients share marginals $p_i$ with $\sum_i p_i\le c<m$.
Some user $i$ therefore has a positive-$H$-measure set $J$ with
$a_i(b):=\E_{H^{m-1}}(1-p_i(b,\mathbf B_{-i}))^k>0$.

A singleton at $j\notin T$ occurs on path when rivals route to $T$ and the focal user chooses $j$ and $k-1$ members of $T$.
Its other recipients jointly reject with positive probability for almost every $b\in J$.
Bayes' rule therefore gives the singleton a positive exclusive receipt, forcing acceptance.
This holds $H$-almost everywhere because the component is absolutely continuous with respect to the observed bid law.
The original action has positive success, so individual rationality gives $w=v-g\ge b$.
Equality is impossible: almost every $b\in J$ has a smaller regular $b'\in J$, whose one-route accepting-singleton event gives positive utility.

Sending the same bid to all proposers preserves old observations and inclusions.
When all rivals route to $T$ and its recipients reject, a new recipient accepts the singleton.
With $C=\binom nk$, success increases by at least $(s/C)^{m-1}a_i(b)/C>0$.
Its utility gain is strictly positive since $w>b$, contradicting optimality.
\end{proof}

Positive submissions now form a common list; atomlessness excludes visible zeros.
Conditional on this list, $K_i\sim\operatorname{Bin}(n,p_i)$ and $\sum_i p_i\le c_m$.
Individual rationality gives $p=0$ on $\{V\le g\}$.
Conditional Jensen, followed by $K(K-1)\le n(K-1)_+$, yields
\[
 \begin{gathered}
 Q_m\le\Pr(V>g)\left[1-\left(1-\min\left\{\frac{c_m}{m\Pr(V>g)},1\right\}\right)^n\right],\\
 D_m\ge(n-1)\E p^2\ge\frac{n-1}{\Pr(V>g)}(\E p)^2
 \ge\frac{n-1}{n^2\Pr(V>g)}Q_m^2.
 \end{gathered}
\]
These give the claimed coverage gap and positive duplication at every positive-coverage limit.

\section{Ample capacity and vanishing bids}
\label{app:ample}

We prove Theorem~\ref{thm:ample} and its interior coefficient bound.
Values are atomless and finite almost surely. Write $w=v-g$, $W=V-g$,
and $U_m(w)$ for net-value utility; nonpositive values abstain.
Retain Section~\ref{sec:model}'s equilibrium class, with atomless positive
bids, and suppose $c_m/m\to\rho$ and
$\Pr(V>g)<\beta=n\rho<n\Pr(V>g)$ until the boundary subsection.
PPI proposers select their top $c_m$ positive bids. The proof therefore
also applies with observed route counts and permits correlated congestion.

\subsection{Service, base charges, and a sufficient interior coefficient}

\begin{lemma}[Full service and vanishing bids]
\label{ample-service}
Suppose $\beta>\Pr(V>g)$. Under PPI, or along balanced rules with
$a_m=\min_r A_m(r)/r\to1$, every fixed $w>0$ and every supported
best response satisfy
\[
 U_m(w)\to w,\qquad S_m(w)\to1,\qquad b_m(w)\to0,
 \qquad
 g\E[(G_m(K)-1)\mathbf1_{\{K\ge1\}}\mid w]\to0.
\]
The utility convergence is uniform on compact positive intervals.
The same conclusions hold if $a_m\ge a>0$ and
$\beta a^2>\Pr(V>g)$.
\end{lemma}

\begin{proof}
It suffices to prove the claim for $a_m\ge a>0$ with
$\beta a^2>\Pr(V>g)$: if $a_m\to1$, choose such a fixed $a<1$.
Use the quantiles of Lemma~\ref{lem:balanced-order} and extract their
limit $p$ as the quantile tolerance tends to zero. Put $P=p/a$.
A one-route priority deviation gives $U(w)\ge(w-P)_+$.

If $p>0$, the capacity and truncated-payment argument in
Section~\ref{aff-bounds} gives $L_m\to\beta$ and
$T_b\ge ap\beta=a^2P\beta$. That argument uses positive bid
quantiles and bounded bid truncations. Let $T_g$ denote excess base payment. The user identity and
$U\ge\max(0,w-P)$ give
$T_b(w)+T_g(w)=wS-U\le PS$. Thus
$T_b\le P\Pr(V>g)$, a contradiction.

Hence $p=0$ and $U_m(w)\to w$, uniformly on compact positive
intervals by the Lipschitz bound. For every supported action,
\[
 w-U_m(w)=w(1-S_m(w))+b\E A_m(K)
       +g\E[(G_m(K)-1)\mathbf1_{\{K\ge1\}}\mid w].
\]
All three terms are nonnegative, and $\E A_m(K)\ge S_m(w)$.
They give each claimed limit. Every subsequence has the same limit.
\end{proof}

For affine payments, $a=[1+(n-1)\eta_b]/n$; solving
$\beta a^2>\Pr(V>g)$ gives the stated interior coefficient bound.
If $d_g:=\min_{2\le r\le n}g(G(r)-1)>0$, Lemma~\ref{ample-service}
also gives $\Pr(K\ge2\mid w)\to0$, so $Q,L\to\Pr(V>g)$ and
$\mu\to1$.

\begin{lemma}[PPI routing with a positive duplicate base charge]
\label{ample-positive-base}
If $d_g>0$, each fixed positive net value eventually has only one-route
best responses under PPI.
\end{lemma}
\begin{proof}
Let $X\in\{0,1\}^n$ be one rival's exchangeable vector of higher-bid
contacts, with mean $\alpha$ per coordinate. Marginal acceptance bounded
below implies pair acceptance bounded below. Only $\alpha\to\rho$
needs proof: a smaller limit gives acceptance tending to one, and a
larger limit violates the marginal bound. For $n\ge3$,
$\operatorname{Corr}(X_1,X_2)\ge-1/(n-1)>-1$.
For $n=2$, $\Pr(X_1+X_2>0)\le\Pr(V>g)$ gives
$\Pr(X_1=X_2=1)\ge2\alpha-\Pr(V>g)$; thus correlation plus one
is bounded below in the limit by
\[
 \frac{3\beta-2\Pr(V>g)-\beta^2}{2\rho(1-\rho)}>0.
\]
Positivity follows by splitting $\Pr(V>g)<\beta<2$ at $\beta=1$.
The bounded triangular-array bivariate CLT therefore gives a positive
lower-left quadrant probability: the marginal cutoff is bounded below
and correlation stays above $-1$. Correlation $1$ and diverging positive
cutoffs preserve this conclusion.

Full service makes each fixed positive type's marginal acceptance at
least $1/n-o(1)$. A multiroute best reply would then incur duplicate
base payment with probability bounded below, contrary to
Lemma~\ref{ample-service}. Thus each type eventually uses one route.
Bounded convergence gives $\E k\to\Pr(V>g)<\beta$; binomial
concentration makes every positive copy accepted except with probability
$O(e^{-\kappa m})$, uniformly in bid and routes. Arbitrarily small
singleton bids give $U_m(w)\ge w(1-O(e^{-\kappa m}))$, and the
payment identity gives $b\le Cwe^{-\kappa m}$ for every supported action.
\end{proof}

\subsection{Strict submitted slack with free duplicate base charges}

Suppose $g(G(r)-1)=0$ for $r\ge1$. Put $\ell_m=\E k/n$ and
$f_m(w)=w-U_m(w)$. Let $r_{\mathrm{all},m}$ be the probability that all $n$ proposers
have at least $c_m$ positive rival submissions. A bid tending to zero
on all routes gives utility at least $w(1-r_{\mathrm{all},m})$.
Thus every supported positive action satisfies
\[
 d_m(b,k):=\Pr(K=0\mid b,k)\le r_{\mathrm{all},m},\qquad
 f_m(w)=w d_m(b,k)+bkq_m(b).
 \tag{AM.1}\label{ample-outside}
\]
The envelope also makes $f_m$ nonnegative, increasing, and concave,
with $f_m(0)=0$. In particular,
\[
 \min\{1,w/z\}f_m(z)\le f_m(w)\le\max\{1,w/z\}f_m(z)
 \quad(w,z>0).
 \tag{AM.2}\label{ample-deficit-comparison}
\]

\begin{lemma}[Positive route fractions and strict slack]
\label{ample-slack}
Along every such equilibrium sequence, there are $a_1,a_2,\eta>0$
such that eventually
\[
 \Pr(k=1)\ge a_1,\qquad \Pr(k\ge2)\ge a_2,
 \qquad \ell_m\le\rho-\eta.
\]
Consequently, $\Pr(K\ne k\mid b,k)\le ne^{-\kappa m}$ uniformly
over positive bids and routes, and $b\le2we^{-\kappa m}$ on supported
actions, for some $\kappa>0$ and all large $m$.
\end{lemma}

\begin{proof}
First suppose singleton mass vanishes along a subsequence. Participation
tends to $\Pr(V>g)$, so some $k_0\ge2$ has positive mass. Choose a
bottom $h$ bid mass of that count with at least $3h$ higher mass of the
same count. A positive part has values in $[\epsilon,H_0]$.
The first $c_m$ rivals can all have higher bids and the target's subset,
an event of probability at least $s^{c_m}$, $s=h/\binom n{k_0}>0$,
which rejects all its copies. Choose a top bid mass $\delta<h$ with
$D_{\mathrm{KL}}(\rho\Vert\delta)>-\rho\log s+1$.
It contains a supported multiroute bid $b_H$ at bounded value $w_H\le H_1$.
Its marginal rejection $r_m$ has this larger rate. Dropping to one route
at the same bid gives
\[
 b_H(k_H-1)q_H\le w_H(S_H-q_H)\le H_1r_m,
 \qquad b_H\le H_1r_m/(1-r_m).
\]
A lower-group user's gain from one route at $b_H$ is at least
$\epsilon s^{c_m}-H_0r_m-H_1r_m/(1-r_m)>0$ eventually.
This proves a singleton floor $a_1>0$. The argument also holds at
$\rho=1$ if $c_m<m$, using rate $\log(1/\delta)$.

Write $\pi_{1,m}=\Pr(k=1)$ and let $R_m$ be a rival's route count.
Taking supported bids to the singleton bid law's lower endpoint in
\eqref{ample-outside}, atomlessness gives
\[
 \Pr\{\operatorname{Bin}(m-1,\pi_{1,m}/n)\ge c_m\}
 \le r_{\mathrm{all},m}
 \le\Pr\left\{\sum_{i=1}^{m-1}R_{m,i}\ge nc_m\right\}.
 \tag{AM.3}\label{ample-lowest-singleton}
\]
If multiroute mass vanished, binomial asymptotics and Chernoff would give
\[
 \sup_{\theta\ge0}\{\beta\theta-
 \log[1-\Pr(V>g)+\Pr(V>g)e^\theta]\}
 \le D_{\mathrm{KL}}(\rho\Vert\Pr(V>g)/n).
\]
For $\beta>1$ the left side is infinite. Otherwise it is the minimum
relative entropy for all $n$ singleton-contact means to exceed $\rho$;
the right side constrains only one. The unique one-coordinate minimizer
raises that mean to $\rho$ and lowers the others below $\rho$ because
$\beta>\Pr(V>g)$. Strict entropy convexity gives the opposite strict
inequality. Thus multiroute mass also has a floor $a_2>0$.

If $\ell_m\ge\rho+\epsilon$, a bottom bid mass $\delta<\epsilon/2$
has higher-bid visibility at least $\rho+\epsilon/2$. Concentration
makes its success vanish, contradicting full service on a positive
part in a compact positive value interval. Hence $\limsup\ell_m\le\rho$.
Choose $b_H$ in the upper half of a fixed multiroute bid mass at bounded
value $w_H\le H_1$. Its higher-bid visibility is at most
$\ell_m-a_2/n\le\rho-a_2/(4n)$ eventually. Thus, for some $J>0$,
its rejection is at most $e^{-Jm}$ and the one-route comparison gives
$b_H\le H_1e^{-Jm}/(1-e^{-Jm})$.

Suppose $\ell_m\ge\rho-\delta$. At a bottom-$\delta$ bid, a rival's
higher-bid route count $X$ has $\E X/n\ge\rho-2\delta$,
$\Pr(X=1)\ge a_1-\delta$, and $\Pr(X\ge2)\ge a_2-\delta$.
Under the tilt $e^{\theta X}$, $0\le\theta\le1$, these masses stay
positive, giving $\Var_\theta(X)\ge v_*>0$ uniformly. Choose
$\theta=4n\delta/v_*<1$ with $n\theta<J/2$.
Uniform recipient subsets are preserved; each tilted coordinate mean
is at least $\rho+2\delta$. Hoeffding and a union bound make all
proposers congested with tilted probability at least $1/2$ eventually.
The reverse product density is at least $e^{-n\theta(m-1)}$, since
$\E e^{\theta X}\ge1$. A positive part of the bottom mass has values
in $[\epsilon,H_0]$ and gains at least
\[
 \tfrac\epsilon2e^{-n\theta m}-H_0e^{-Jm}
 -H_1e^{-Jm}/(1-e^{-Jm})>0
\]
by taking one route at $b_H$. This proves strict slack. Binomial
concentration for positive arrivals gives uniform acceptance. Finally,
\eqref{ample-outside} gives $bkq_m(b)\le wr_{\mathrm{all},m}$ with
$r_{\mathrm{all},m}\le e^{-\kappa m}$ and $kq_m(b)\ge1/2$ eventually.
\end{proof}

\subsection{Finite congestion rates and the limiting route system}

For a uniform size-$j$ rival route and a fixed set of $k$ recipients,
define the probability generating polynomial
\[
 P_{kj}(z)=\sum_{a=0}^k
 \frac{\binom ka\binom{n-k}{j-a}}{\binom nj}z^a,
\]
with invalid binomial coefficients equal to zero. A rival law
$\lambda=(\lambda_0,\ldots,\lambda_n)$ includes zero routes for rivals
whose bids are below the target's bid. Put
\[
 \mathcal I_k(\lambda)=\sup_{\theta\ge0}
 \left\{k\rho\theta-\log\left[1+\sum_{j=1}^n
 \lambda_j(P_{kj}(e^\theta)-1)\right]\right\}.
 \tag{AM.4}\label{ample-rival-rate}
\]

\begin{lemma}[Simultaneous congestion rate]
\label{ample-event-rate}
Suppose the marginal mean of the rival law is below $\rho$ and the
maximum number of its contacts within the selected $k$ recipients
exceeds $k\rho$. For iid rivals with route laws converging to this law,
the probability that every selected recipient has at least $c_m$
higher bids has logarithmic rate $\mathcal I_k(\lambda)$.
The rate and the probability estimates are continuous and locally
uniform in this domain.
\end{lemma}
\begin{proof}

Let $Y_k$ count a rival's selected contacts. Chernoff applied to their
total gives the upper bound. Tilt each rival vector by $e^{\theta Y_k}$.
Exchangeability makes selected means equal; a tilt slightly above the
finite optimizer makes each exceed $\rho$. Intersect simultaneous
congestion with concentration of the selected total under the tilted
law, then reverse the exact density and decrease the tilt to the
optimizer. This gives the lower rate. Finite support gives local
uniformity and continuity.

Multinomial type bounds also identify the rate with minimum relative
entropy subject to the selected coordinate means being at least $\rho$,
including boundary constraints. Symmetrization reduces it to the scalar
rate. We will use the following strict form: with positive singleton
mass, adding a coordinate constraint strictly raises a positive rate
whose first optimizer is finite. Indeed, swap an inside coordinate
with an outside coordinate. Their original weights agree; the positive
tilt weights the inside contact more, strictly so on singleton routes.
Thus each outside tilted mean is below the inside mean $\rho$.
The unique first minimizer violates the added constraint; compactness
and strict entropy convexity give the strict increase, or infinite
rate if the enlarged feasible set is empty.
\end{proof}

\begin{lemma}[Broadcast mass and the common deficit rate]
\label{ample-common-rate}
Every subsequential route-law limit $\pi$ has $\pi_0=1-\Pr(V>g)$,
$\pi_1,\pi_n>0$, and $\sum_jj\pi_j<\beta$. Define
\[
 J=\sup_{\theta\ge0}\left\{\beta\theta-
 \log\left[1-\Pr(V>g)+\sum_{j=1}^n\pi_je^{j\theta}\right]\right\}.
 \tag{AM.5}\label{ample-full-rate}
\]
It is positive and finite, with a unique finite positive optimizer.
For every fixed $w>0$,
\[
 -m^{-1}\log r_{\mathrm{all},m}\to J,\qquad
 -m^{-1}\log f_m(w)\to J.
\]
The latter convergence is uniform on compact positive value intervals.
\end{lemma}
\begin{proof}

Strict slack gives every asserted property of $\pi$ except $\pi_n>0$.
By \eqref{ample-lowest-singleton} and Chernoff,
$J\le D_{\mathrm{KL}}(\rho\Vert\pi_1/n)<\infty$.
If $\pi_n=0$, the largest occupied count $r<n$ must satisfy $r\ge\beta$.
For every $k<n$, the maximum selected count is
$\min(k,r)>k\rho$; Lemma~\ref{ample-event-rate}'s strict constraint
comparison therefore gives partial rate below $J$, including $r=\beta$.
At a sufficiently small bottom bid mass, higher-rival laws differ from
$\pi$ by arbitrarily small total variation. The rate lemma makes the
failure of its positive mass of $k<n$ actions exponentially larger than
the full-law Chernoff bound, contrary to \eqref{ample-outside}.
Hence $\pi_n>0$, and the full event has a finite optimizer and rate $J$.

The broadcast deviation gives $f_m(w)\le wr_{\mathrm{all},m}$.
For the reverse rate, take a bottom bid mass whose higher-rival full
event has rate at most $J+\epsilon$, uniformly by the rate lemma.
Choose one supported action there in a compact positive value interval;
its failure bounds its deficit below by a constant times
$e^{-m(J+\epsilon)}$. Equation~\eqref{ample-deficit-comparison}
transfers this bound uniformly to compact positive intervals.
Send $\epsilon\downarrow0$.
\end{proof}

For a positive vector $\pi=(\pi_1,\ldots,\pi_n)$, let
\[
 M_k(\theta;\pi)=1+\sum_{j=1}^k\pi_j(P_{kj}(e^\theta)-1),\qquad
 I_k(\pi)=\sup_{\theta\ge0}\{k\rho\theta-\log M_k(\theta;\pi)\}.
 \tag{AM.6}\label{ample-prefix-rate}
\]
The probability space retains classes $j>k$ with zero route contribution.

\begin{theorem}[Unique aggregate route law]
\label{ample-route-law}
The finite system
\[
 \pi_j>0,\qquad \sum_j\pi_j=\Pr(V>g),\qquad
 \sum_jj\pi_j<\beta,\qquad I_1(\pi)=\cdots=I_n(\pi)=J
 \tag{AM.7}\label{ample-system}
\]
has exactly one solution, the route and inclusion law in
Theorem~\ref{thm:ample}(ii).
Let $w_0=\operatorname*{ess\,inf}(V-g\mid V>g)$ and fix $z>w_0$.
For $C_m=f_m(z)$,
\[
 -m^{-1}\log C_m\to J,\qquad f_m(w)/C_m\to1\quad(w>w_0),
 \qquad kB_m/C_m\to1
\]
where the final convergence is in probability conditional on $V>g$.
This aggregate solution is determined by $n,\rho,\Pr(V>g)$.
\end{theorem}

\begin{proof}

Fix a route-law convergent subsequence and set
$a_m=\Pr(B_m>0)\to\Pr(V>g)$. The ascending rank
$u_m(b)=\Pr(0<B_m\le b)$ is Lebesgue distributed on $(0,a_m)$.
Its joint measures $\nu_{j,m}$ with route count $j$ are dominated by
Lebesgue measure. Extract weak limits $\nu_j$, with
$\sum_j\nu_j=du$ on $(0,\Pr(V>g))$. Their tails $p_{j,m}(u)$ converge
uniformly to $p_j(u)=\nu_j((u,\Pr(V>g)])$, by the common Lipschitz bound.

\emph{Ranks and normalized deficits.}\label{ample-ranks}\label{ample-tiers}
At $\nu_k$-almost every used rank, $p_k(u)>0$; strict slack gives its
higher-rival rate $\mathcal I_k(u)$ a finite positive optimizer.
For $\theta>0$, coupling a uniform $j$-subset with its first member gives
$P_{kj}(e^\theta)-1\ge(k/n)(e^\theta-1)$.
Removing positive rank mass therefore reduces the generating function
at the earlier optimizer: the rate strictly increases.
Equation~\eqref{ample-outside} and locally uniform event rates give
$\mathcal I_k(u)\ge J$ almost everywhere, since a violating rank
neighborhood would have failure rate below $J$. Strict increase and
atomless ranks give $\mathcal I_k(u)>J$ almost everywhere.
Exhaust the used-rank measures by compact neighborhoods with fixed
positive rate gaps. Uniform tail convergence preserves their finite
Chernoff gaps. The omitted mass tends to zero, so the common deficit
rate gives $d_m(B_m,k)/C_m\to0$ in probability conditional on $W>0$.

In a positive-probability value band between $w_0$ and $\min(w,z)$,
bounded away from zero, choose a supported action at $t_m$ with failure
$o(C_m)$. Its touching affine loss bound gives
$f_m(x)\le f_m(t_m)+(x-t_m)d_m$. Monotonicity then yields
$|f_m(w)-f_m(z)|\le(\max(w,z)-t_m)d_m=o(C_m)$, uniformly on compact
intervals above $w_0$. Discarding small lower and upper value tails gives
$f_m(W)/C_m\to1$ in probability. Bounded-value truncation and
\eqref{ample-outside}, with uniform $q_m\to1$, give $kB_m/C_m\to1$.

\emph{Prefix rates and missing classes.}\label{ample-occupied}\label{ample-missing-classes}
The bid levels $1/j$ put class $k$ in rank interval
$(\sum_{j>k}\pi_j,\sum_{j\ge k}\pi_j)$ up to vanishing exceptional mass. At its lower edge the higher-rival law is
the prefix $j\le k$. Approaching that edge through used ranks gives
$I_k(\pi)\ge J$. If strict, a normalized bid $x<1/k$ in the next tier
gap has failure $o(C_m)$ by a fixed Chernoff tilt, and total loss
$kxC_m+o(C_m)<f_m(w)$, a contradiction. Thus each occupied rate equals $J$.
If an intermediate $k$ is absent, let $r<k<\ell$ be its neighboring
occupied counts. The common prefix through $r$ and $k$ has positive
singleton mass. Lemma~\ref{ample-event-rate}'s strict added-constraint
comparison gives $\mathcal I_k>\mathcal I_r=J$.
Choose $x<1/k$ in $(1/\ell,1/r)$; the same deviation contradicts
optimality, including an infinite rate by using a sufficiently large
finite Chernoff tilt. Thus every count is positive and satisfies
\eqref{ample-system}.

\emph{Uniqueness.}\label{ample-uniqueness}
Each rate has a finite positive optimizer. The first is
$I_1=D_{\mathrm{KL}}(\rho\Vert\pi_1/n)$, strictly decreasing in $\pi_1$.
For two solutions $A,B$ with $J_A<J_B$, therefore, $D_1>0$, where
$D_k=\sum_{j\le k}(\pi_j^A-\pi_j^B)$.
At a first $D_k\le0$, put $T_j=P_{kj}(e^\theta)-1$.
These are positive and strictly increasing in $j$, by adding a uniform
unused route. Summation by parts gives
\[
 M_k(\theta;\pi^A)-M_k(\theta;\pi^B)
 =D_kT_k-\sum_{j<k}D_j(T_{j+1}-T_j)<0.
 \tag{AM.8}\label{ample-cumulative}
\]
At $B$'s optimizer this contradicts $I_k(\pi^A)=J_A<J_B$.
Thus all $D_k>0$, impossible since $D_n=0$.
Interchanging $A,B$ forces equal rates; strict decrease in each new
component then forces equal vectors. Each optimizer solves $\partial_\theta\log M_k(\theta_k;\pi)=k\rho$.

\emph{Algebraic existence.}\label{ample-system-existence}
Set $J_{\min}=D_{\mathrm{KL}}(\rho\Vert\Pr(V>g)/n)$.
For $J>J_{\min}$ recursively solve $I_k=J$ for $\pi_k>0$, within
total mass budget $\Pr(V>g)$. The zero-new-mass rate exceeds $J$ by
the strict added-constraint comparison; the positive rate decreases
continuously and strictly in the new mass. Thus each root is unique.
Large $J$ is feasible: take $\epsilon<\Pr(V>g)/(2n)$ and
$J>\max_k[-\rho\log(\epsilon/\binom nk)]$.
The trial $\pi_k=\epsilon$ has at most this rate, since the first
$c_m$ rivals can all select the same size-$k$ subset. Every root is
therefore below $\epsilon$.

Feasibility persists for larger rates. Compare feasible vector $A$ at
$J_A$ with a recursive vector $B$ at $J_B>J_A$.
Its first cumulative mass is smaller. At each later step, set $B$'s
trial new mass to equalize cumulative mass with $A$.
Equation~\eqref{ample-cumulative} makes the trial rate below $J_A$
(or zero); hence $B$'s root precedes the trial and fits the budget.

Take feasible rates decreasing to their infimum $J_*$ and extract a
vector limit. If $J_*=J_{\min}$, its first mass is $\Pr(V>g)$ and
its second is zero. The added-constraint rate then exceeds $J_*$,
contrary to lower semicontinuity of the rate, a supremum of continuous
functions. Thus $J_*>J_{\min}$ and the first limiting mass is positive.
The same argument excludes any first later zero component. Interior
rate continuity now gives every rate equal to $J_*>0$ and mean below
$\beta$. If total mass were below budget, continuity and strict decrease
in each new component would extend all roots to a smaller rate.
This proves algebraic existence, separately from finite-equilibrium existence.

Uniqueness gives full convergence; uniform acceptance transfers it to
inclusion laws. The deficit and bid conclusions also hold on every subsequence.
\end{proof}

\begin{proof}[Completion of Theorem~\ref{thm:ample}]
Lemma~\ref{ample-service} and Lemma~\ref{ample-positive-base} prove
part (i). Lemma~\ref{ample-slack} and Theorem~\ref{ample-route-law}
prove part (ii). For pay-once, Theorem~\ref{po-limit} applies because
$\rho<\Pr(V>g)$. Its positive cutoff gives strictly positive failure
at every finite served value, and it fills capacity. Thus
$Q_{\mathrm{once}}<\Pr(V>g)$ and
$\mu_{\mathrm{once}}=\beta/Q_{\mathrm{once}}>\beta/\Pr(V>g)$.
Combine this with part (ii) for part (iii).
\end{proof}

\subsection{The upper capacity boundary}

\begin{theorem}[Vanishing finite capacity shortage]
\label{ample-upper-boundary}
Suppose instead that $c_m/m\to1$ and $c_m<m$ for all sufficiently
large $m$. Retain PPI, free duplicate base charges, and the stated
atomless finite equilibrium class. Every existing sequence satisfies
\[
 \Pr(k=j),\Pr(K=j)\longrightarrow
 \frac{\Pr(V>g)\binom nj}{2^n-1}\quad(1\le j\le n),
 \qquad Q_m\to\Pr(V>g),\qquad
 \mu_m\to\frac{n2^{n-1}}{2^n-1}.
\]
Conditional on participation, this is
$\operatorname{Bin}(n,1/2)$ conditioned to be positive.
For $C_m=f_m(z)$ with any fixed $z>w_0$,
\[
 -m^{-1}\log C_m\to\log\frac{2^n-1}{\Pr(V>g)},\qquad
 f_m(w)/C_m\to1\ (w>w_0),\qquad
 kB_m/C_m\to1
\]
where the last convergence is in probability conditional on $V>g$.
\end{theorem}

\begin{proof}
The service quantile argument and the singleton-floor argument above
apply at $\rho=1$; the latter's event remains feasible because $c_m<m$.
Hence $\Pr(k=1)\ge a_1>0$ and
$\E k/n\le\Pr(V>g)-(n-1)a_1/n<1$ eventually, with an arbitrarily
small error in the participation term. Binomial concentration gives
uniform exponential acceptance. Extract a route law $\pi$ with
$\pi_1>0$ and total positive mass $\Pr(V>g)$.

Let $a_{k,m}$ be the probability that one higher-bid rival contacts
all $k$ selected recipients. Put $M=m-1$ and $d_m=M-c_m\ge0$.
Their simultaneous congestion event $F_{k,m}$ satisfies
\[
 a_{k,m}^{M}\le\Pr(F_{k,m})
 \le\Pr\{\operatorname{Bin}(M,a_{k,m})\ge M-kd_m\}.
 \tag{AM.9}\label{ample-endpoint-tail}
\]
The lower event has every rival contact all recipients; the upper bound
allows at most $kd_m$ rivals missing any contact. Since $d_m/m\to0$,
Chernoff gives rate $-\log a$ when $a_{k,m}\to a\in(0,1)$, locally
uniformly. If $a_{k,m}\to0$, fixed Chernoff tilts followed by arbitrarily
large tilts give infinite rate.
Equation~\eqref{ample-lowest-singleton} therefore forces $\pi_n>0$
and gives full-event rate $J=-\log\pi_n$.
The broadcast deviation bounds deficits above by $wr_{\mathrm{all},m}$.
At a bottom-$\delta$ bid, all rivals can belong to the higher-bid
broadcast mass, at least $\pi_n-2\delta$. A supported action there
in a compact positive value interval has deficit at least a fixed
constant times $(\pi_n-2\delta)^{m-1}$. Apply
\eqref{ample-deficit-comparison} and send $\delta\downarrow0$ to obtain
the common rate $J$, uniformly on compact positive value intervals.

Use the rank measures $\nu_j$ of Theorem~\ref{ample-route-law}. At rank $u$,
\[
 a_k(u)=\sum_{j=k}^n\frac{\binom{n-k}{j-k}}{\binom nj}
 \nu_j((u,\Pr(V>g)]).
\]
At $\nu_k$-almost every used rank, $a_k(u)>0$, and
\eqref{ample-outside} and \eqref{ample-endpoint-tail} imply
$a_k(u)\le\pi_n$. Equality has zero $\nu_k$ measure: $a_k$ is
continuous and nonincreasing, while removing $k$th-class mass reduces
it by that mass divided by $\binom nk$. Thus used-rank failure rates
exceed $J$ almost everywhere. The compact exhaustion, touching affine
loss, and value truncation arguments in Theorem~\ref{ample-route-law}
now give $d_m(B_m,k)/C_m\to0$, $f_m(w)/C_m\to1$, and $kB_m/C_m\to1$.
Their hypotheses are the strict rate gap, common deficit rate, and
uniform acceptance proved above.

Below the $k$th reciprocal bid tier, only classes $j\le k$ remain
above the bid. Their probability of contacting all $k$ recipients is
$\pi_k/\binom nk$. Thus the prefix rate is
$-\log(\pi_k/\binom nk)$, with infinite value when $\pi_k=0$.
Approaching an occupied tier from used ranks makes this rate at least
$J$. If strict, reducing the bid by a fixed fraction within the next
tier gap leaves failure $o(C_m)$ by \eqref{ample-endpoint-tail} and
reduces payment below $C_m$, a contradiction. An absent intermediate
count permits the same deviation with infinite rate, since all its
higher rivals have fewer than $k$ routes except for vanishing mass.
Hence every count is positive and $\pi_k=\binom nk e^{-J}$.
Summing gives $e^{-J}=\Pr(V>g)/(2^n-1)$; then
$\sum_k k\binom nk=n2^{n-1}$ gives multiplicity. Every subsequence
has this limit, and uniform acceptance transfers it to inclusion counts.
\end{proof}

If $c_m\ge m$, every positive bid is accepted; lowering it preserves
allocation and reduces payment. Thus every positive atomless bid profile
admits a profitable deviation.

\end{document}